\documentclass{aa}
\usepackage{natbib}
\usepackage[varg]{txfonts}
\usepackage{graphicx}

\usepackage{lscape}
\usepackage{rotating}
\usepackage{multirow}
\usepackage{hhline}
\usepackage{tabularx}
\usepackage{booktabs}
\usepackage{caption}
\usepackage{pict2e,slashbox}

\usepackage[]{textpos}   
 \usepackage{color} 

\def\Msun{\ifmmode{\mathrm M_\odot}\else{M$_\odot$}\fi}

\newcommand{\re}{\ensuremath{r_\mathrm{e}}}
\newcommand{\hd}{\ensuremath{h_\mathrm{d}}}
\newcommand{\nc}{\ensuremath{n_\mathrm{c}}}
\newcommand{\rec}{\ensuremath{r_\mathrm{e,C}}}

\begin{document}


\title{Formation of S0 galaxies through mergers}
\subtitle{Bulge-disc structural coupling resulting from major mergers }

\author{M.~Querejeta\inst{1,2}
\and M.~C.~Eliche-Moral\inst{2}
\and T.~Tapia\inst{3}
\and A.~Borlaff\inst{2}
\and C.~Rodr\'{i}guez-P\'{e}rez\inst{2}
\and J.~Zamorano\inst{2}
\and J.~Gallego\inst{2}}

\institute{Max Planck Institute for Astronomy, K\"{o}nigstuhl 17, D-69117 Heidelberg, Germany,  \email{querejeta@mpia-hd.mpg.de}
  \and Departamento de Astrof\'{i}sica y Ciencias de la Atm\'{o}sfera, Universidad Complutense de Madrid, E-28040 Madrid, Spain
  \and Instituto de Astrof\'{i}sica de Canarias, C/ V\'{i}a L\'{a}ctea, E-38200 La Laguna, Tenerife, Spain}

\date{Received ..... / Accepted .....}

\abstract {Observations reveal a strong structural coupling between bulge and disc in S0 galaxies, which seems difficult to explain if they have formed from supposedly catastrophic events such as major mergers.} 
{We face this question by quantifying the bulge-disc coupling in dissipative simulations of major and minor mergers that result in realistic S0s. }
{We have studied the dissipative $N$-body binary merger simulations from the GalMer database that give rise to realistic, relaxed E/S0 and S0 remnants (67 major and 29 minor mergers). We simulate surface brightness profiles of these S0-like remnants in the $K$ band, mimicking typical observational conditions, to perform bulge-disc decompositions analogous to those carried out in real S0s. Additional components have been included when needed. The global bulge-disc structure of these remnants has been compared with real data.} 
{The S0-like remnants distribute in the $B/T$ -- \re\ -- \hd\ parameter space consistently with real bright S0s, where $B/T$ is the bulge-to-total luminosity ratio, \re\ is the bulge effective radius, and \hd\ is the disc scalelength. Major mergers can rebuild a bulge-disc coupling in the remnants after having destroyed the structures of the progenitors, whereas minor mergers directly preserve them. Remnants exhibit $B/T$ and $\re / \hd$ spanning a wide range of values, and their distribution is consistent with observations.
Many remnants have bulge S\'{e}rsic indices ranging $1<n<2$,
flat appearance, and contain residual star formation in embedded discs,
a result which agrees with the presence of pseudobulges in real S0s. 
} 
{Contrary to the popular view, mergers (and in particular, major events) can result in S0 remnants with realistically coupled bulge-disc structures in less than $\sim 3$\,Gyr. The bulge-disc coupling and the presence of pseudobulges in real S0s cannot be used as an argument against the possible major-merger origin of these galaxies.
}

\keywords{galaxies: bulges -- galaxies: evolution -- galaxies: formation -- galaxies: interactions -- galaxies: structure}

\titlerunning{Bulge-disc structural coupling in S0s resulting from major mergers}
\authorrunning{Querejeta et al.}

\maketitle

\section{Introduction}
\label{Sec:introduction}

Traditionally regarded as a transition class between ellipticals and spirals, S0 galaxies have deserved relatively marginal attention for decades, but have recently made their way to centre stage in the context of astronomical research. It was \citet{1936rene.book.....H} who classified these lentil-shaped passive galaxies (lenticulars) as some kind of primordial spirals devoid of spiral arms and star formation; therefore, in his well-known tuning-fork diagram S0s are depicted in between elliptical and spiral galaxies. As part of an increasing interest in these astronomical objects, three independent studies have drawn attention to the fact that lenticulars constitute a heterogeneous family of galaxies, with structural and rotational properties more similar to those of spirals than to ellipticals
\citep[see][]{2010MNRAS.405.1089L,2011MNRAS.416.1680C,2012ApJS..198....2K}. 

Strenuous effort has been devoted in the last few decades, via observations and modelling, to try to understand the processes that give rise to and transform these galaxies. In any case, there are observational constraints that any evolutionary mechanisms should preserve. \citet[L10 hereafter]{2010MNRAS.405.1089L} have recently studied the photometric scaling relations in a sample of 175 early-type galaxies containing almost 120 S0s. They have found noticeable correlations between the bulge and disc photometric parameters in S0s (in particular, between their scalelengths and magnitudes), clearly pointing to a strong bulge-disc structural coupling in these galaxies. This result imposes a strong constraint to any mechanisms that try to explain the origin and evolution of lenticulars.

Observations support the idea that gas stripping due to ram pressure in clusters can effectively transform spirals into S0s \citep{2005AJ....130...65C,2006ApJ...649L..75C, 2008AJ....136.1623C, 2008MNRAS.388.1245R, 2008A&A...491..455V, 2009A&A...496..669V, 2012A&A...537A.143V, 2010ApJ...717..147S, 2010MNRAS.405.1624M, 2011AJ....141..164A}. If we also consider that the fraction of lenticular galaxies decreases with redshift, whereas the fraction of spirals increases, it is tempting to conclude that lenticulars come out of spirals which have had their gas stolen \citep[see e.g.][]{2006A&A...458..101A}. 

Nevertheless, according to \citet{2009ApJ...692..298W}, S0s are at least as common in groups as they are in clusters, and clearly more common in groups than in the less dense field. Moreover, cluster S0s are usually located within groups in the cluster, and usually exhibit traces of past mergers \citep{2009ApJ...699.1518R,2010ApJ...720..569R,2010ApJ...715..972J,2013ApJ...764L..20M}. The dominant galaxy evolution mechanims in groups are tidal interactions and mergers \citep{2014ApJ...782...53M,2014AdSpR..53..950M}, so the gas stripping mechanism can only account for the origin of a fraction of S0s. 

Similarly, internal secular evolution is often invoked to explain how S0s emerge out of spirals. This is supported by the fact that many S0s host pseudobulges \citep[][L10]{2006AJ....132.2634L, 2013pss6.book...91G} i.e. bulges with bulge-to-total luminosity ratios and concentrations typical of late-type spiral galaxies ($B/T<0.2$ and $n\sim 1$), which often contain embedded discs, inner spiral patterns, nuclear bars, some star formation, and which present rotational support more similar to spiral galaxies than to classical bulges
\citep[see][]{2004ARA&A..42..603K}. These properties seem incompatible with the smooth spheroidal structure that usually results from major merger events. So, they are usually attributed to the evolution induced by bars and other internal components in the galaxies, which direct material towards the centre, enhancing the bulge component  \citep{1990ApJ...363..391P}. However, the bulges that emerge out of gas-free bar models do not show concentrations and sizes compatible with the parameters observed in real S0s \citep[see the comparison in][EM12 and EM13 henceforth]{2012A&A...547A..48E,2013A&A...552A..67E}.

Current hierarchical models of galaxy formation assume that the bulges of S0s formed by means of a major merger of early discs or by a sequence of minor merger events, followed by a later disc rebuilding out of the left-over gas and stripped stars \citep{1999MNRAS.310.1087S}. This scenario seems compatible with the fact that most S0s reside in groups and with observations reporting merging relics in many S0s (see references above and in EM12).

Concerning the minor merger mechanism (mass ratios above 7:1), these are known to induce gentle transformations to the global structure of the progenitor \citep{2001A&A...367..428A,2006A&A...457...91E,2011A&A...533A.104E,2010MNRAS.403.1009M}. In fact, recent simulations of gas-free intermediate and minor mergers onto S0s show that these events can preserve or even enhance the structural bulge-disc coupling by triggering internal secular evolution in the surviving disc (EM12; EM13). So, in principle, minor mergers are consistent with bulge-disc coupling.

On the contrary, the bulges and discs of S0s formed through major mergers (mass ratios below 4:1) are expected to be 
structurally decoupled, as the encounters must destroy the original structure of the progenitors, rebuilding the remnant bulges and discs through independent processes (bulges from the merger and discs from later material reaccretion). This expected bulge-disc decoupling  in major mergers directly contradicts the observations of nearby S0s commented above \citep[see][L10]{2009ApJ...692L..34L}. Nevertheless, recent observational and theoretical studies support the idea that major mergers must have been relevant for the evolution of present-day massive E-S0s since $z\sim 1.5$ -- 2 \citep{2010A&A...519A..55E,2010arXiv1003.0686E,2011MNRAS.412..684B,2011MNRAS.412L...6B,2011MNRAS.411.1435T,2011ApJ...743...87W,2013MNRAS.428..999P}. Therefore, the question is whether this popular view of major mergers as catastrophic events that destroy any bulge-disc coupling in a galaxy is realistic or not. Recent studies show that discs can survive even 1:1 mergers, mostly depending on the initial gas content \citep{2009ApJ...691.1168H}, but whether major mergers can account for the bulge-disc structural coupling observed in real S0s has not yet been explored.

We address this question using \textit{N}-body dissipative simulations of major galaxy mergers provided by the GalMer project \citep{2010A&A...518A..61C}. We have also analysed the minor merger simulations onto an S0 progenitor available in the database. These simulations consider different morphological types and mass ratios for the colliding galaxies, covering a wide range of orbital parameters. We have identified the encounters that end up in relaxed remnants with realistic S0-like morphology, to determine whether the photometric structures of these merger-built S0s are compatible with those observed for real S0s or not. In a forthcoming paper (Querejeta et al., in prep.), we will analyse in detail whether our S0 remnants can additionally reproduce the photometric scaling relations reported by L10 for real S0s.

A brief description of the models is presented in Sect.\,\ref{Sec:models}. The methodological approach is summarised in Sect.\,\ref{Sec:methodology}. In Sect.\,\ref{Sec:identification} we explain the criteria used to identify what encounters result in E/S0 or S0 galaxies. Then, we simulate realistic 1D surface brightness profiles of the remnants in the $K$ band and perform multi-component decompositions (Sects.\,\ref{Sec:profiles} and \ref{Sec:decompositions}). In Sect.\,\ref{Sec:results}, we compare the characteristic photometric parameters of the bulges and discs derived from the decompositions of the S0-like remnants with the parameters of real S0s. Model limitations are commented in Sect.\,\ref{Sec:limitations}. The discussion and final conclusions derived from this study are presented in Sects.\,\ref{Sec:discussion} and \ref{Sec:conclusions}, respectively. We assume a concordance cosmology \citep[$\Omega_{M}=0.3$, $\Omega_{\Lambda}=0.7$, $H_{0}=70$~km\,s$^{-1}$\,Mpc$^{-1}$, see][]{2007ApJS..170..377S}, and 
magnitudes are provided in the Vega system.

\section{Description of models} 
\label{Sec:models}

Part of the HORIZON collaboration, GalMer\footnote{GalMer project: http://galmer.obspm.fr} is a public database containing $\sim 1000$ hydrodynamic $N$-body simulations of galaxy mergers with intermediate resolution, sampling various mass ratios, morphological properties of the progenitors, and orbital characteristics. The project and the database are thoroughly described in \citet{2010A&A...518A..61C}, so we will just provide a brief summary here. 

A representative sample of galaxy morphologies is considered, ranging from giants to dwarfs (\textit{g, i, d}) and from ellipticals to spirals (E0, S0, Sa, Sb, Sd). The orbits differ in the relative orientation of the spins of the progenitors with respect to the orbital angular momentum, the pericentral distance, and the initial motion energy. The stellar mass ratios of the encounters range from 1:1 to 20:1 depending on the progenitors (see Table\,\ref{Tab:massratios}). Up to date, the database contains 876 giant--giant interactions and 126 gS0--dwarf encounters. As we are interested in S0-like remnants, we have initially considered all the merger experiments within the database as possible candidates to give rise to an E/S0 or S0. We will comment on the selection of the dynamically relaxed S0-like remnants in Sect.\,\ref{Sec:identification}.

The progenitor galaxies are modelled using spherical non-rotating dark-matter haloes, with optional stellar and gaseous discs, and central non-rotating bulges. The E0 progenitor lacks any stellar or gaseous discs, the S0 initial model does not have a gaseous disc, and the Sd progenitor is bulgeless. The bulge-to-disc ratios are 2.0, 0.7, and 0.4 for the S0, Sa, and Sb progenitors, respectively; the ratio of dark to baryonic matter ranges from 0.43 (gE) to 3 (gSd). Haloes and bulges are constructed using Plummer spheres with characteristic mass and radius $M_\mathrm{Bulge}$ and $r_\mathrm{Bulge}$ for the bulge and $M_\mathrm{Halo}$ and $r_\mathrm{Halo}$ for the dark matter halo, as indicated in Table\,\ref{tab:morph_params}. The discs follow \citet{1975PASJ...27..533M} density profiles, with masses and vertical and radial scalelengths as described in the same Table. 
The total number of particles is 120,000 for each giant galaxy and 48,000 for the dwarfs, distributed among each galaxy component depending
on the morphology, except for the gS0 progenitor, which has 480,000 particles. This means that we have a total of 240,000
particles in the major merger experiments and 528,000 in minor merger ones (see Table\,\ref{tab:morph_params}). The gS0 progenitor is barred at the start of the simulation, whereas the other giant progenitors are not. Total stellar masses in the giant progenitors range $\sim 0.5 \mbox{--} 1.5\times 10^{11}\,\Msun$. Therefore, the final stellar mass of the remnants ranges between $\sim 1 \mbox{--} 3\times 10^{11}\,\Msun$ in the major mergers and $\sim 1.2 \mbox{--} 1.3\times 10^{11}\,\Msun$ in the minor ones, depending on the efficiency of the star formation induced by the encounter and on the masses of the progenitors. This will be relevant when comparing to real data in Sect.\,\ref{Sec:results}. 

\begin{table}
\caption{Mass ratios of the GalMer merger experiments}
\label{Tab:massratios}
{\normalsize
\begin{tabular}{ccccccc}
\hline
\\[-0.2cm]  & & \multicolumn{5}{l}{Major mergers (giant -- giant)} \\ \\[-0.3cm]
\toprule
\backslashbox[8mm]{Type 1}{Type 2}    &     & gE0 & gS0 & gSa   & gSb & gSd\\
\midrule
gE0    &      &      1:1     & --       & 1.5:1  & 3:1 & 3:1\\
gSa    &     &        --    & --       & 1:1   & 2:1   & 2:1\\
gSb    &     &        --    & --       & --   & 1:1      & 1:1\\
gSd    &     &        --    & --       & --   &  --        & 1:1\\
\bottomrule
\\[-0.2cm]  & & \multicolumn{5}{l}{Minor mergers (gS0 -- dwarf)} \\ \\[-0.3cm]
\toprule
\backslashbox[8mm]{Type 1}{Type 2}    &     & dE0 & dS0 & dSa   & dSb & dSd\\
\midrule
gS0    &    &        7:1      & 10:1       & 10:1  & 20:1        & 20:1\\
\bottomrule
\end{tabular}
\tablefoot{The present study has considered all merger simulations available from the GalMer database up to February 2014 (876 major events and 126 minor ones, 1002 experiments in total). The stellar mass ratios of the encounters depend on the morphological type of the progenitors, as indicated in this table.
}
}
\end{table}

The simulations make use of a TreeSPH technique, using the code described in \citet{2002A&A...388..826S}. Gravitational forces are calculated using a hierarchical tree method \citep{1986Natur.324..446B}, the resulting forces are then softened to a Plummer potential, and finally use smooth particle hydrodynamics to follow gas evolution \citep{1977AJ.....82.1013L,1982JCoPh..46..429G}. The softening length is fixed to $\epsilon = 280$\,pc in the giant--giant encounters, and to $\epsilon = 200$\,pc in giant--dwarf runs.  All experiments have been evolved for a total time period of 2.95 -- 3.50\,Gyr.  

One of the most important features of these simulations is that they take into account the effects of gas and star formation. Gas is modelled as isothermal ($T_\mathrm{gas} = 10^4$~K), and star formation is implemented using the method described in \citet{1994ApJ...437..611M}: a prescription for the star formation rate (SFR) is first defined, assuming the Schmidt-Kennicutt law, and then the formation of stars out of the gaseous component is implemented via hybrid particles, which contain a gas and stellar fraction that vary with time (depending on the conditions in the surroundings). Enrichment of the interstellar medium (ISM), stellar mass loss, metallicity changes, and energy injection out of supernova explosions are also considered.

For each combination of progenitors, a wide set of orbits is sampled. The initial distance between galaxies is always set equal to 100~kpc, but initial velocity is allowed to take up the discrete values 200, 300, 370 and 580 km~s$^{-1}$. Secondly, for each case, both retrograde and prograde orbits are taken into account. Orbits with different pericentre distances are also simulated (8, 16, and 24\,kpc). Finally, six possible values of inclination are considered: $i=0$, 33, 45, 60, 75, and 90$^{\circ}$.

We refer to the simulations using a notation that aims to make the initial conditions of the encounter explicit. We first refer to the morphology of the primary galaxy, followed by the morphology of the secondary (\textit{g, i, d} for giant, intermediate and dwarf; E0, S0, Sa, Sb, Sd for the corresponding Hubble types); then an ``o'' followed by the numerical identifier of the orbit used in the GalMer database, which is unique for each combination of orbital parameters under consideration. For example, gS0dSao6 means that a dwarf Sa (galaxy 2) is accreted by a giant S0 (galaxy 1), following the orbit identified as number 6 in the database (i.e. the one with an orbital inclination of 33$^{\circ}$, pericentre of 16~kpc, initial motion energy of 2.5$\times10^{4}$~km$^{2}$~s$^{-2}$, and retrograde spin-orbit coupling). 

\begin{figure*}[t!]
\center
\includegraphics[width = 0.99\textwidth]{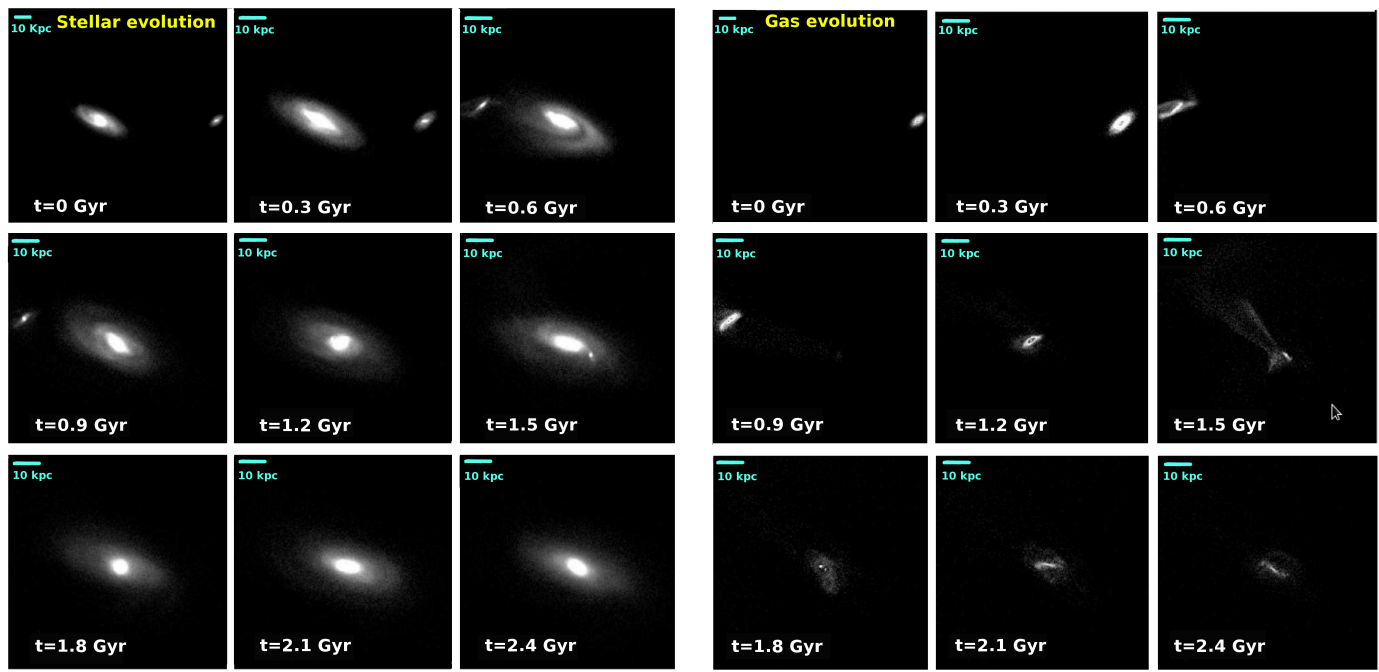}  
\caption{Time evolution of the stellar and gaseous material in the minor merger model gS0dSbo99 (left and right panels, respectively). Time is shown at the bottom left corner of each frame. At each snapshot, the line of sight has been set to $\theta = 30^\mathrm{o}$ and $\phi = -90^\mathrm{o}$ and nearly centred on the giant S0. The field of view has been increased in the frames corresponding to $t=0$ to show the original location of the dSb satellite. In this experiment, only the dwarf galaxy contains gas at the start of the simulation. This simulation has been run for a total period of 3\,Gyr. The snapshots have been obtained with the previewer of the  GalMer database.
}
\label{fig:mergerevol}
\end{figure*}

The GalMer database provides the simulations in FITS binary tables, one per stored time step of the total computed time period (intervals of 50\,Myr are considered). The tables contain the mass, position, velocity, and other relevant properties for each particle in the simulation at each time. Figure\,\ref{fig:mergerevol} represents the time evolution of the stellar and gaseous component of the accretion of a dwarf Sb by a giant S0 galaxy, obtained using the snapshots previewer of the GalMer database. 

We stress that the GalMer database only contained giant -- giant major merger encounters and minor mergers over a giant S0 progenitor when the present study was carried out.

\begin{table*}
\caption{Masses, scalelengths, scaleheights, and number of particles of the different components in the progenitor galaxies}
\label{tab:morph_params}
{\normalsize
\begin{center}
\begin{tabular}{llrrrrrrrrrrrrr} 
\toprule
\multicolumn{2}{c}{Characteristic parameters} & gE0 & gS0 & gSa & gSb & gSd & dE0 & dS0 & dSa & dSb & dSd\\
\midrule
(a)& $M_\mathrm{Bulge}$ [$2.3\times 10^{9} M_\odot$] & 70 & 10 & 10 & 5 & 0   & 7 & 1 & 1 & 0.5 & 0\\
& $M_\mathrm{Halo}$ [$2.3\times 10^{9} M_\odot$] & 30 & 50 & 50 & 75 & 75   & 3 & 5 & 5 & 7.5 & 7.5\\
& $r_\mathrm{Bulge}$ [kpc] & 4 & 2 & 2 & 1 & --                                                      & 1.3 & 0.6 & 0.6 & 0.3 & --\\
& $r_\mathrm{Halo}$ [kpc] & 7 & 10 & 10 & 12 & 15                                                 & 2.2 & 3.2 & 3.2 & 3.8 & 4.7
\\\vspace{-0.2cm}\\\hline \vspace{-0.2cm}\\
(b)& $M_\mathrm{\star,Disc}$ [$2.3\times 10^{9} M_\odot$] & 0 & 40 & 40 & 20 & 25  & 0 & 4 & 4 & 2 & 2.5\\
& $M_\mathrm{g,Disc}/M_\mathrm{\star,Disc}$ & 0 & 0 & 0.1 & 0.2 & 0.3                              & 0 & 0 & 0.1 & 0.2 & 0.3\\
& $a_\mathrm{\star,Disc}$ [kpc] & -- & 4 & 4 & 5 & 6                                                         & -- & 1.3 & 1.3 & 1.6 & 1.9\\
& $h_\mathrm{\star,Disc}$ [kpc] & -- & 0.5 & 0.5 &0.5 & 0.5                                              & -- & 0.16 & 0.16  & 0.16  & 0.16\\
& $a_\mathrm{g,Disc}$ [kpc] & -- & -- & 5 & 6 & 7                                                               & -- & -- & 1.6 & 1.9 & 2.2\\
& $h_\mathrm{g,Disc}$ [kpc] & -- & -- & 0.2 & 0.2 & 0.2                                                      & -- & -- & 0.06 & 0.06 & 0.06
\\\vspace{-0.2cm}\\\hline \vspace{-0.2cm}\\
(c) & $N_\mathrm{hybrid}$ & \multicolumn{1}{r}{--}    & \multicolumn{1}{r}{--}    & 20,000 & 40,000 & 60,000                                          &\multicolumn{1}{r}{--}    &\multicolumn{1}{r}{--}   & 8,000 & 16,000 & 24,000 \\   
     & $N_\mathrm{stellar}$ & 80,000 & 320,000 & 60,000 & 40,000 & 20,000                     & 32,000 & 32,000 & 24,000 & 16,000 & 8,000\\   
     & $N_\mathrm{DM}$ & 40,000 & 160,000 & 40,000 & 40,000 & 40,000                       &  16,000 & 16,000 & 16,000 & 16,000 & 16,000\\
\bottomrule
\end{tabular}
\begin{minipage}[t]{0.93\textwidth}{\vspace{0.2cm} \emph{Rows}: (a) Plummer sphere parameters used to model the bulges and haloes of the progenitor galaxies, as a function of the considered morphological types: $M_\mathrm{Bulge}$ is the (stellar) mass of the bulge, $M_\mathrm{Halo}$ is the total mass of the dark matter halo, while $r_\mathrm{Bulge}$ and $r_\mathrm{Halo}$ are their corresponding effective radii. (b) Parameters of the Miyamoto-Nagai density profiles used to model the gaseous and stellar discs of the different progenitors: the sub-index $\star$ denotes the stellar disc, and the sub-index $g$ refer to the gaseous disc, $M$ is mass, $a$ is effective radius, and $h$ is the vertical scalelength. (c) Number of hybrid particles (initially fully gaseous), collisionless stellar particles, and dark matter particles used for each progenitor galaxy.
}
\end{minipage}
\end{center}
}
\end{table*}

\section{Methodology}
\label{Sec:methodology}

We have first identified all the merger simulations that end up in a dynamically-relaxed remnant with morphology, structure, kinematics, SFRs and gas content typical of E/S0 and S0 galaxies (S0-like galaxies hereafter). The selection of the sample of S0-like relaxed remnants will be presented in detail in Eliche-Moral et al.\,(in prep.), so we only provide a short summary in Sect.\,\ref{Sec:identification}. 

We have simulated realistic surface brightness profiles in the $K$ band for these S0-like relaxed remnants, mimicking the observing conditions of recent samples of nearby S0s with which we have compared the products of our merger simulations (see Sect.\,\ref{Sec:profiles}). We then worked out structural decompositions for such remnants (Sect.\,\ref{Sec:decompositions}), to ultimately perform a detailed comparison of the resulting photometric parameters with those obtained for real galaxies (Sect.\,\ref{Sec:results}). This will allow us to assess whether a major merger origin of S0s is compatible with the observational constraints imposed by real S0 data or not. We have also included in this analysis the relaxed S0 remnants resulting from the minor merger experiments available from the GalMer database (gS0 -- dwarf encounters).

\subsection{Identification of S0 remnants}
\label{Sec:identification}

\subsubsection{Preselection of relaxed, apparently disc-like remnants}

From the initial sample of 1002 merger simulations available from GalMer, we rejected all merger experiments that do not result in a one-body remnant at the end of the simulation. Using the previewer of the database, three co-authors independently identified visually the models that resulted in final stellar remnants with disc components and apparently relaxed morphologies. A total of 215 major merger experiments and 72 minor merger ones were selected as candidates to give rise to relaxed S0 remnants at the end of the simulation. 
We then determined which remnants have really reached a relaxed dynamical state, obtaining two final subsamples of 173 major merger and 29 minor merger models. Time periods from full merger to the end of the simulation range between $\sim 1$ and $\sim 2$\,Gyr.

\subsubsection{Simulation of realistic images of the remnants}

Visual classification has proven to be more reliable than quantitative criteria when it comes to identifying faint structures \citep[such as external discs or spirals, see][]{1999MNRAS.308..569A,2009MNRAS.393.1324B}, so we have performed a visual morphological classification of the final remnants. 
Our intention was to make sure that we have selected the remnants that would be classified as E/S0 or S0 types by observers, to incur a fair comparison with real data. Therefore, we simulated realistic photometric images of the remnants in several broad bands ($B$, $V$, $R$, $I$, and $K$) mimicking 
 mean properties of current observational surveys of nearby galaxies, and then classify them morphologically attending to these images. In Eliche-Moral et al.\,(in prep.), we show several examples of how relevant accounting for the observational effects is to distinguish between E and S0 remnants.

The stellar mass of each particle ($\sim (3.5 -- 20.0) \times 10^5\Msun$)
was converted into light flux in the different photometric bands considering the $M/L$ ratio in the band of a stellar population with the average age and metallicity of the stellar content within the particle at each time. 
 We considered star formation histories (SFH) characteristic of each morphological type according to observations, and using the stellar population synthesis models by \citet{2003MNRAS.344.1000B}. A Chabrier initial mass function and the Padova 1994 evolutionary tracks have been used \citep{1994A&AS..106..275B}.

For the collisionless stellar particles, we have assumed SFHs that are characteristic of real galaxies of the same morphological type as the progenitor that initially hosted the particle (E, S0, Sa, Sb, or Sd), according to the parametrisations described in \citet[]{2010A&A...519A..55E}. Old stellar particles do not have an assigned age in the GalMer simulations, so we have assumed the typical age of the old stellar population located in the outer discs of nearby S0 galaxies \citep[$\sim 10$\,Gyr, see][]{2012MNRAS.427..790S,2013MSAIS..25...93S}, independently of the type of the progenitor the particle initially belonged to. The SFH experienced by the hybrid particles is different for each particle. We have approximated the SFH of each one by a simple stellar population model (SSP) with the average age and metallicity of the mass in stars contained in the particle. 

We have also simulated the effects of the typical observing conditions of current surveys of nearby S0 galaxies in our photometric images, mimicking their characteristic limiting magnitudes, signal-to-noise, spatial resolution, and seeing values in each band.  We have added photonic noise considering that the limiting magnitude in each band corresponds to $S/N=3$ in the reference observational samples.
A distance of 30~Mpc has been considered to all our remnants, as it is the average distance of the S0 galaxies within the Near-InfraRed S0 Survey\footnote{More information on NIRS0S available at: http://www.oulu.fi/\-astronomy/\-nirs0s/} \citep[NIRS0S][]{2011MNRAS.418.1452L}, which is the reference observational sample that we will base our comparisons on (see Sect.\,\ref{Sec:results}). Assuming this distance, we have transformed intrinsic physical lengths in the remnants into sky projected angular ones and we have implemented the effects of the cosmological dimming, assuming a concordance $\Lambda$CDM cosmology. We have not simulated dust extinction effects in these artificial images.

\subsubsection{Visual identification of S0-like remnants}

It is often difficult to distinguish between ellipticals and face-on lenticulars, due to the absence of prominent spiral arms; in such cases, only a break in the surface-brightness gradient can allow us to conclude that we are dealing with an S0 \citep{2009ApJ...692..298W}. Therefore, we have simulated images in each band for face-on and edge-on views to make the identification of S0s easier. By assumption, the face-on view of each remnant corresponds to the direction of the total angular momentum of its baryonic material. We have defined the edge-on view as the perpendicular direction contained in the XY plane of the original coordinates system of the simulation. Some examples of the artificial face-on and edge-on images in the $K$ band for some remnants are shown in Fig.\,\ref{fig:sbrfits}. 

The classification was performed visually by five co-authors independently. The morphological type assigned to each remnant is the median value of the five classifications. A complete agreement between all classifiers was obtained in 85\% of the major merger remnants, ensuring the robustness of the classification. In the sample of 173 relaxed major merger remnants with possible detectable discs, we finally identified 106 Es, 25 E/S0s, and 42 S0s, which correspond to the following percentages: 61.3\% of Es, 14.4\% of E/S0, and 24.3\% of S0s. We note that the elliptical galaxies in this subsample harbour a disc component detectable in their density maps, but not in realistic broad band images. All remnants from the minor merger simulations (gS0 -- dwarf encounters) are still S0s after the encounter, according to all co-authors.

The morphology of the remnants has been analysed in detail.
None of the S0-like major-merger remnants exhibits a strong bar. All minor mergers result in barred galaxies, but the progenitor gS0 has already a strong bar at the start of the simulation. The E/S0 and S0 remnants usually have lenses, ovals, and inner discs detectable in the images. We confirmed the morphological classification by analysing realistic simulations of their 1D surface brightness profiles in the $V$, $R$, and $K$ bands (for more information, see Sect.\,\ref{Sec:profiles}). All of them presented clear bulge-disc structures. 
 We have also derived the final rotational support, star formation rates and final gas content of these S0-like remnants, and we have compared them to typical values measured in real S0s. We find remnants to be consistent with S0s also according to these properties.


 Our S0-like remnants have a median (B-R) colour of $\sim 0.9$, whereas local ETGs of similar masses typically have (B-R) colours of $\sim 1.5$ \citep{2001ApJ...550..212B}. This difference in the (B-R) colour is due to the youth of the starbursts induced by the encounters in the centres of the remnants. This makes the average $K$-band $M/L$ ratios used in our models to be $\sim 2$ times smaller than the typical ones in quiescent early-type galaxies, and thus our remnants are $\sim 1$\,mag brighter in $K$-band than real S0s with similar stellar masses. In any case, blue central structures are quite common in nearby E-S0 \citep[in fact, they are usually considered as evidence of recent merging, see ][]{2009AJ....138..579K,2010A&A...515A...3H,2010ApJ...708..841W}, and the remnants would exhibit analogous colours to those observed in present-day S0s with $\sim 1$ -- 2\,Gyr of additional passive evolution. There are other sources of uncertainty in the determination of the $M/L$ ratios used in the models, which are inherent to the assumptions adopted to estimate them, such as the age assigned to the collisionless stellar particles, the SFHs, or the considered IMF. 
We will take this into account for the interpretation of the results in Sect.\,\ref{Sec:results}.

\begin{figure*}[!t]
\center
\includegraphics[width=0.92\textwidth]{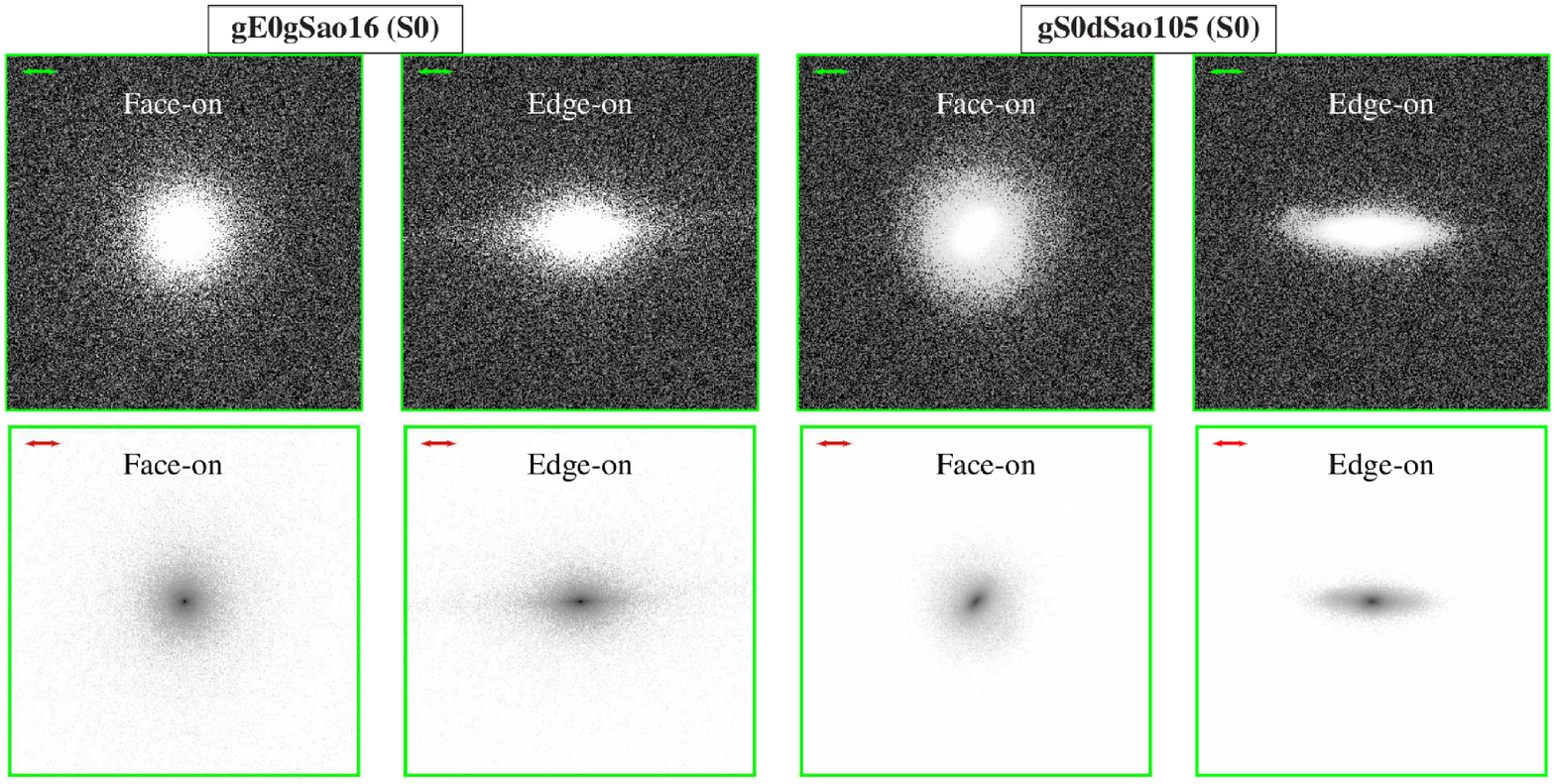}
\includegraphics[width=0.40\textwidth, bb = 40 170 458 640, clip]{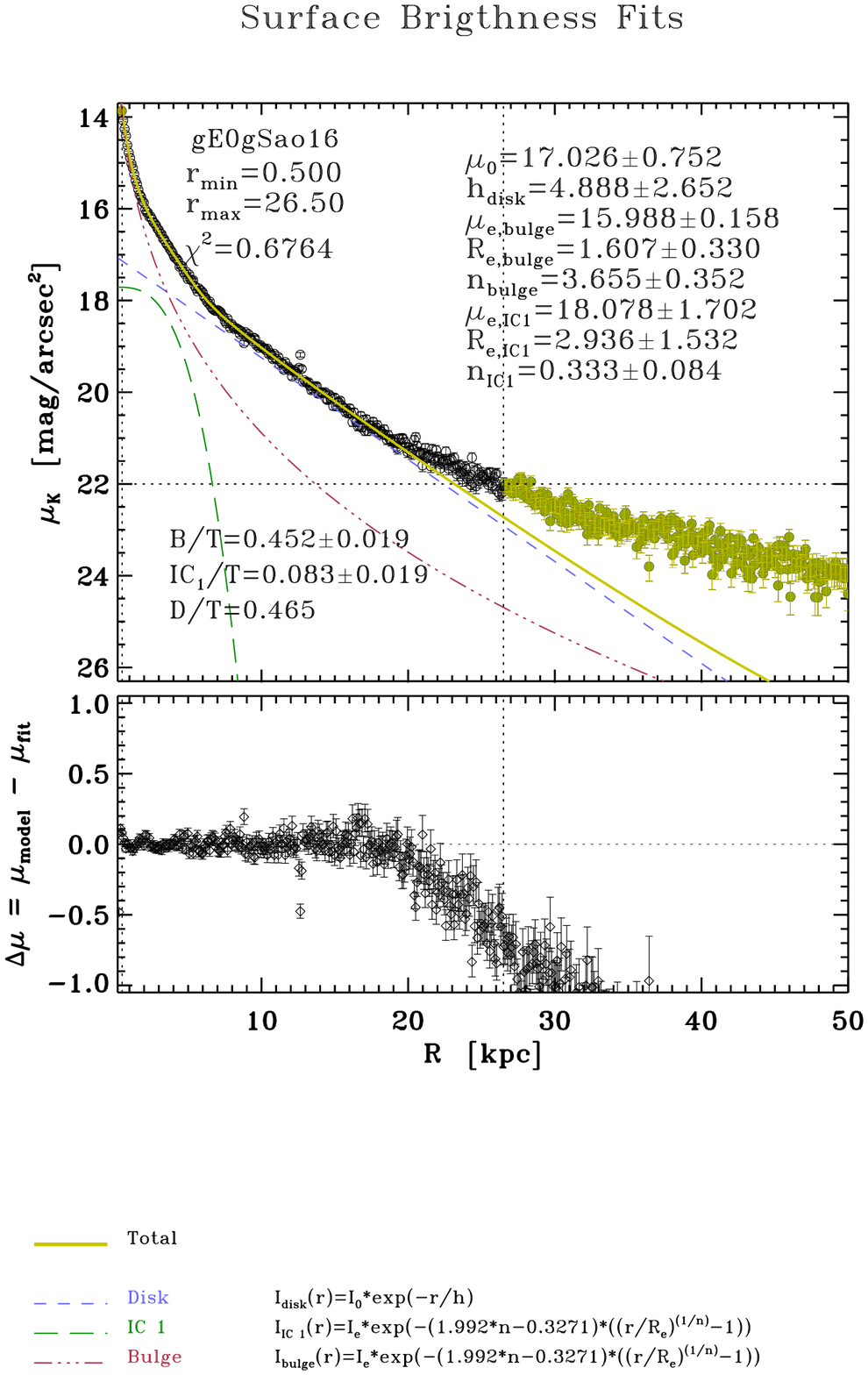}
\includegraphics[width=0.40\textwidth, bb = 40 170 458 640, clip]{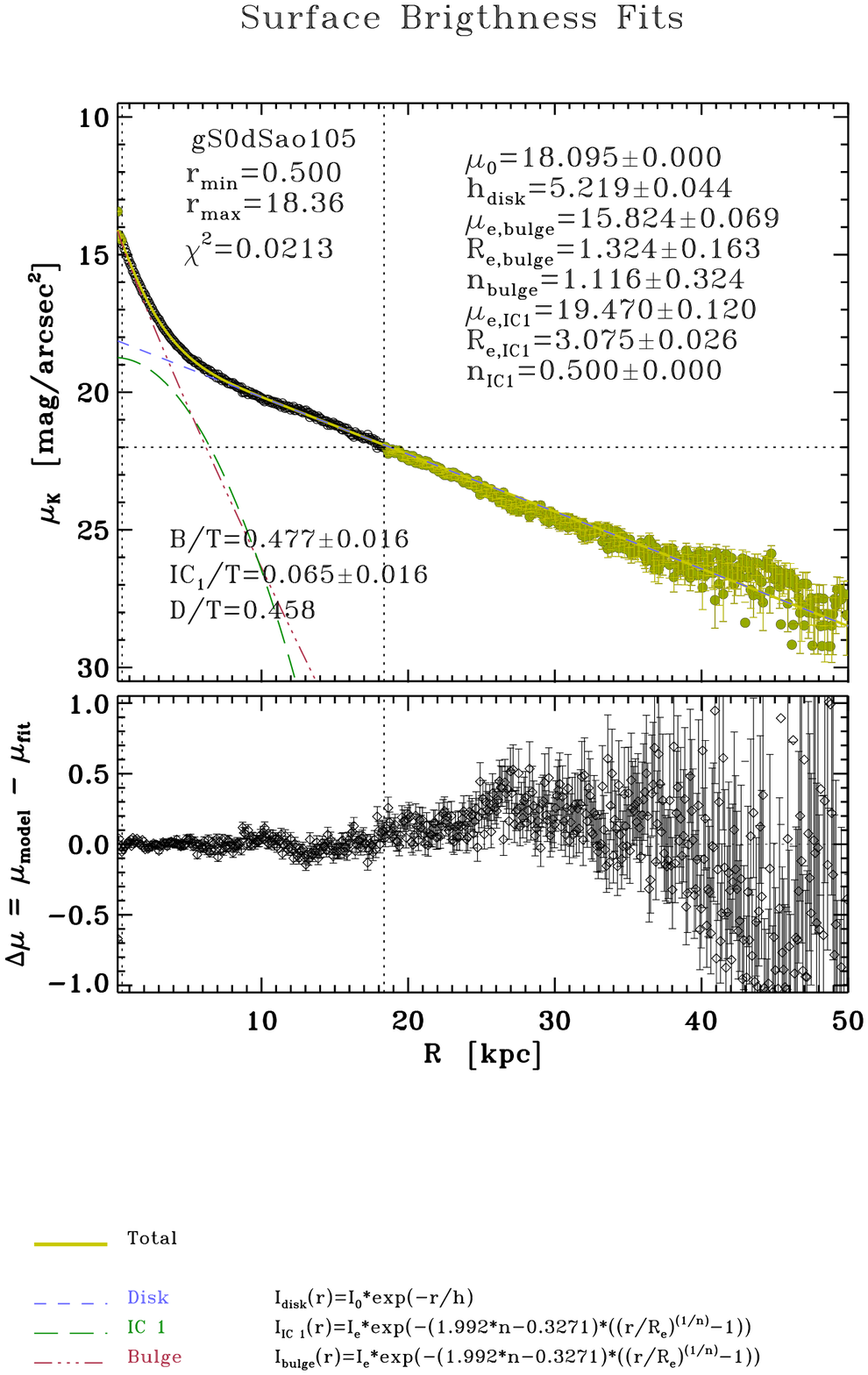} 
\caption{Simulated $K$-band images and radial surface brightness profiles of some S0-like remnants from our sample of major and minor mergers. The simulations assume $D = 30$\,Mpc, $\mu_{K,lim}= 22$\,mag\,arcsec$^{-2}$ for $S/N=3$, and a spatial resolution of 0.7\arcsec. \textbf{First two rows of panels}: Simulated $K$-band images of the final remnants for face-on and edge-on views. The horizontal arrow at the top left of each panel represents a physical length of 5\,kpc. The field of view is 50\,kpc~${\times}~$50\,kpc. We have used different logarithmic greyscales to highlight the structure of the outer discs (first row of panels) or of the central bulges (second row of panels). \textbf{Third row of panels}: Simulated radial $K$-band surface brightness profiles and multicomponent decompositions performed to them. \emph{Dotted horizontal line}: Limiting surface brightness of the images. \emph{Dotted vertical lines}: Minimum and maximum radii considered in the fit. \emph{Black empty circles}: Data 
considered in the fit. \emph{Green filled circles}: Data excluded from the fit. \emph{Red dotted-dashed line}: Fitted S\'{e}rsic bulge. \emph{Blue dashed line}: Fitted exponential disc.  \emph{Green long-dashed line}: Additional S\'{e}rsic component required in the fit (representing components such as ovals, bars, lenses, or embedded inner discs). \emph{Solid light green line}: Total profile resulting from the fit. \textbf{Fourth row of panels}: Residuals of the fits as a function of radial location in the galaxy. 
[\emph{A colour version is available in the electronic
edition.}]
}
\label{fig:sbrfits}
\end{figure*}

\begin{figure*}[!]
\center
\includegraphics[width=0.92\textwidth]{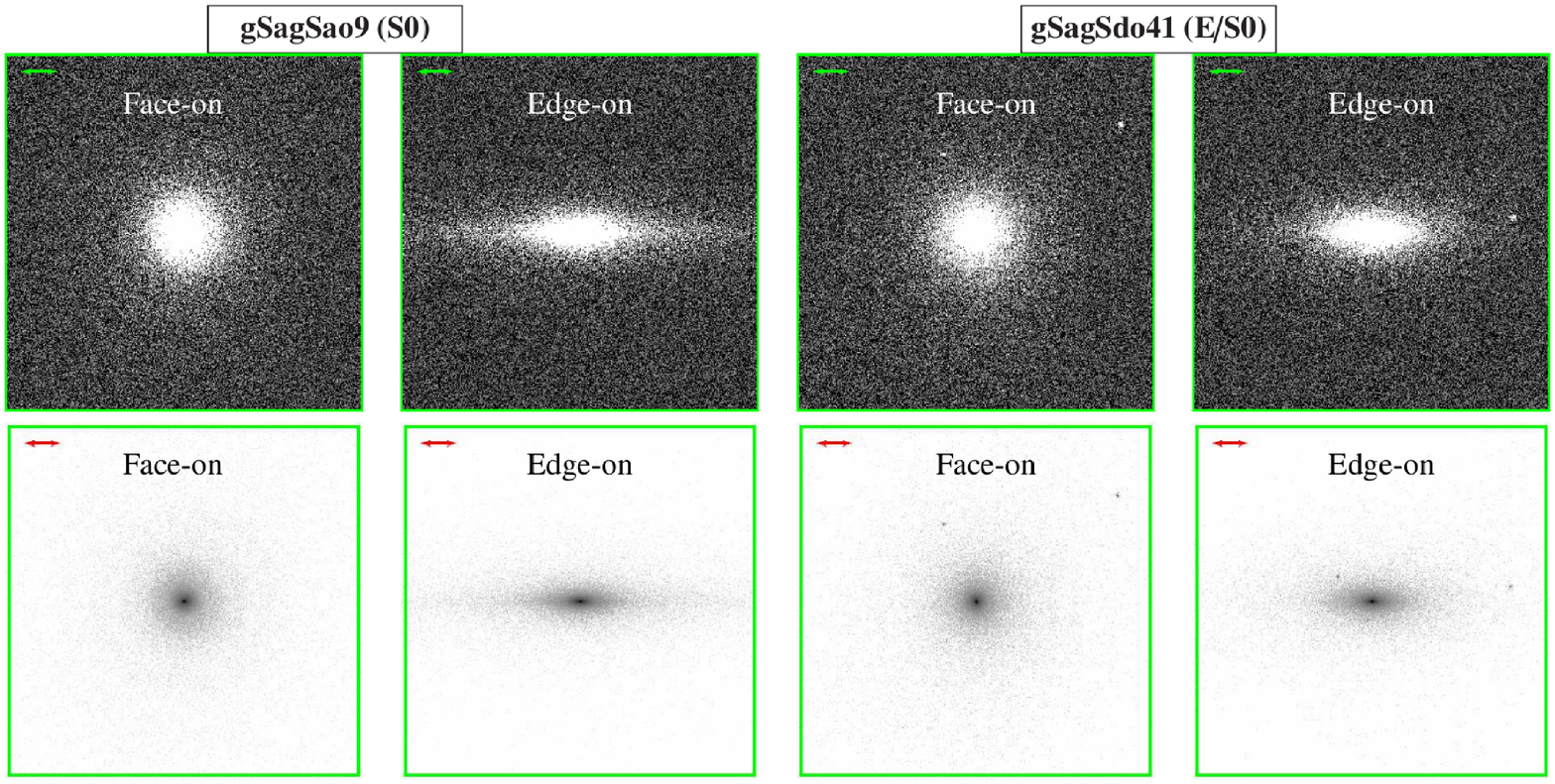}
\includegraphics[width=0.40\textwidth, bb = 40 170 458 640, clip]{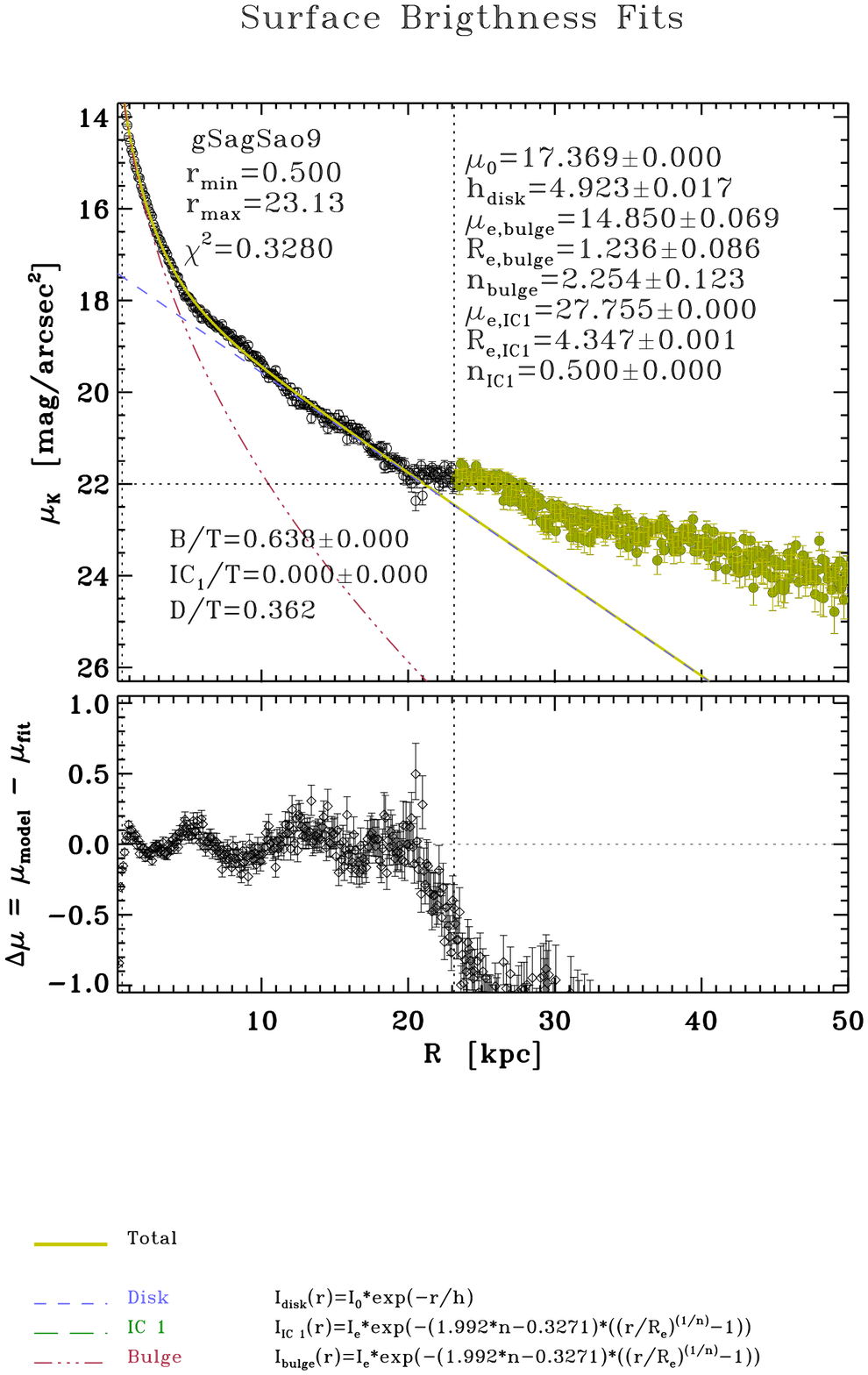}
\includegraphics[width=0.40\textwidth, bb = 40 170 458 640, clip]{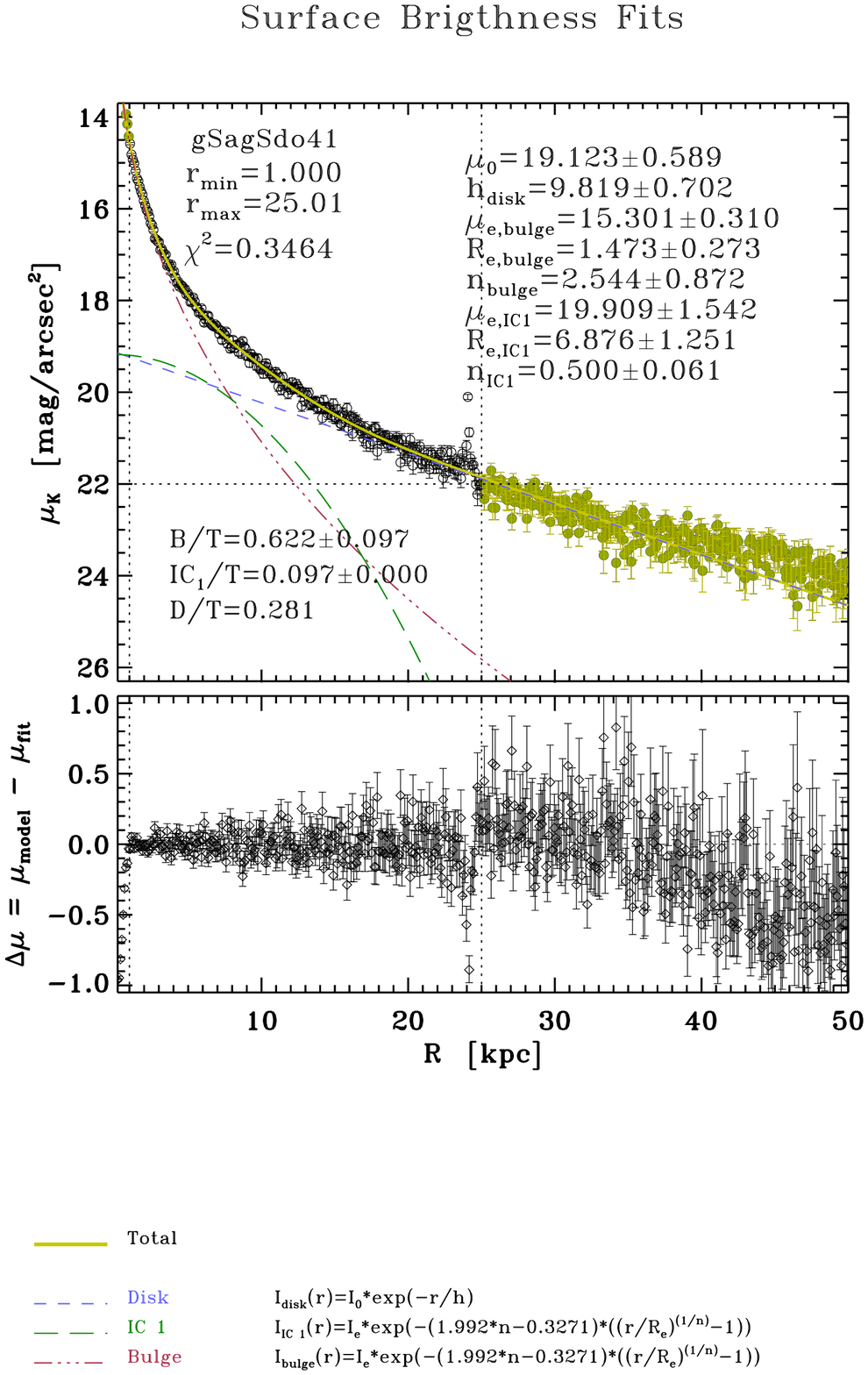} 

\addtocounter{figure}{-1}
\vspace{0.2cm}
\caption{Continued.}
\end{figure*}

\begin{figure*}[!]
\center
\includegraphics[width=0.92\textwidth]{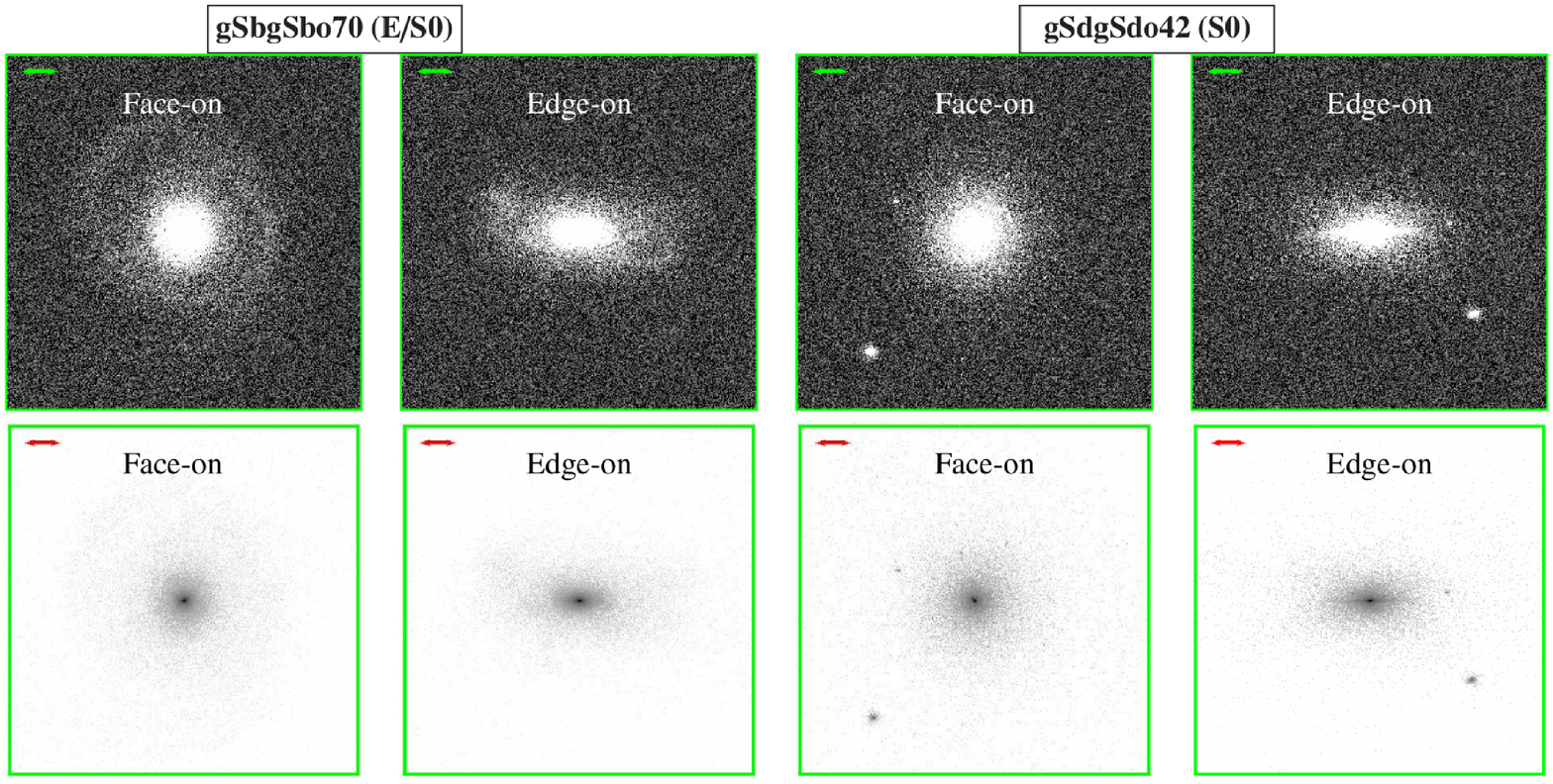}
\includegraphics[width=0.40\textwidth, bb = 40 170 458 640, clip]{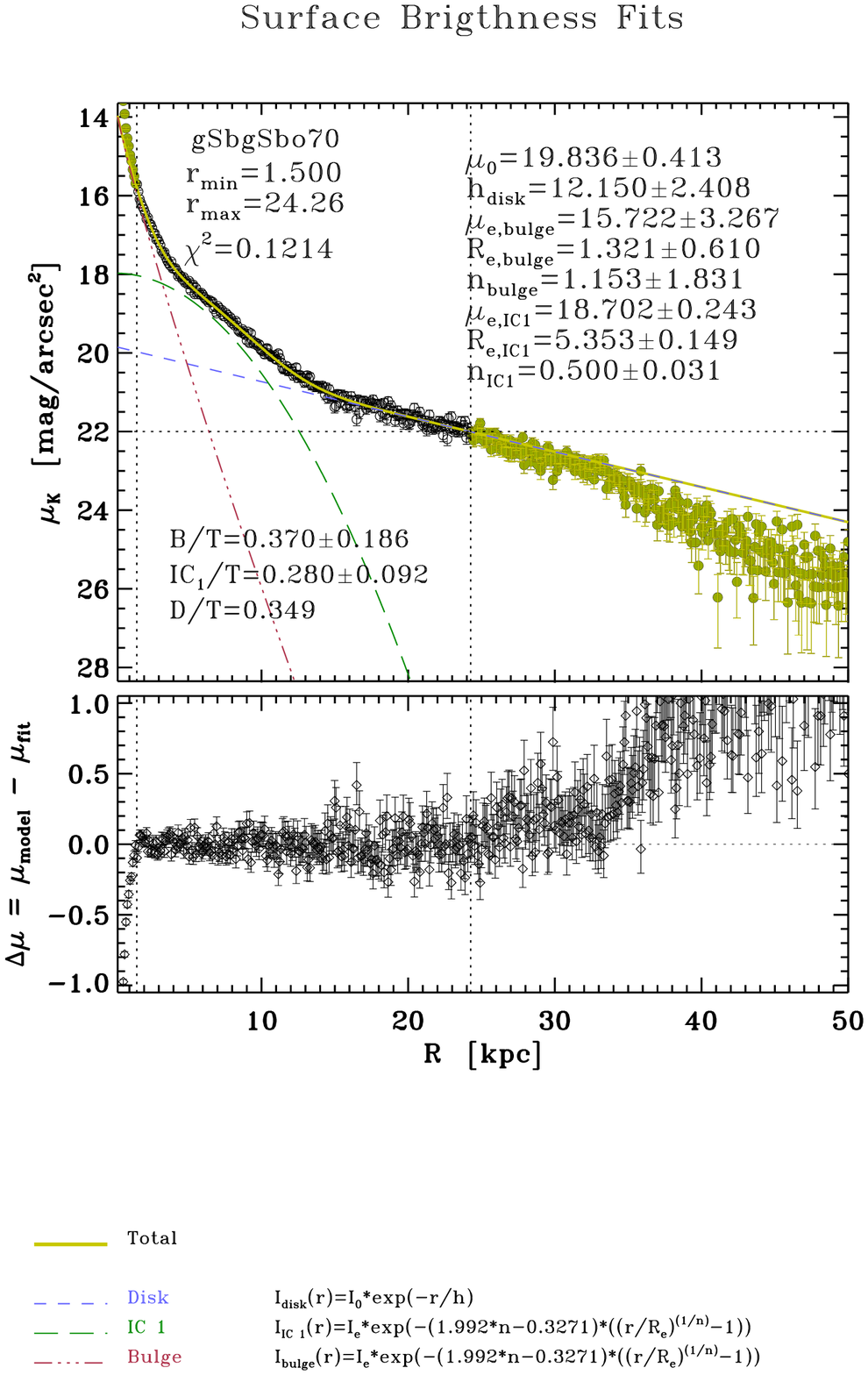}
\includegraphics[width=0.40\textwidth, bb = 40 170 458 640, clip]{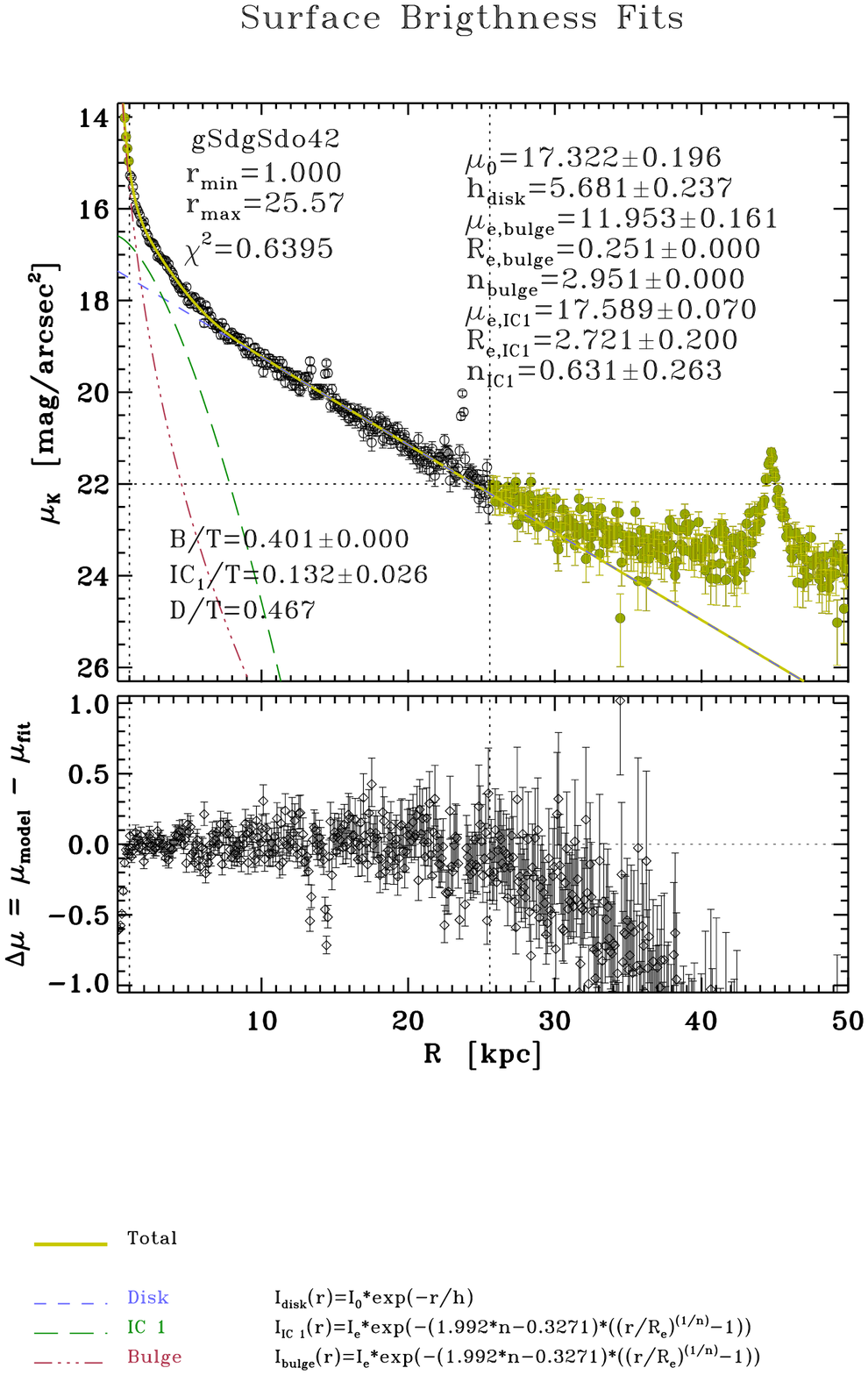} 
\addtocounter{figure}{-1}
\vspace{0.2cm}
\caption{Continued.}
\end{figure*}

\subsection{Simulation of realistic surface brightness profiles}
\label{Sec:profiles}

\subsubsection{Simulating the conditions of NIRS0S}

We have simulated realistic surface brightness profiles of our S0-like remnants reproducing the typical observing conditions in the photometric band of the NIRS0S data, in order to compare the parameters of the bulges and discs in our remnants with those exhibited by real S0 galaxies contained in that sample (L10). 

The $K$-band images in NIRS0S have an average depth of $\mu_K\sim 22$\,mag\,arcsec$^{-2}$ for a limiting signal-to-noise ratio of $S/N =3$. We have adopted this limiting surface brightness for the same $S/N$ to simulate the surface brightness profiles. The projected spatial resolution has been set to the average of NIRS0S data ($\sim 0.7$\arcsec). It is equivalent to $\sim 100$\,pc for the distance considered for our remnants ($D=30$\,Mpc, see Sect.\,\ref{Sec:identification}), which is even lower than the softening length used in the simulations ($\epsilon = 280$\,pc in the major mergers and $\epsilon = 200$\,pc in minor events). Therefore, the minimum spatial resolution is ultimately set by $\epsilon$ in each case. As commented in Sect.\,\ref{Sec:identification}, we have also included photonic noise to the data considering the limiting magnitude indicated above for $S/N=3$. No dust extinction effects have been included, as they are expected to be negligible in the $K$ band.

L10 obtained their parameters from 2D multi-component decompositions to deep $K$-band images of the galaxies, not from 1D surface brightness profiles.  We will discuss next why this strategy is problematic when applied to the simulated images from GalMer, and will justify our decomposition method based on 1D azimuthally-averaged surface brightness profiles. In any case, as we will see in \ref{Sec:1Dprofiles}, the 1D profiles clearly show a `three-zone' structure very similar to that found in observations like NIRS0S.

\subsubsection{Optimal decomposition method: 1D vs 2D}

We initially reproduced the procedure from L10 by carrying out multi-component decompositions to our artificial $K$-band images for face-on views of the remnants using  GALFIT\footnote{GALFIT home page: http://users.obs.carnegiescience.edu/\-peng/\-work/\-galfit/\-galfit.html}, a highly efficient algorithm for 2D fitting of analytic functions to digital images of galaxies \citep{2002AJ....124..266P,2010AJ....139.2097P}. However, we found several problems when comparing with real data. 

\begin{figure*}[t!]
\center
\includegraphics[width = 0.92\textwidth]{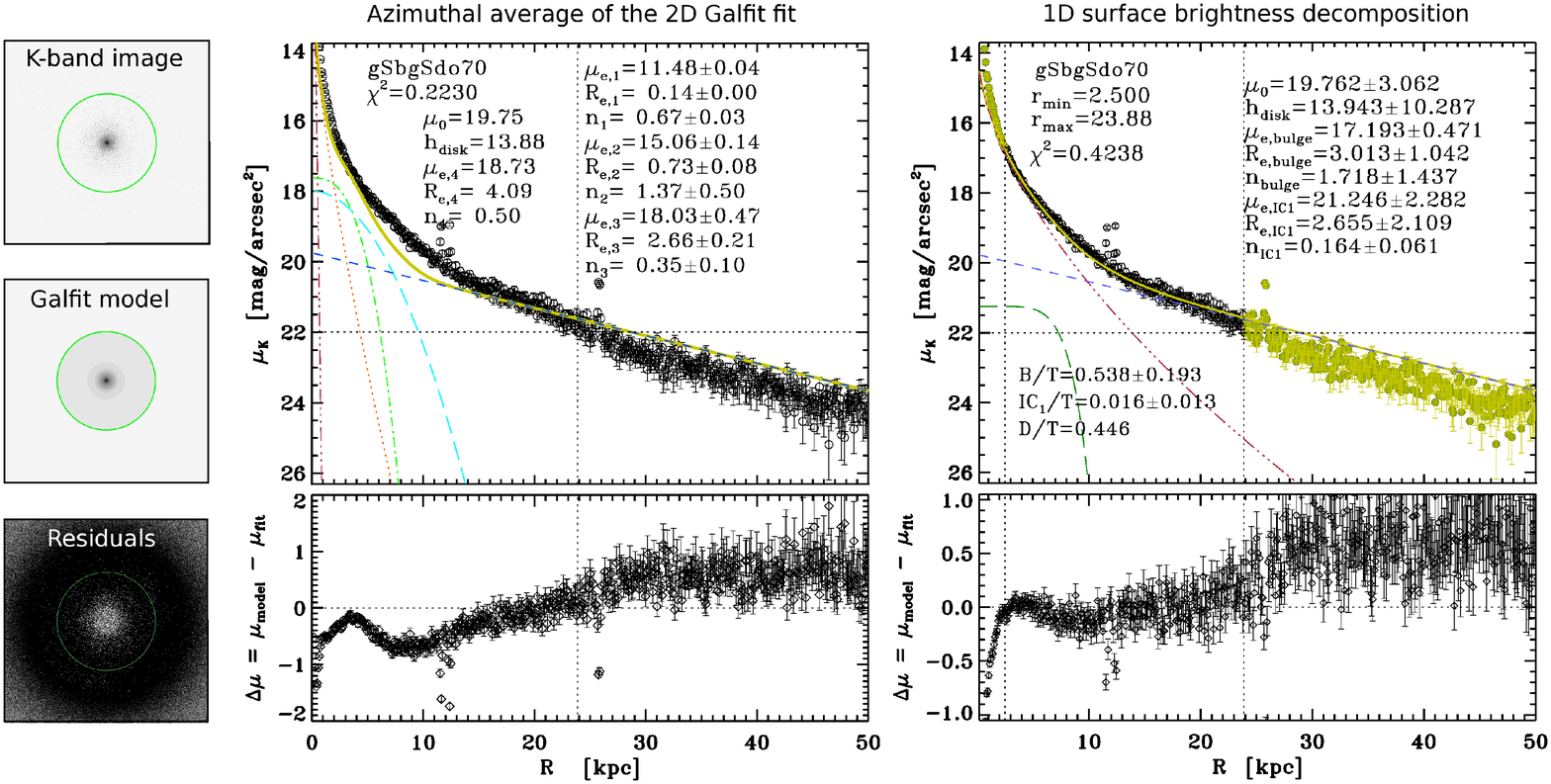}  
\caption{Comparison between the best GALFIT multi-component decomposition of the 2D $K$-band image of the remnant gSbgSdo70 and the corresponding 1D decomposition.  \textbf{Left panels}: Artificial $K$-band image of the remnant for a face-on view (\emph{top}), 5-component model obtained with GALFIT (free of noise, \emph{middle}), and residuals resulting from the subtraction of the model to the image (\emph{bottom}). The same logarithmic greyscale is used in the two upper panels, whereas the linear greyscale in the bottom panel emphasises the
background noise at the outskirts. The field of view corresponds to 100\,kpc$\times$100\,kpc, and the green circle represents $R=25$\,kpc in the galaxy. \textbf{Middle panels}: Azimuthal average of the GALFIT model (\emph{green solid line}) compared to the 1D surface brightness profile (\emph{circles}). The five axisymmetric S\'{e}rsic components required by the modelling are plotted in the top panel: a nuclear component (component 1, \emph{red dotted-dashed line}), a thin inner disc with $n = 1.37$ (component 2, \emph{orange dotted line}), a lense with $n = 0.35$ (component 3, \emph{green dotted-dashed line}), a fixed extended lense component with $n = 0.5$ (component 4, \emph{light blue long-dashed line}), and the outer exponential disc, fixed to the result obtained in the 1D fit (\emph{dark blue dashed line}). The residuals are plotted as a function of radius in the bottom panel. \textbf{Right panels}: For comparison, 1D decomposition obtained fitting the azimuthally-averaged $K$-band surface brightness profile. The legend is the same as in Fig.\,\ref{fig:sbrfits}. [\emph{A colour version is available in the electronic edition.}]}
\label{fig:galfit}
\end{figure*}

One of the problems was that the decompositions required too many components in the centre to reproduce the original image realistically ($\sim 3$--5 subcomponents beside the bulge and the disc, see an example in Fig.\,\ref{fig:galfit}), whereas the decompositions performed to the S0 galaxies in NIRS0S required $\sim 2$ at most (L10). Real S0s exhibit a more diffused and smooth appearance than our remnants in $K$-band images, in the sense that they look less structured in inner components at the central regions. However, we know that the basic structure of the bulges and discs in our S0-like remnants are consistent with those observed in real S0s in optical bands (Eliche-Moral et al., in prep.),  so the problem was not that the central structure in our remnants was unrealistic, but that the bulge substructures were much more noticeable in our simulated $K$-band images than in real cases. 
Therefore, the flux of the bulge component in the remnants got divided into several subcomponents in our 2D decompositions. As commented in Sect\,\ref{Sec:identification}, this effect would probably disappear by simply allowing the remnants to evolve passively for a few Gyr more.

The fact that we have not included dust extinction in our simulated images may also have an effect here, as it would contribute to blur the appearance of the bright and young inner components formed at the centre of the remnants. In fact, dust lanes are frequent central features in nearby S0s \citep[][]{2010A&A...519A..40A,2010MNRAS.407.2475F}. Additionally, present-day massive S0s must have evolved passively for much longer periods of time than our remnants \citep{2009MNRAS.393.1467F,2010A&A...519A..55E,2012MNRAS.427..790S,2013A&A...558A..23D,2013MNRAS.428..999P,2014arXiv1403.4932C}; so nearby S0s tend to be dynamically more relaxed and mixed, and their young stellar populations are dimmer. Consequently, we would expect present-day S0s to look more fuzzy than our S0-like remnants. 

In addition, the simulated remnants present an inherent grainy appearance at the galaxy outskirsts due to the high mass of the stellar particles in the simulation ($\sim (3.5 -- 20.0) \times$\,$10^5\Msun$).
At large radii, light accumulates at the pixels where the particles are located, instead of displaying a homogeneous distribution in space (see Fig.\,\ref{fig:sbrfits}). Since light distributed in disjointed pixels cannot be identified as a unique component by fitting codes such as GALFIT, the scalelength of the outer discs in our remnants is significantly underestimated.

Fig.\,\ref{fig:galfit} illustrates this problem for the remnant of gSbgSbo70. The original $K$-band image, the best GALFIT model, and its residuals are shown on the left panels. We used five S\'{e}rsic components to generate a model with a similar appearance to the original image, trying to minimise the residuals. However, we had to fix the outer disc parameters to those obtained from the exponential fit to the 1D surface brightness profile, because the light (mass) was distributed in isolated pixels towards the outskirts, making it difficult for GALFIT to fit a continuous light distribution there. This is noticeable in the pixels with positive residuals at large radii and the extended circular region with negative residuals (which is tracing the modelled disc). This problem also affected the inner disc region (fitted with the fourth GALFIT component). The central panels of Fig.\,\ref{fig:galfit} show the 1D profile derived by azimuthally averaging the 2D GALFIT model, compared to the 1D surface brightness profile of the remnant. The model clearly loses light in the inner disc (traced by component 4 in the GALFIT model). The problem of the too complex structure of the bulges in the simulated images is also illustrated by this panel. The GALFIT model loses light in the centre compared to the 1D surface brightness profile (we note also the positive residuals at the centre of the remnant, in the leftmost panel).

We found that these problems could be solved by performing multi-component decompositions based on 1D face-on azimuthally-averaged surface brightness profiles instead of performing them on 2D photometric images. The 1D profiles already average the 2D spatial information at each radius, blurring the complex central structure of the remnant bulges. They also improve the signal-to-noise at the outskirts compared to 2D surface brightness maps, avoiding the problem of the granularity of the discs in the artificial images.
 In the right panels of Fig.\,\ref{fig:galfit}, we show how the decomposition performed directly on the 1D surface brightness profile of model gSbgSbo70 overrides the problems of the structured appearance of the bulge and the grainy structure of the disc in the 2D artificial images.

Therefore, we have simulated realistic surface brightness profiles of our S0-like remnants in the $K$ band, mimicking the observing conditions of the NIRS0S data, to perform 1D multi-component decompositions.

Our major merger remnants are quite axisymmetric,
and 1D and 2D decompositions have been proven to provide similar photometric parameters in such case \citep[within typical observational errors, see][]{1996A&A...313...45D,1996ApJ...457L..73C,2003ApJ...582..689M,2010AJ....139.2097P}. This ensures that the comparison between the photometric decompositions performed to NIRS0S and to our remnants is fair. However, all remnants in the minor merger experiments are still barred at the end of the simulation (in fact, the original gS0 progenitor is already barred). In these cases, the 1D decompositions provide a description of the azimuthally-averaged light distribution of the bar.

\subsubsection{Constructing 1D surface brightness profiles}
\label{Sec:1Dprofiles}

We have converted the projected radial mass density profiles of the remnants into surface brightness profiles in the $K$ band, following an analogous procedure to the one described in Sect.\,\ref{Sec:identification} for simulating photometric images (see details there). We have derived azimuthally-averaged 1D surface density profiles of the stellar remnants in face-on views, so we do not have to apply any correction for galaxy inclination to the obtained profiles.

We plot the azimuthally-averaged surface brightness profiles of some remnants in Fig.\,\ref{fig:sbrfits}. All the remnants exhibit typical bulge-disc structures. The majority of them clearly show `three-zone' profiles proving the existence of additional subcomponents in the centre, such as the lenses in models gE0gSao16 and gSbgSbo70, the nuclear bar in experiment gS0dSao105, or the inner disc in model gSdgSdo42 (compare their profiles with their images in the figure). This is very frequent in real S0s \citep[L10;][]{2011MNRAS.418.1452L,2011MNRAS.414.3645S}. 
The remnants coming from gas-rich progenitors usually have several tidal satellites orbiting around, which are responsible for the peaks that appear in their surface brightness profiles (as it occurs in gSdgSdo42). 

\begin{figure}[!th]
\center
\includegraphics[width=0.49\textwidth]{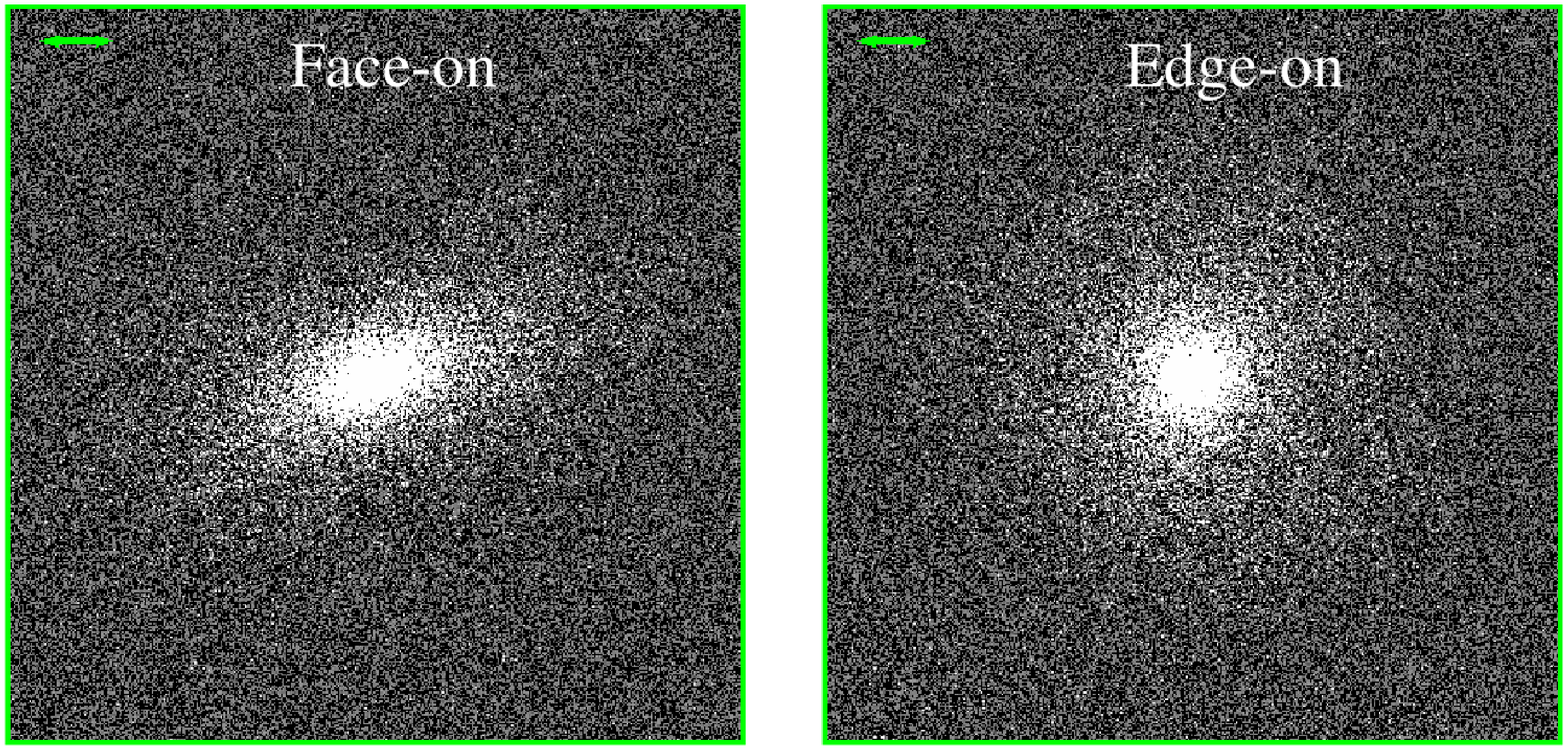}
\caption{Artificial $K$-band photometric images of remnant gSdgSdo51 (E/S0), for the face-on and edge-on views defined on the basis of the total stellar angular momentum (see Sect.\,\ref{Sec:identification}). The remnant contains a high-rotating embedded disc, very inclined with respect to the main outer disc, which dominates the total stellar spin of the galaxy. This is biasing the ``face-on view'' towards a face-on view of this embedded inner disc, instead of a face-on view of the outer disc. The field of view in all panels corresponds to 50\,kpc$\times$50\,kpc. The arrows indicate a physical length of 5\,kpc.
[\emph{A colour version is available in the electronic
edition.}]
}
\label{fig:edgeon_faceon_change}
\end{figure}

In some merger experiments with very inclined orbits, the remnant contains high-rotating inner discs, completely embedded in the light distribution of the bulge, which are very inclined with respect to the orientation of the main disc of the remnant. In these cases, the face-on view derived from the total angular momentum of the stars in the galaxy (as commented in Sect.\,\ref{Sec:identification}) was biased to show a face-on view of this inner disc, not of the external remnant disc. In these cases, the line of sight was corrected to provide a true face-on view of the main remnant disc. One example is shown in Fig.\,\ref{fig:edgeon_faceon_change}.

\subsection{Photometric decompositions of the remnants} 
\label{Sec:decompositions}

We have performed multi-component decompositions to the simulated 1D surface brightness profiles in the $K$ band to compare with the results by L10. The majority of remnants had profiles that required an additional component besides the bulge and the disc to be adequately modelled (see Sect.\,\ref{Sec:1Dprofiles}). To account for this, we performed bulge$+$disc (B$+$D) and bulge$+$[inner component]$+$disc (B$+$C$+$D) decompositions for all S0-like remnants, and selected the one that reproduced better the total galaxy profile in each case.

\subsubsection{Fitting functions} 

For the bulges, we assumed the traditional S\'{e}rsic profile \citep{1968adga.book.....S,1993MNRAS.265.1013C,2001A&A...367..405P,2011A&A...525A.136B},

\begin{equation} \label{eq:Sersic}
I_\mathrm{B}(r)=I_\mathrm{e,B}\, \exp\,\left[- b_{n}\, \left( \left(\frac{r}{\re}\right) ^{1/n} - 1\right) \right] ,
\end{equation} 

\noindent where \re\ is the bulge effective radius, $I_{\mathrm{e,B}}$ is the surface brightness at \re, and $n$ is the S\'{e}rsic index. The factor $b_{n}$ is a function of the parameter $n$, which may be approximated by $b_{n}=1.9992\, n-0.3271$ in the range $1<n<10$ with an error $<0.15$\% \citep{2001MNRAS.326..543G}. We had to constrain $n$ to the observational range of values during the fitting ($1\leq n\leq 4$, see L10), because many remnants had small nuclear bars and ovals embedded in the bulges  (which biased $n$ towards $n\sim 0.2$ -- 0.5) or central cusps resulting from the merger-induced starbursts (which led to unrealistically high $n$ values $n\gtrsim 5$). In two models (gSdgSdo1 and gSdgSdo74), we had to force $n=4$ since the beginning to obtain reasonable decompositions. 

The exponential law adequately describes the global radial profiles of most remnant discs down to the limiting magnitude considered (see Fig.\,\ref{fig:sbrfits}), but many discs in our remnants exhibit breaks at deeper magnitudes \citep[see][]{2014arXiv1407.5097B}. We have thus adopted a simple Freeman exponential profile to model the discs in the remnants  \citep[][]{1970ApJ...160..811F},
 
\begin{equation} \label{eq:disc}
I_\mathrm{D}(r)=I_{\mathrm{0,D}}\,\exp\left( -\frac{r}{\hd}\right) , 
\end{equation} 

\noindent where $h_{\mathrm{D}}$ is the disc scalelength (radius at which the surface brightness is reduced by $1/e$) and $I_{\mathrm{0,D}}$, its central surface brightness. 

The additional inner components have been fitted including another S\'{e}rsic component,

\begin{equation} \label{eq:Component}
I_\mathrm{C}(r)=I_{\mathrm{e,C}}\,\exp\,\left[- b_{n,C}\, \left( \left( \frac{r}{\rec}\right) ^{1/\nc}-1\right) \right] , 
\end{equation}

\noindent where \rec\ is the effective radius of the additional inner component, $I_{\mathrm{e,C}}$ is its surface brightness at \rec, and $\nc$ is its S\'{e}rsic index. According to observations, lenses, ovals, and bars have S\'{e}rsic profiles with $\nc<1$, while embedded inner discs typically have $\nc\sim 1$ \citep[][ L10]{2005MNRAS.362.1319L,2009ApJ...692L..34L,2009IAUS..254..173S}. We left the S\'{e}rsic index $\nc$ as a free parameter in the fits and checked that the fitted values agreed well with the morphology of the inner components visible in the artificial photometric images. 




\subsubsection{Fitting strategy} 

We performed the B$+$D and B$+$C$+$D fits (depending on the profile) using a Levenberg-Marquardt nonlinear fitting algorithm to locate the $\chi^2$ minimum solution by iterative changes to the parameters in eqs.\,\ref{eq:Sersic}--\ref{eq:Component}. We initially consider a minimum fitting radius of $r_\mathrm{min}=0.3$\,kpc (which is approximately the highest softening length used in the simulations). However, most remnants had nuclear compact sources or discs embedded within the spheroidal light distribution of the bulge, which have been produced by merger-induced nuclear starbursts and which are still very bright in $K$ band at the end of the simulation (see some examples in the second row of panels in Fig.\,\ref{fig:sbrfits}). These nuclear components have their observational counterparts in real early-type galaxies \citep[see e.g.][]{2002AJ....124...65E,2007ApJ...665.1084B,2009ApJ...692L..34L}. Compact sources steepen the profile at the centre (artificially raising $n$ of the fitted bulge), while young nuclear discs bias $n$ towards $n\sim 1$ in the central regions. We have thus excluded the innermost regions from the profiles affected by these nuclear components by raising $r_\mathrm{min}$ in the fit, only if there was another component that clearly dominated the profile in the centre and extended beyond this minimum radius (i.e. if the compact source or nuclear disc was embedded in a spheroidal bulge-like component). Some experiments also give rise to core-type profiles in the centres of the remnants, that deviated downwards the inward extrapolation of the bulge S\'{e}rsic profile \citep[observational analogues can be found in][]{2009ApJS..182..216K,2013ApJ...768...36D}. We have excluded these regions from the fits by raising $r_\mathrm{min}$ too.

The maximum radius ($r_\mathrm{max}$) in all fits has been set to the radius of the isophote corresponding to the limiting surface brightness in the images ($\mu_K\sim 22$\,mag\,arcsec$^{-2}$). In some major mergers involving a gE0 progenitor, an outer spheroidal envelope of stellar material from the gE0 remains in the remnant, dominating the surface brightness at the external radii over the disc profile (cf. gE0gSao16 in Fig.\,\ref{fig:sbrfits}). Sometimes, these outskirts biased the slope of the fitted disc towards a shallower solution, so we slightly decreased $r_\mathrm{max}$ in these cases to avoid that problem. A similar procedure was adopted when a tidal satellite induced a peak in the profile near $r_\mathrm{max}$ that was clearly biasing the fitted slope of the disc.

We performed several tests changing $r_\mathrm{min}$, $r_\mathrm{max}$, and the initial guesses of the parameters by up to a factor of 10 to check the robustness of the obtained solutions. We found that the B$+$D fits were reasonably stable, whereas the addition of an extra component in the centre strongly degenerated the solutions, in the sense that there were two to four different sets of components that adequately reproduced the global surface brightness profile and provided similar minima of $\chi ^2$. The strong degeneracy of multi-component decompositions is also usual when dealing with real galaxies (see e.g. L10). However, it was easy to discard many of these solutions and select the most appropriate one simply by considering the edge-on morphology of the galaxy  and the radial profile of the residuals of each fit (see Fig.\,\ref{fig:sbrfits}). We consider the fits with the lowest $\chi^2$ values and inspect visually the results and the artificial images, which makes it relatively straightforward to select the most appropriate decomposition. When a given model had two feasible solutions, the scalelengths, characteristic surface brightness, and the total magnitudes of the bulges and discs differed by $\sim 10$--20\% at most, lying within the typical observational errors. This means that the fitted values of these parameters can be considered robust, as well as the bulge-to-total ($B/T$) and disc-to-total ($D/T$) ratios derived from them. On the contrary, $n$ changed noticeably between the different possible solutions. The errors of the fitted parameters were obtained through the bootstrap method \citep{Efron93,Press94} accounting for the errors of the surface brightness profiles. We performed Monte Carlo simulations ($N=100$) of the surface brightness profiles considering the errors associated with each data point (related to particle counting) and performed a B$+$D or B$+$C$+$D decomposition to each realisation. We then estimated the standard deviation of the photometric parameters obtained from the $N$ fits with respect to the nominal values derived by fitting the original profile. We used 3$\sigma$ rejection in this computation to discard outliers resulting from some failed automatic fits. This procedure of estimating errors for the fitted photometric parameters includes to some extent the uncertainties due to the degeneracy of very similar solutions. As commented above, this degeneracy affected noticeably the bulge S\'{e}rsic index. This is the reason why we have obtained relatively large errors for $n$ in some models.

All our S0-like remnants were well described by a B$+$D or B$+$C$+$D profile. Figure\,\ref{fig:sbrfits} shows the multi-component decompositions performed to the radial surface brightness profiles of six S0-like remnants in the $K$ band. We plot the residuals of each fit as a function of radius in the galaxy at the bottom panels.  Very few remnants were better reproduced by a B$+$D fit than by a B$+$C$+$D decomposition (one example of B$+$D fit is model gSagSao9 in Fig.\,\ref{fig:sbrfits}). The additional component included in the B$+$C$+$D fits may represent a lense or oval (as in model gSbgSbo70 in Fig.\,\ref{fig:sbrfits}), an embedded inner disc (as in model gSdgSdo42, see the same figure), or an azimuthally-averaged bar (see model gS0dSao105 in the figure). 

In Table\,\ref{tab:parameters} we list the bulge and disc photometric parameters derived from the multicomponent decompositions performed to our S0-like remnants that have been analysed in the present study, as well as the main characteristics of these decompositions: number and type of components included (i.e. whether it is B$+$D or B$+$C$+$D), $r_\mathrm{min}$ and $r_\mathrm{max}$ considered in each case, and $\chi^2$ of the fit in mag$^2$. $\chi^2$ is typically below $\sim 1$\,mag$ ^{2}$ in total for the whole fitted radial range in most cases. It rises up to $\sim 3$\,mag$^2$ in the cases with tidal satellites within the fitted radial profile. The decompositions performed to the original progenitors are also listed in Table\,\ref{tab:parameters}.

\section{Results}
\label{Sec:results}

In order to quantify to what extent major mergers destroy, preserve, or rebuild bulge-disc coupling, here we compare the photometric parameters derived for our relaxed S0-like remnants with real observational data. We have also included the minor merger models that give rise to a relaxed S0 remnant in the comparison. In particular, we check the overlap of our parameters in various photometric planes with those obtained by \citet[L04 hereafter][]{2004MNRAS.355.1251L}, \citet[W09 henceforth]{2009ApJ...696..411W}, and L10 from near-infrared observations of spirals and S0s. For reference, we also compare our results with the dry minor mergers simulated by \citet[A01 hereafter]{2001A&A...367..428A} and \citet[EM06 henceforth; EM12; EM13]{2006A&A...457...91E}.

\begin{figure}[th]
\begin{center}
\includegraphics[trim=0.6cm 0cm 0.8cm 0cm, clip=true,width = 0.49\textwidth]{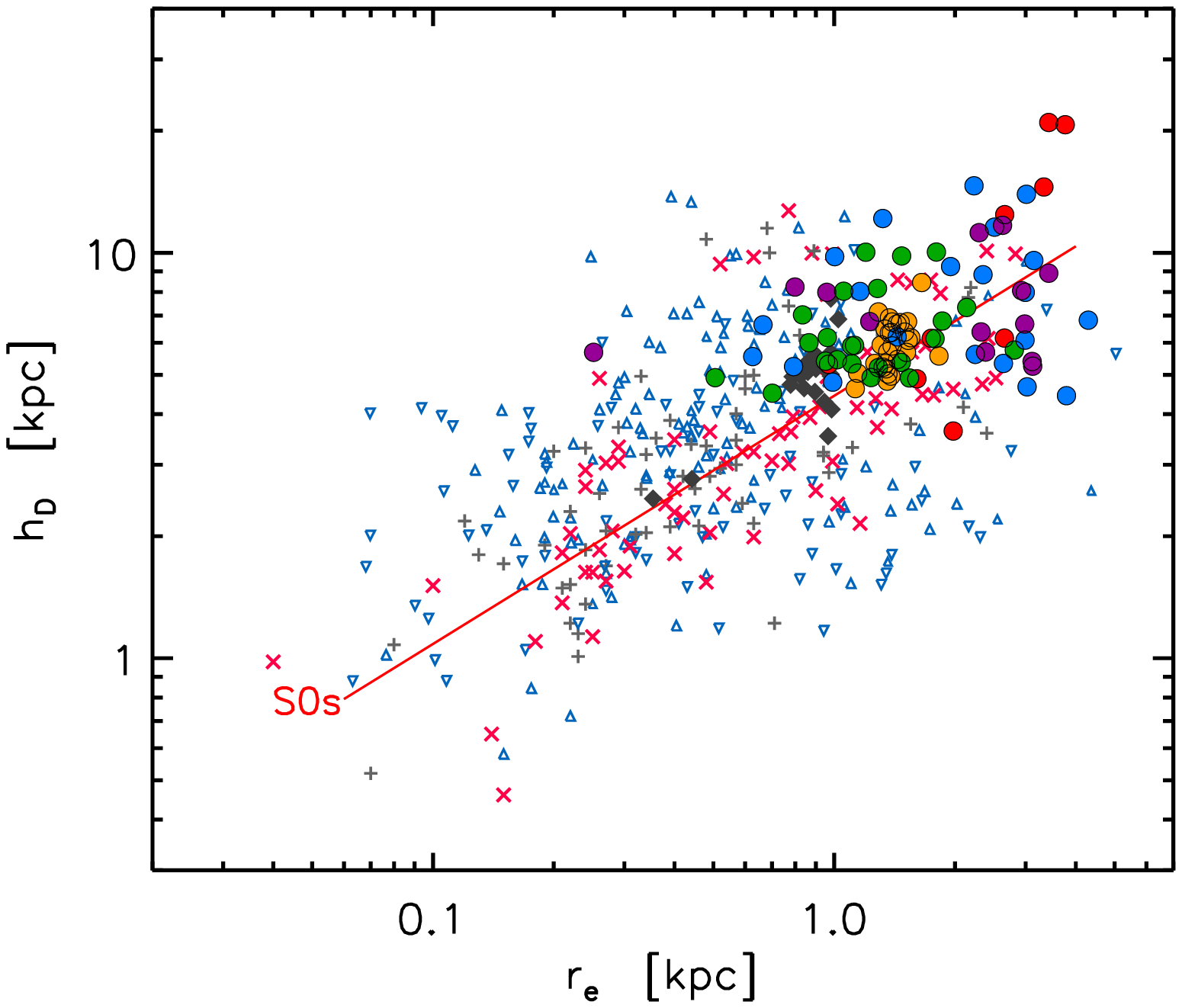} 
\includegraphics[trim=0.8cm 8.7cm 1.65cm 0cm, clip=true,width = 0.49\textwidth]{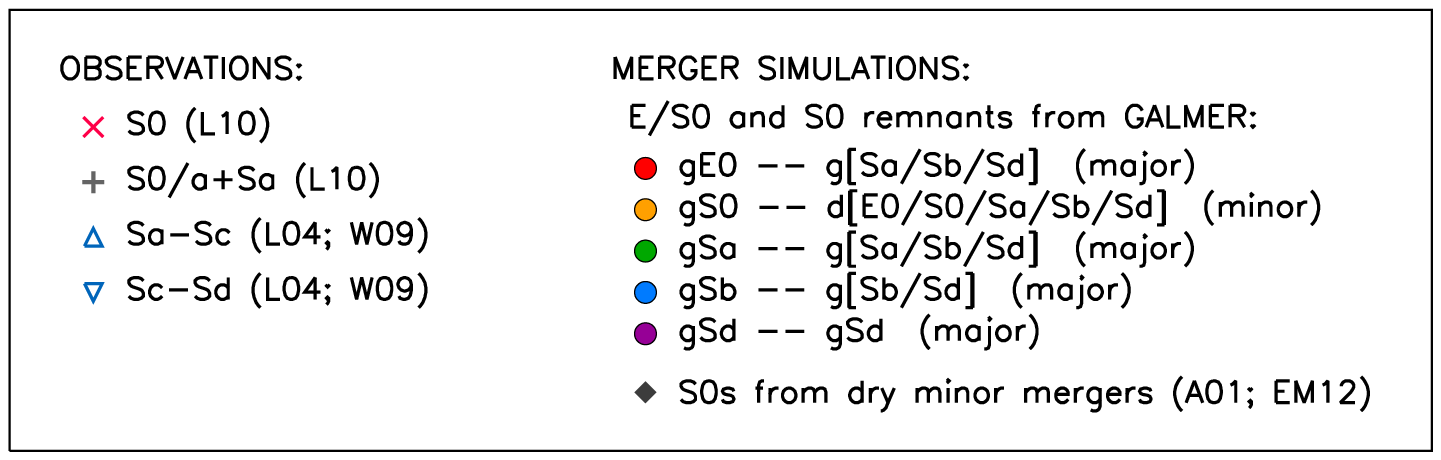} 
\end{center}
\caption{Location of our S0-like remnants in the $\log (\hd)-\log (\re)$ plane, compared to the observational results (L04; W09; L10) and previous simulations of dry minor mergers (A01; EM06; EM12; EM13). Details on the symbols and colour-coding used can be found on the legend. The linear fits performed to the observational distributions of S0 and spiral galaxies are overplotted in the diagram with solid lines only when Pearson's correlation coefficient is greater than 0.5, in this case only for S0s ($\rho_\mathrm{S0}=0.59$, $\rho_\mathrm{Sp}=0.21$).
[\emph{A colour version is available in the electronic
edition.}]}
\label{fig:h_re}
\end{figure}

\begin{figure}[ht]
\begin{center}
\includegraphics[trim=0.5cm 0.3cm 0.7cm 0cm, clip=true,width = 0.49\textwidth]{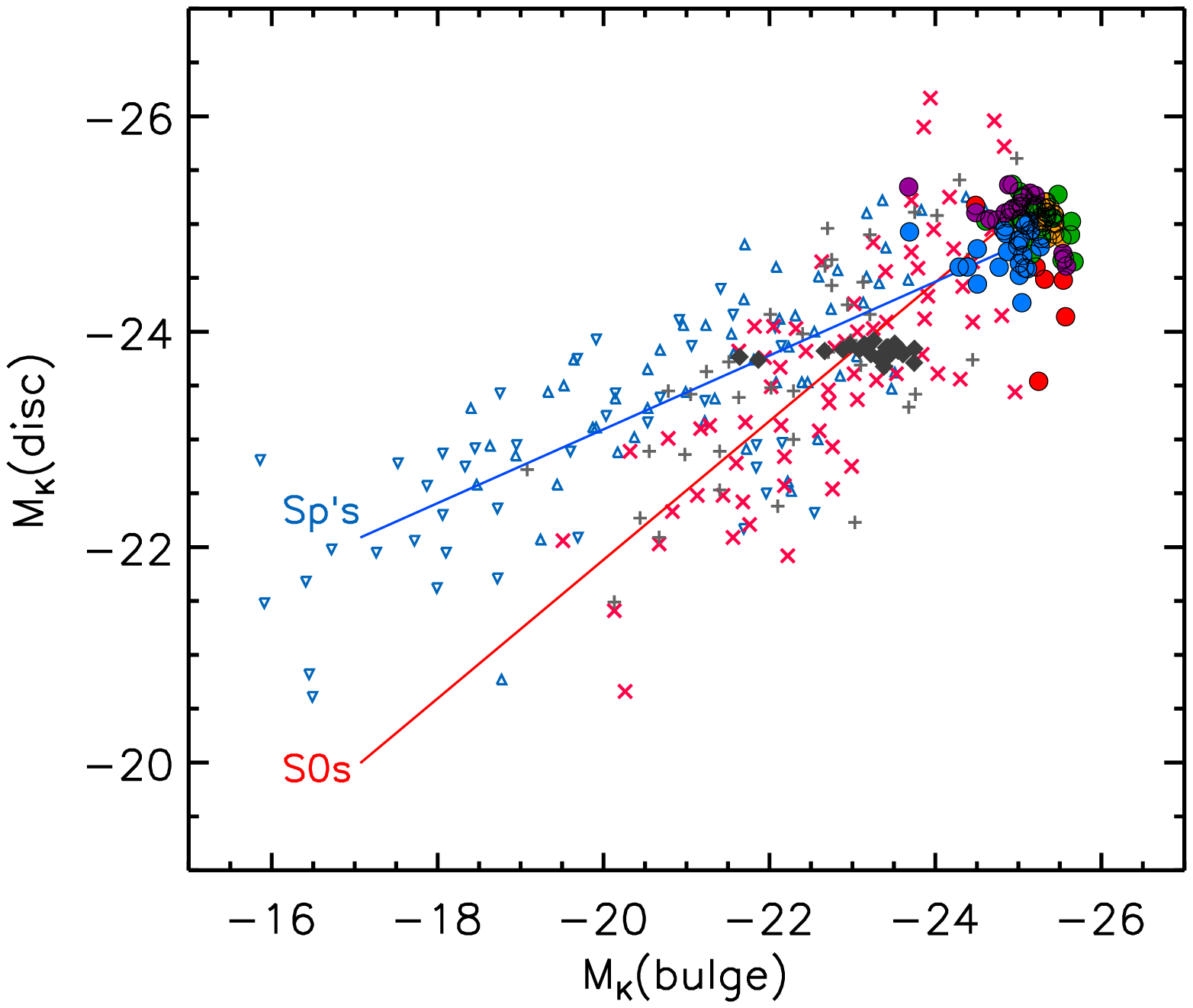} 
\end{center}
\caption{Distribution of our S0-like remnants in the $M_K(\mathrm{disc})$-$M_K(\mathrm{bulge})$, compared to the observational distributions of nearby S0s and spirals (L04; W09; L10) and to previous simulations of dry minor mergers (A01; EM06; EM12; EM13). The linear fits performed to the observational distributions of S0 and spiral galaxies are overplotted in the diagram with solid lines (Pearson $\rho_\mathrm{S0}=0.76$, $\rho_\mathrm{Sp}=0.70$). The symbols represent the same models and observations as in the previous figure; consult the legend in Fig.~\ref{fig:h_re}.
[\emph{A colour version is available in the electronic
edition.}]}
\label{fig:mkdisco_bulge}
\end{figure}

\begin{figure*}[ht]
\begin{center}$
\begin{array}{cc}
\mathrm{MAJOR \; MERGER} & 
\mathrm{MINOR \; MERGER} \\
\includegraphics[trim=0cm 0.3cm 0cm 0cm, clip=true,width = 0.49\textwidth]{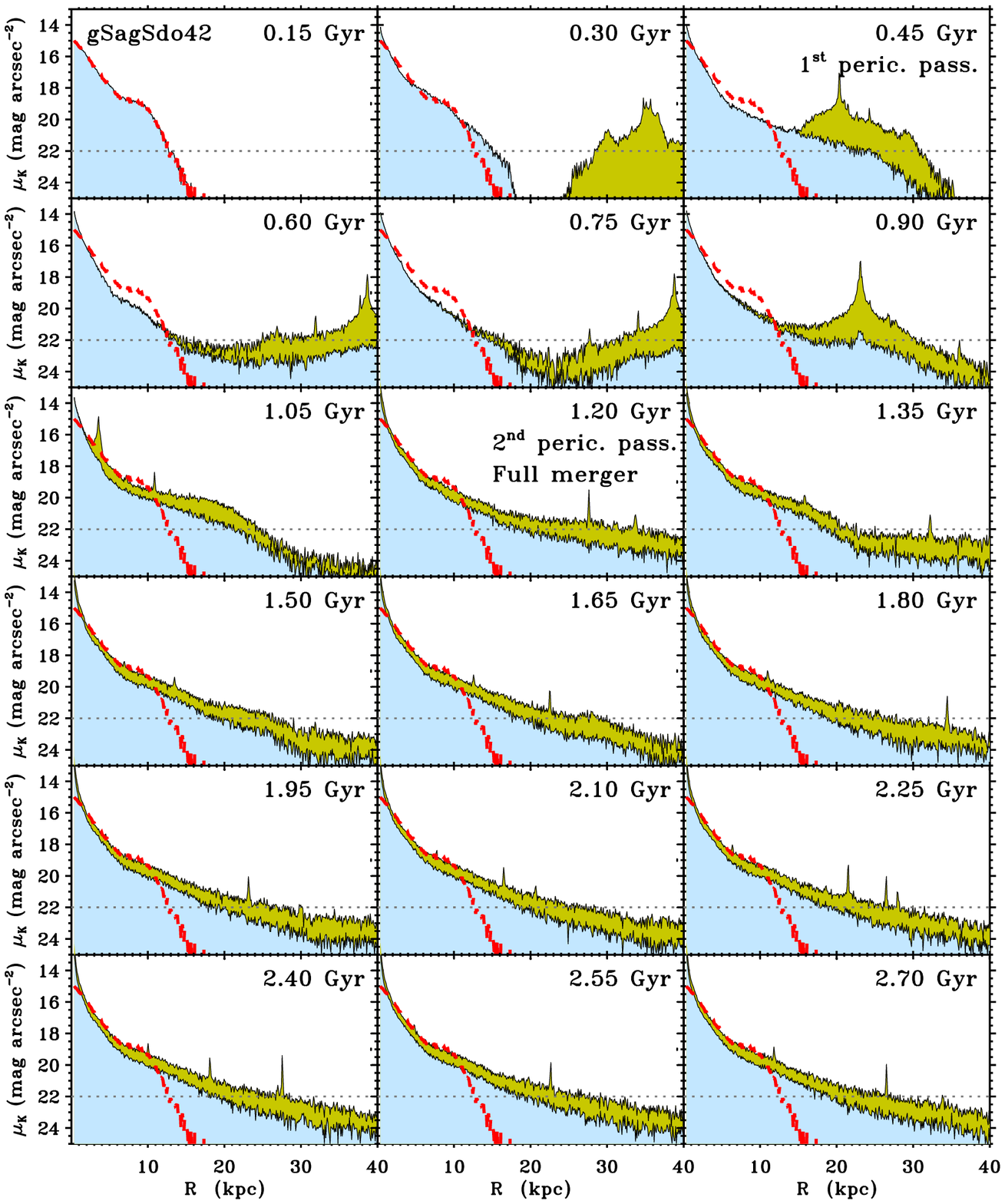} &
\includegraphics[trim=0cm 0.3cm 0cm 0cm, clip=true,width = 0.49\textwidth]{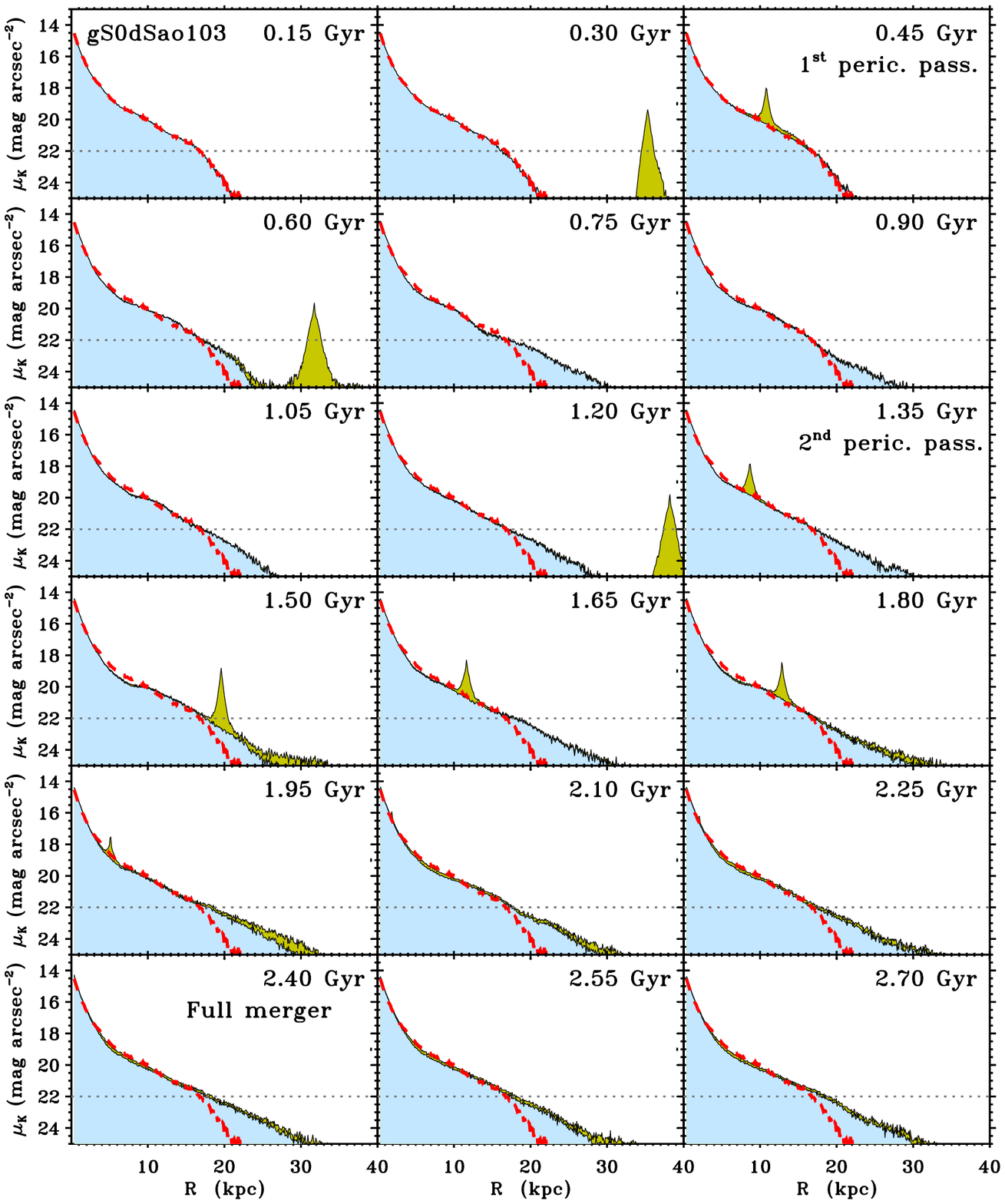} 
\end{array}$
\end{center}
\caption{
 Time evolution of the surface brightness profiles of the stellar material in two models that result in an S0-like remnant, centred on the mass centroid of the most massive (primary) progenitor at each time. \emph{Left panels}: For the major merger gSagSdo42. \emph{Right panels}: For the minor merger gS0dSao103. The contribution of the stellar material of each progenitor to the total profile at each time is marked with a different colour (\emph{blue}: primary progenitor; \emph{green}: secondary progenitor). We have plotted with red dashed lines the original surface brightness profile of the primary progenitor in all panels, to stress that minor merger events essentially preserve the profile, whereas major encounters completely rebuild the bulge and disc profiles in the remnants out of material from both progenitors. The limiting surface brightness that we consider is shown with a horizontal dotted line. We also indicate the first and second pericentre passages, as well as the moment when the full merger is reached. [\emph{A colour version is available in the electronic edition.}]
}
\label{fig:sbrt_major_minor}
\end{figure*}

\subsection{Bulge-disc structural coupling} 
\label{Sec:bulge-disc}

We start by analysing the coupling between the scalelengths of discs and bulges. Fig.~\ref{fig:h_re} shows the distribution of the final S0-like remnants in the $\log (\hd)-\log (\re)$ plane. Each filled circle represents the fate of a different merger event, with the colour code going from red to purple reflecting that later Hubble-type progenitors are involved (and, consequently, it roughly corresponds to a sequence of increasing gas content). With less saturated colours and smaller symbols, at the background, the observational results from L04, W09, and L10 are depicted. For reference, the remnants of the collisionless simulations of minor mergers onto an S0 are also shown with grey diamonds (A01; EM06; EM12; EM13). These collisionless models are scalable, i.e. they can be moved diagonally in this diagram just considering a different length unit. We have also overplotted the least-squares linear fits performed to the observational distributions of spirals and S0s, whenever the Pearson correlation coefficient $\rho$ is larger than 0.5.

Observationally, all disc galaxies, from S0s to spirals, populate a diagonal region in the plane $\log (\hd)-\log (\re)$, as it can be seen in Fig.~\ref{fig:h_re}. Scatter exists, but it is still remarkable that bulges of a certain size predominantly exist in galaxies with a given disc scalelength. Moreover, it is important to emphasise that the distributions of the different types of galaxies overlap: a similar increasing tendency seems to hold for all kinds of disc galaxies, with a steeper slope in the case of S0s.

The merger remnants that we are studying cluster towards the upper-right corner of the plane (Fig.~\ref{fig:h_re}), but this is because we are dealing with remnants which have masses similar to the most massive S0s in the NIRS0S sample, at the upper end of the sizes and luminosities of observed lenticulars. This is natural, as the masses of the progenitors ranged $\sim 0.5$ -- $1.5\times10^{11}\Msun$ (Sect.\,\ref{Sec:models}). The distribution of scalelengths for the bulges and discs of our remnants overlaps with that of the largest S0s; even the scatter introduced by the different types of encounter is compatible with the scatter of real galaxies. 

For the largest bulge effective radii, from Fig.~\ref{fig:h_re} it seems like, on bulk, lenticulars have slightly larger disc scalelengths than spirals. It is precisely towards this upper side of the observational distribution of S0s where our remnants are located. The S0 remnants of our minor mergers cover a rather small area of the $\log (\hd)-\log (\re)$ plane, and they lie close to the results from previous studies based on dry minor merger simulations (A01; EM12). The different sets of major mergers, however, span a much larger range of values, reproducing well the scatter observed in real S0s. Finally, it is also interesting to note that the experiments with the largest gas fractions are the ones that show the largest deviations in positions on the plane. Our S0-like remnants exhibit a bulge-disc coupling in terms of sizes that is consistent with the one observed in real S0s.

In Fig.\,\ref{fig:mkdisco_bulge}, we plot the distribution of real and simulated S0s in the $M_K(\mathrm{disc})$ -- $M_K(\mathrm{bulge})$ plane. The linear fits performed to the distributions of real S0s and spirals indicate that the total $K$-band magnitudes of their bulges and discs correlate linearly, although the linear trend of the S0s is tilted with respect to the one of spirals. 
All our remnants accumulate towards the upper end area covered by the brightest real S0s. More importantly, the remnants from both minor and major mergers fulfill well the observational constraint of bulge-disc coupling also in terms of luminosity: the magnitude of the disc takes up a value which is proportional to the bulge magnitude within some scatter (shared both by simulations and observations). 

We have again overplotted the collisionless minor merger models by A01 and EM12 in Fig.\,\ref{fig:mkdisco_bulge} for comparison. They can be moved diagonally in the plane considering a different mass unit (i.e., luminosity unit), up to the region where our dissipative minor mergers are located. Therefore, the inclusion of gas and star formation effects does not seem to be relevant to preserve the bulge-disc coupling in satellite accretions onto gas-poor progenitors. 

As commented in Sect.\,\ref{Sec:identification}, our remnants are brighter than real S0s of analogous masses in the NIRS0S sample by a factor of $\sim 2$, due to recent merger-driven starbursts. If the remnants were allowed to relax passively for an additional period of $\sim 1$ -- 2\,Gyr, their colours would become more similar to those of nearby S0s and the remnants would experience a dimming of $\sim 1$\,mag in the $K$ band.
 Assuming that the average $M/L$ ratios of the bulge and the disc are similar, the remnants would move diagonally towards fainter magnitudes up to $\sim 1$\,mag in Fig.\,\ref{fig:mkdisco_bulge}, nearly following the line fitted to the distribution of real S0s. Therefore, a global dimming of the remnants by $\sim 1$\,mag in $K$-band would keep the agreement between real and simulated S0s in the $M_K(\mathrm{disc})$ -- $M_K(\mathrm{bulge})$ diagram.

Figures~\ref{fig:h_re} and \ref{fig:mkdisco_bulge} prove that the S0-like remnants resulting from major and minor mergers present a bulge-disc coupling consistent with observations in terms of scalelengths and luminosities. In the minor merger experiments, the global structure of the main progenitor disc is mostly preserved at the end of the simulation in all cases. In the right panels of Fig.\,\ref{fig:sbrt_major_minor}, we show the time evolution of the stellar surface brightness profile of the gS0 progenitor in the minor merger experiment gS0dSao103 (\textit{blue}). We have overplotted the additional contribution from accreted stellar material coming from the dSa satellite at each time, as the disruption process evolves (\textit{green}). The original surface brightness profile of the gS0 progenitor is shown in all panels for comparison (\textit{red dashed line}). At the end of the simulation, the total profile due to the stars from the gS0 and the dSa in the remnant is very similar to the original profile of the gS0 progenitor at all radii down to the limiting magnitude under consideration. Therefore, in minor mergers, the changes experienced by the bulge and disc structures are small in general, mostly driven by internal secular processes induced by the encounters (see EM13), so it is not surprising that bulge and disc preserve their coupling at the end of the accretion. The original bar that the gS0 progenitor has at the start of the simulation in the minor merger experiments is strengthened at each pericentre passage of the satellite that takes place before the full merger (see  Fig.\,\ref{fig:mergerevol}). 

 The bulges and discs in the S0-like remnants resulting from major mergers exhibit a realistic structural coupling, even if the progenitor discs are basically destroyed in such major encounters. In the left panels of Fig.\,\ref{fig:sbrt_major_minor}, we show the time evolution of the stellar surface brightness profile for the gSa progenitor in the major merger model gSagSdo42. Here, we also highlight with different colours the contribution to the total profile at each time of the stars coming from each progenitor, as well as the original profile of the gSa galaxy (\textit{red dashed line}). At the end of the simulation, the distribution of the stellar mass coming from the gSa progenitor (\textit{blue}) is completely different at all radii from its original profile. There has been considerable mass migration from intermediate radii to the core and to the outskirts. The material originally belonging to the gSd has also been accreted at different radii, rebuilding a new bulge$+$disc structure. Then, in major mergers, the processes after the bulge relaxation and the disc rebuilding seem to force both components to keep a structural connection. This suggests that the bulge-disc coupling of all disc galaxies (and not just of S0s) may arise from fundamental physics. This would explain why both S0 and spiral galaxies exhibit a similar bulge-disc coupling in terms of sizes and luminosities (despite having different evolutionary pathways), as well as why the S0 galaxies resulting from events as violent as major mergers do still fulfill this coupling.  

These results confirm, both from the sizes point of view and from the perspective of the luminosity, that the analysed S0-like remnants show a similar coupling between their bulges and discs to that observed in real S0s. This proves that, contrary to the widespread belief, major mergers can produce S0 remnants with coupled bulge-disc structures analogous to those observed in real S0s.


\begin{figure}[!th]
\center
\includegraphics[width = 0.47\textwidth, bb = 10 59.2  474 396,
clip]{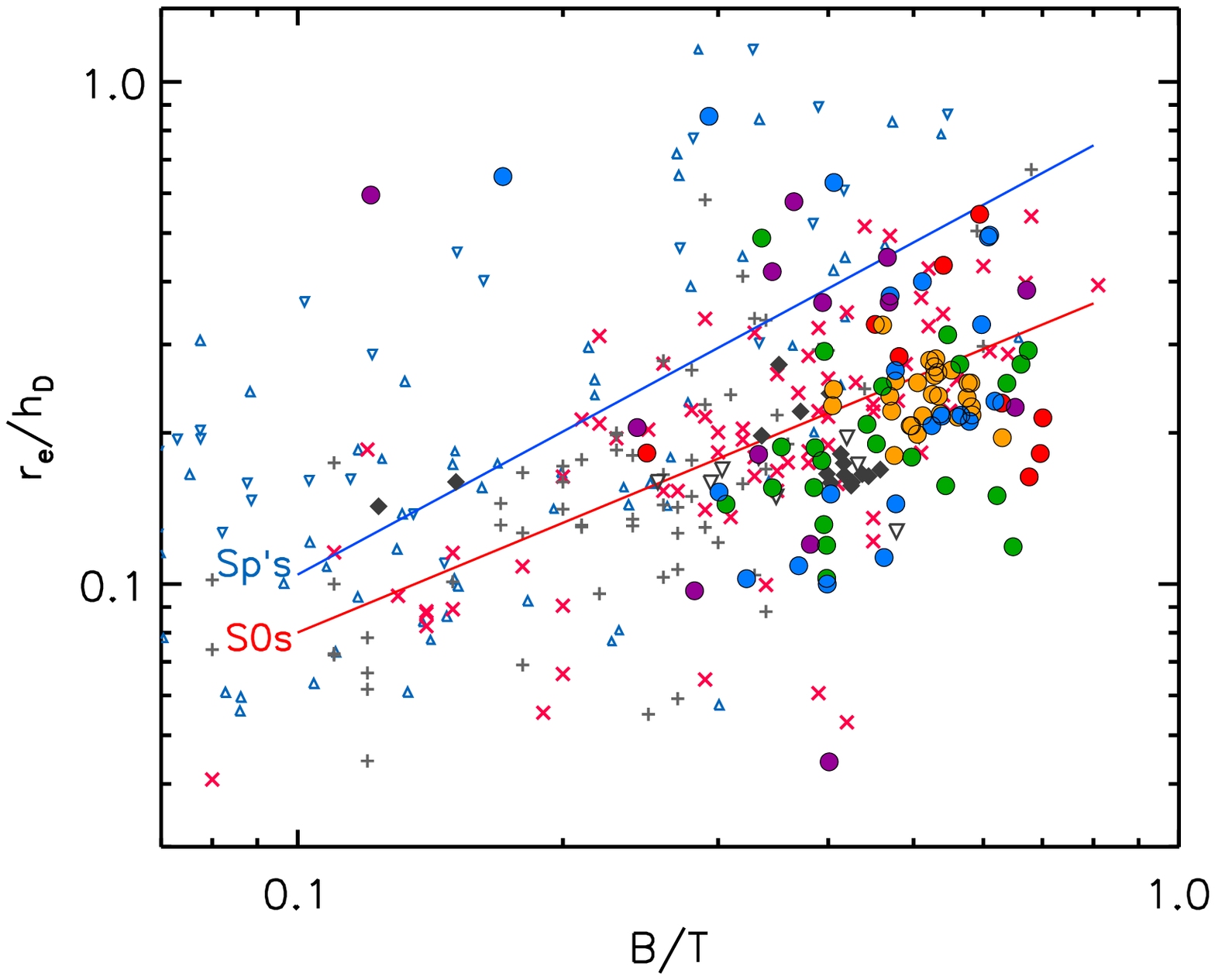}
\includegraphics[width = 0.47\textwidth, bb = 10 59.2  474 385,
clip]{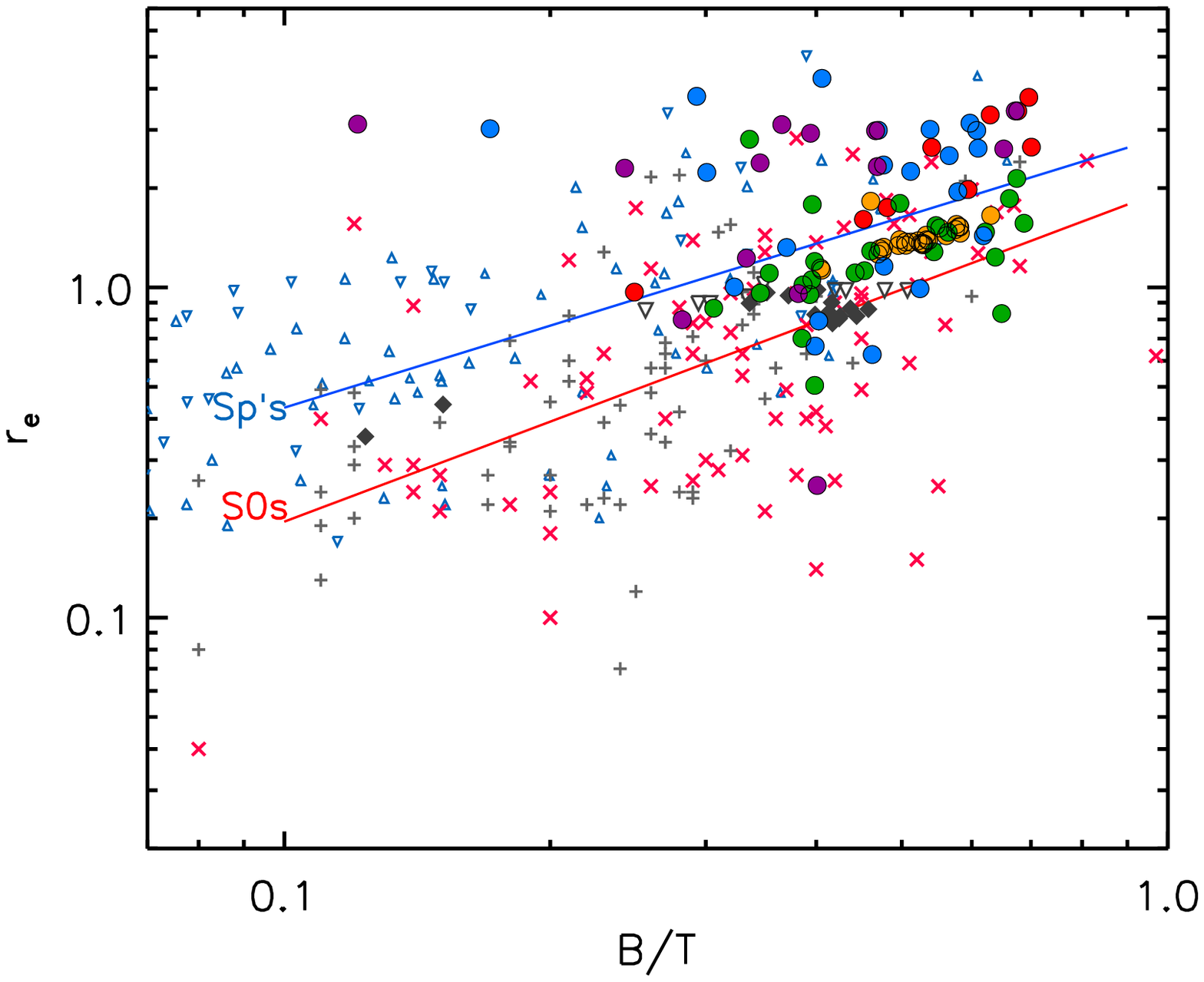} 
\includegraphics[width = 0.47\textwidth, bb = 10 12  474 385,
clip]{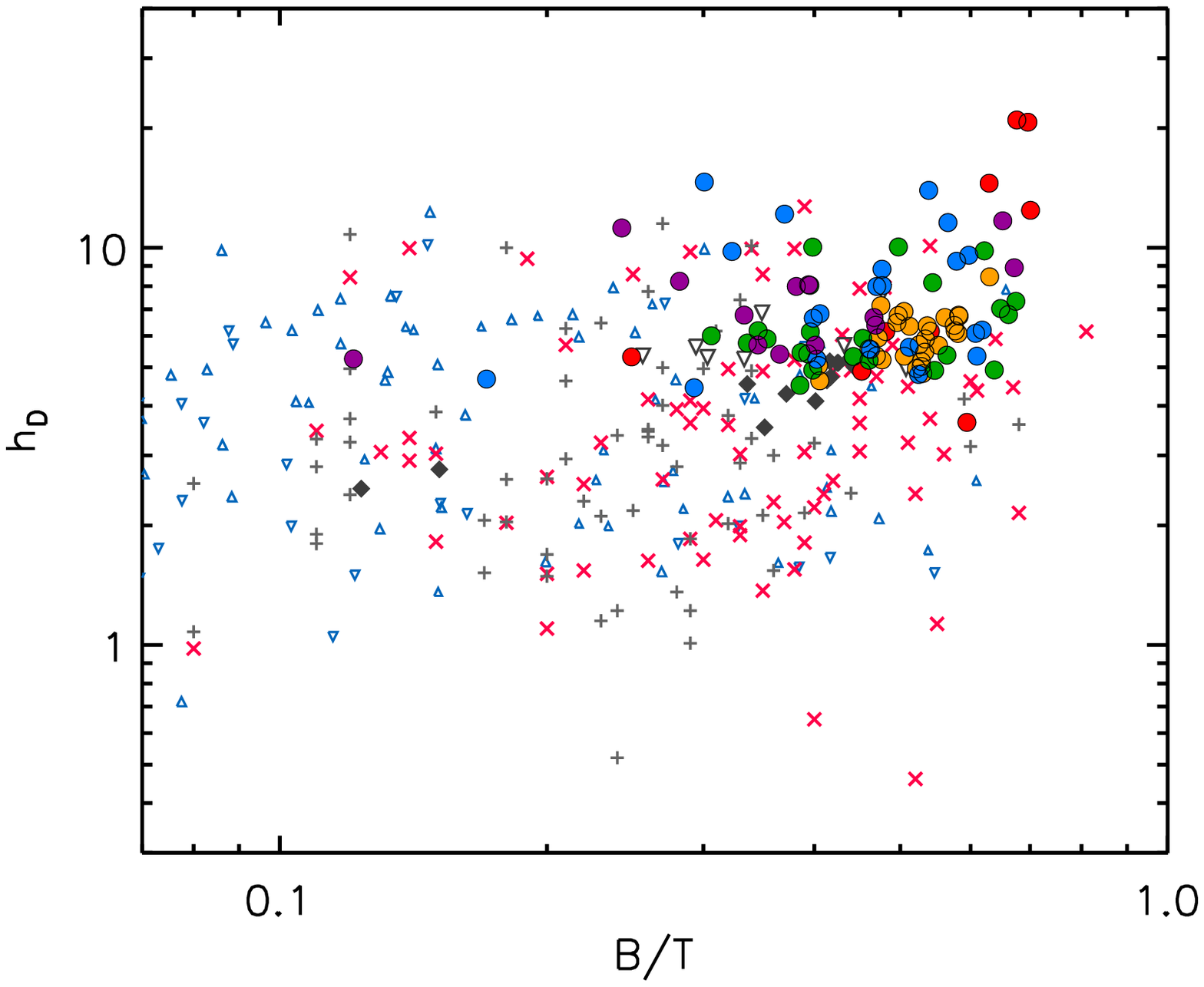} 
\caption{Distribution of  $\re / \hd$, \re, and \hd\  versus $B/T$ for our S0-like remnants resulting from major and minor mergers, compared to real observations of S0s and spirals (L04; W09; L10) and to previous collisionless simulations of minor mergers (A01; EM06; EM12; EM13). The linear fits to the observational distributions of S0 and spiral galaxies are overplotted with solid lines only when Pearson's correlation coefficient exceeds 0.5 (here, in the top and middle panels: \textit{top panel}: $\rho_\mathrm{S0}=0.64$, $\rho_\mathrm{Sp}=0.63$; \textit{middle panel}: $\rho_\mathrm{S0}=0.60$, $\rho_\mathrm{Sp}=0.55$; \textit{bottom panel}: $\rho_\mathrm{S0}=0.0021$, $\rho_\mathrm{Sp}=-0.20$).
The symbols represent the same models and observations as in the previous figures; consult the legend in Fig.~\ref{fig:h_re}.
[\emph{A colour version is available in the electronic
edition.}]}
\label{fig:rebt_hdbt}
\end{figure}

\subsection{Trends of the bulge and disc scalelengths with $B/T$} 
\label{Sec:trends}

The correlation between $\re/\hd$ and $B/T$ or $n$ in spirals has undergone significant debate in the last couple of decades: the apparent lack of observed correlation in the earlier studies lead some authors to claim that the Hubble sequence was scale-free \citep{1996A&A...313...45D, 1996ApJ...457L..73C, 2001MNRAS.326..543G,  2003ApJ...582..689M, 2007ApJ...665.1104B}; on the contrary, others found a slight increasing trend in those planes, suggesting that earlier Hubble types (with higher $B/T$) tend to host bulges of relatively larger sizes \citep{1999ApJ...524L..23G,2009MNRAS.393.1531G,2010MNRAS.401..559M}. L10 found a slight decreasing trend of $\re/\hd$ with the morphological galaxy type $T$ \citep{1991trcb.book.....D} from $T=-3$ (SO$^{-}$) to $T=2$ (Sa), which becomes essentially constant from $T=2$ to $T=6$ (Sa to Sc; see their Fig.\,5).

Independently of whether $\re/\hd$ exhibits significant trends with $B/T$ and $n$ or not, we have studied if our S0-like remnants overlap with real S0s in the $\re/\hd$ -- $B/T$ and $\re/\hd$ -- $n$ planes. In Fig.~\ref{fig:rebt_hdbt} we first plot $\re/\hd$ as a function of the $B/T$ ratio, compared to observational data (top panel). From the linear fit to observational S0s  and late-type spirals we notice that, in spite of the large scatter, there is a similar increasing trend in both galaxy types, but with a systematic offset upwards in the case of spirals. When we overplot our remnants in this plane, it becomes clear that we are mostly reproducing the S0s, preferentially overlapping the area around the linear fit to lenticulars rather than the late-type spirals. There are only three outliers exhibiting the highest $\re/\hd$ values, corresponding to gas-rich major mergers. In any case, the scatter in the resulting $\re/\hd$ of the remnants as a function of $B/T$ is large, but the distribution of the simulated S0s is consistent with the increasing trend that the observational S0s seem to follow (within the scatter). 

It is also remarkable that the S0-like remnants resulting from our major merger experiments span the observational ranges of $\re/\hd$ and $B/T$ values in the top panel of Fig.~\ref{fig:rebt_hdbt}, although their stellar masses cover a relatively narrow range ($\sim 1$ -- $3\times10^{11}\Msun$). This means that major mergers can give rise to S0 galaxies with very different global properties starting from similar progenitors, just depending on the initial conditions of the encounters. 

The intermediate and bottom panels of Fig.~\ref{fig:rebt_hdbt} can shed some light on the origin of the systematic shift between spirals and S0s in the $\re/\hd$ -- $B/T$ plane, and explain why our remnants end up covering the area they do. The intermediate panel shows that spirals tend to have systematically larger bulge effective radii for a given bulge-to-total luminosity ratio. The distributions overlap, but most of our remnants tend to align where most S0s lie, on average corresponding to smaller $\re$ than their spiral counterparts. It is also interesting to note that remnants with the largest dispersions (which still lie close to some of the S0 outliers in the plane) are those with the highest gas fractions. For completeness, the bottom panel shows the relation between $\hd$ and $B/T$, but no linear fit to observations is attempted here, since the Pearson correlation coefficients are low. In any case, it is reassuring to find that the merger remnants that we are studying populate an area which is observationally covered by lenticulars.

 In conclusion, the S0-like remnants resulting from major and minor mergers are consistent with the distribution of real S0s in the $\re/\hd$, \re, and \hd\ versus $B/T$ planes.

\subsection{Pseudobulges resulting from major mergers} 
\label{Sec:bulges}

\begin{figure*}[ht]
\begin{center}
\includegraphics[trim=-0.18cm 0.5cm 0.87cm 0cm, clip=true,width = 0.49\textwidth]{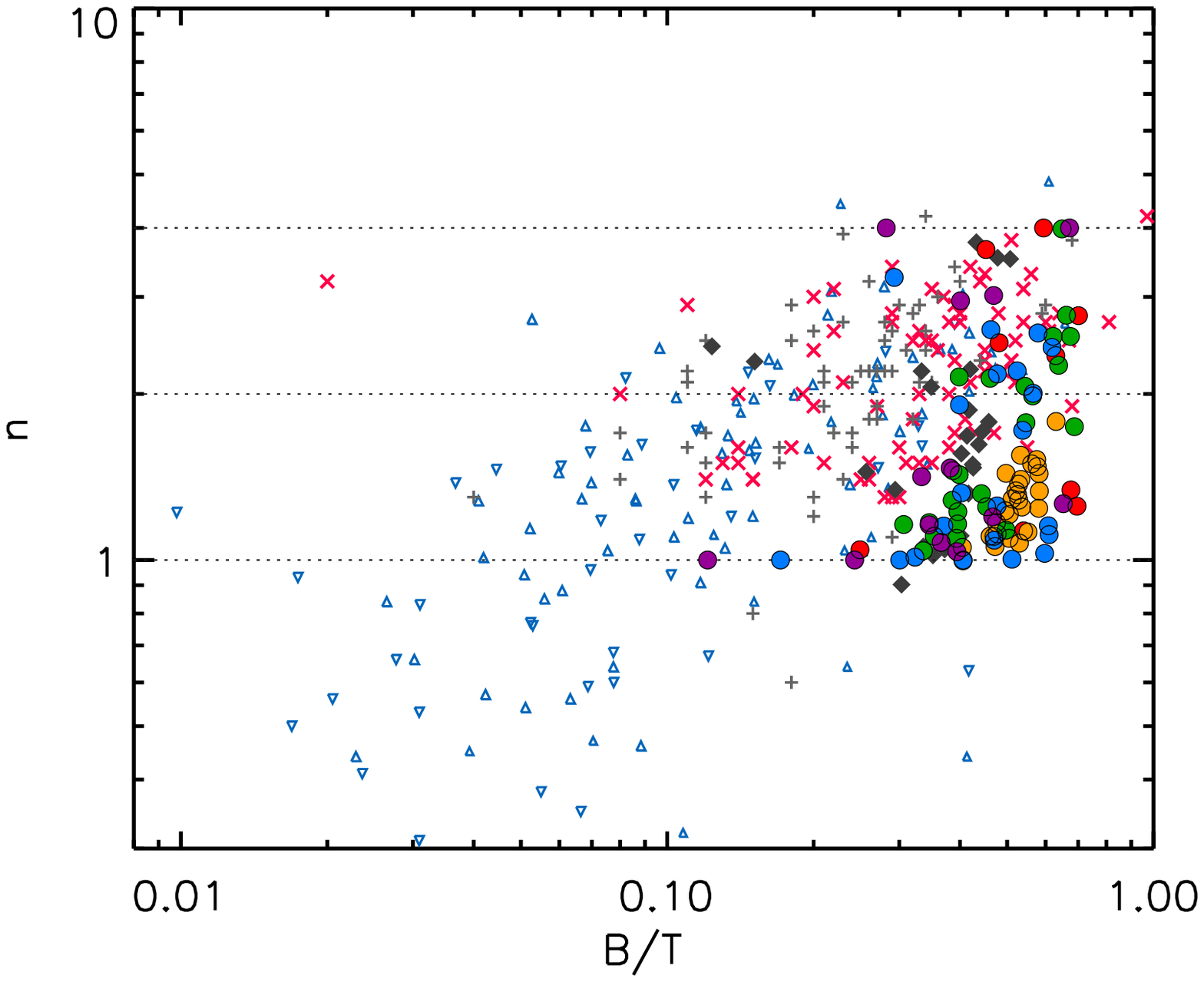} 
\includegraphics[trim=0.41cm 0.5cm 0.18cm 0cm, clip=true,width = 0.49\textwidth]{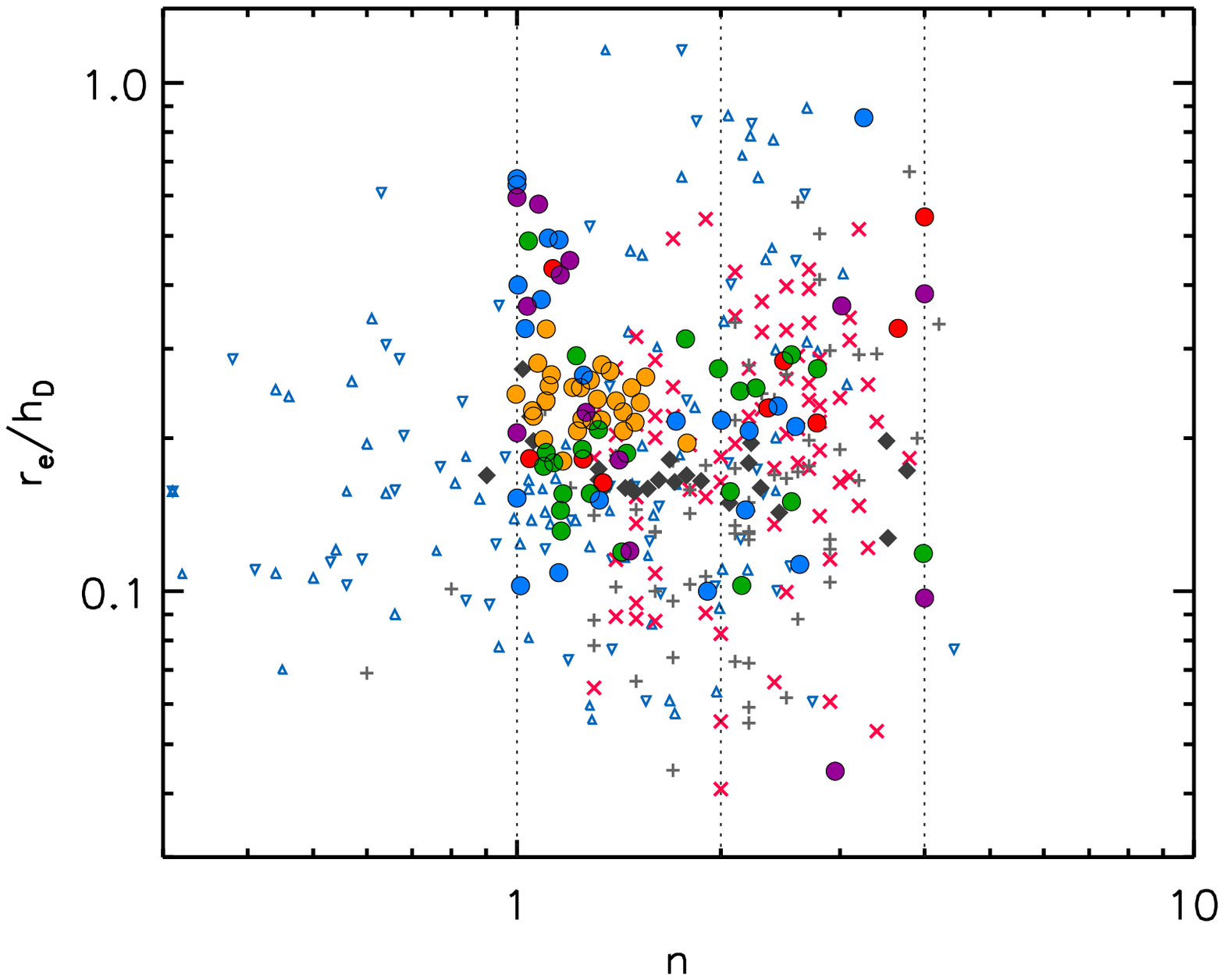}
\end{center}
\caption{Distribution of S\'{e}rsic indices $n$ with $B/T$ and $\re/\hd$ in our S0-like remnants (left and right panels, respectively), compared to the observational parameters from real S0s and spirals (L04; W09; L10). We also represent the locations of previous simulations of dry minor mergers (A01; EM06; EM12; EM13). The dotted lines indicate the location of $n=1$, 2, and 4 in each diagram. The symbols represent the same models and observations as in the previous figures; consult the legend in Fig.~\ref{fig:h_re}.
[\emph{A colour version is available in the electronic
edition.}]}
\label{fig:n_BT}
\end{figure*}

Figure~\ref{fig:n_BT} shows the distribution of our S0-like remnants in the $n$ -- $B/T$ and $\re/\hd$ -- $n$ planes (left and right panels, respectively), compared to the distributions of real S0s and spirals and to previous simulations of dry minor mergers. The left panel indicates that real galaxies distribute diagonally in the $n$ -- $B/T$ plane, with earlier types tending to accumulate towards higher $B/T$ and $n$ values. Half of our major merger remnants exhibit bulges with concentrations ($n$) and light contents relative to the total ($B/T$) compatible with the observations of S0 galaxies, whereas the other half exhibit too low $n$ values for their $B/T$ ratios. All minor merger remnants present less concentrated bulges (i.e., lower $n$) than real S0s with similar $B/T$. These models accumulate in a clump below the diagonal distribution of real galaxies at $n\sim 1$ and $0.3<B/T<0.8$. L10 showed that real S0s tend to exhibit lower $n$ values than popularly thought (they usually have $n\lesssim 2$, instead of the widespread belief of typical $n>3$ bulges). However, although our remnants tend to exhibit $n<2$ bulges accordingly to L10 results, many of them are too displaced from the location of real S0s in the $n$ -- $B/T$ diagram (see Fig.~\ref{fig:n_BT}). 

The low $n$ values of the remnants coming from a minor merger are an artefact of the initial conditions. The original gS0 is already offset from the location of S0s in this diagram ($B/T=0.2$ and $n=1$, see Table\,\ref{tab:parameters}). Therefore, even though all minor mergers onto this progenitor induce an increment of the concentration ($n$) and relative luminosity of the bulge ($B/T$), this bulge growth is not enough to counteract the initial conditions. So, these minor merger experiments would probably give rise to S0s with higher $n$ for similar $B/T$ ratios if the gS0 progenitor already had a more realistic S\'{e}rsic index. The dry minor mergers simulated by A01 and EM12 provide more realistic $n$ values for their $B/T$, just because the original S0 progenitor used in these encounters laid onto the observational cloud since the beginning.

\begin{figure}[ht]
\begin{center}
\includegraphics[trim=0cm 0cm 0cm 0cm, clip=true,width = 0.49\textwidth]{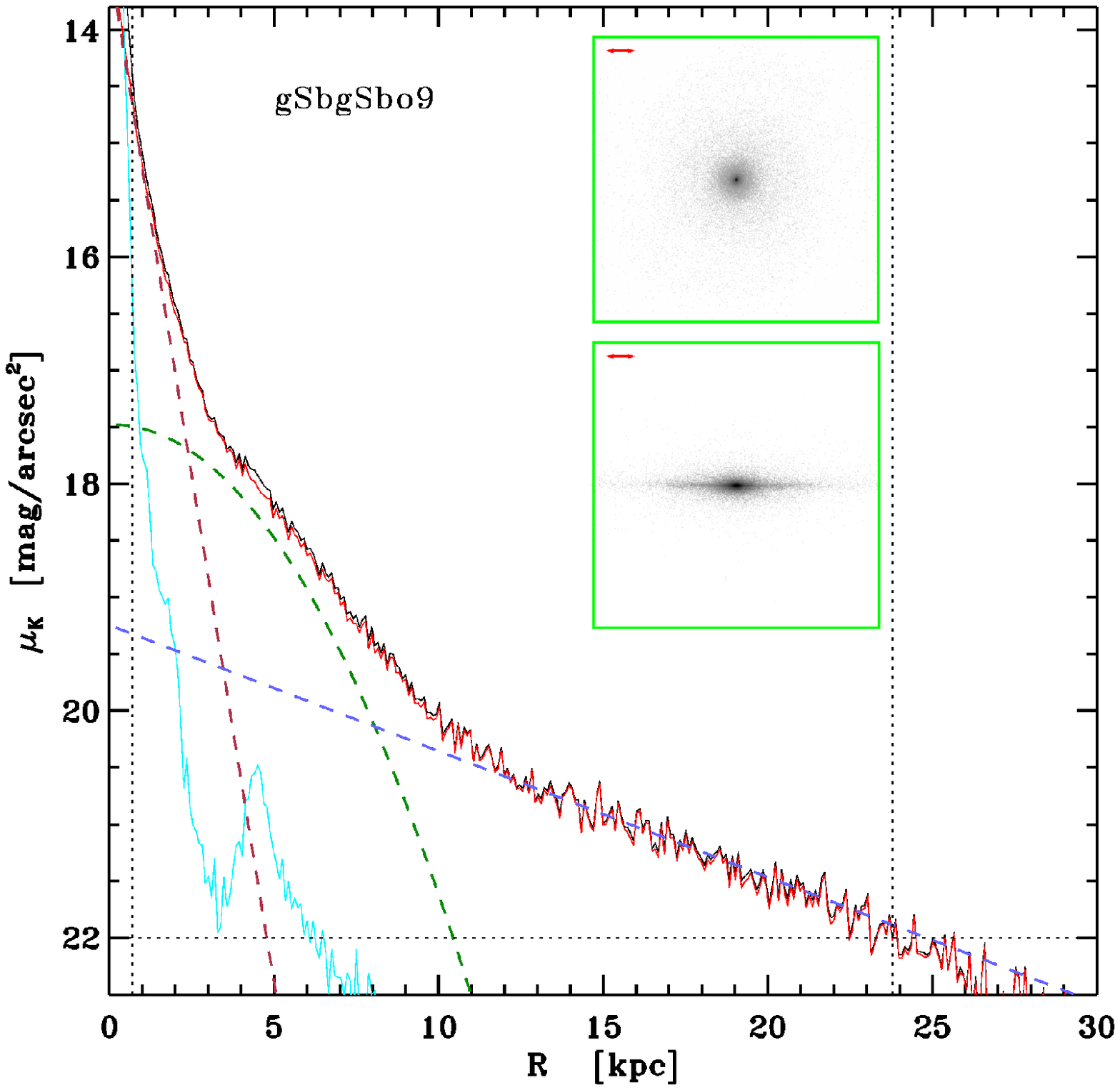} 
\end{center}
\caption{
Bulge-lense-disc decomposition performed to an S0-like remnant with an $n\sim 1$ bulge (model gSbgSbo9). \emph{Solid black line}: Total surface brightness profile in the $K$ band. \emph{Red solid line}: Contribution of the old stars to the total profile. \emph{Light blue solid line}: Contribution of the young stars to the total profile. \emph{Red dashed line}: Fitted bulge ($n\sim 1.0$). \emph{Green dashed line}: Fitted lense component. \emph{Blue dashed line}: Fitted disc. We have overplotted the limiting magnitude and the minimum and maximum radii considered in the decomposition (\emph{dotted straight lines}). The subframes of each panel show the artificial $K$-band images of the remnant for face-on and edge-on views, using a logarithmic greyscale to highlight the substructures at the centre (in particular, the flattened, disc-shaped bulge). The field of view is 50\,kpc$\times$50\,kpc. [\emph{A colour version is available in the electronic edition.}]
}
\label{fig:BLDdecomp}
\end{figure}

 In major mergers, the reason for the offset in S\'{e}rsic indices is related to the collapse of gas particles towards the centre. In some cases, gas accumulates at the remnant centre, giving rise to inner discs of $\sim 3$ -- 4\,kpc size made of newborn stars. These flat central structures made of young stars dominate the light distribution and bias the bulge S\'{e}rsic index towards $n \sim 1$. However, in other cases, the inner discs made of young stars are too small ($\sim 2$\,kpc) to explain the $n \sim 1$ bulges which dominate the profile out to $R\sim 5$\,kpc, and which are basically made up of old stars. In these cases, the explanation of the bulge flattening may arise from gas dynamics. Gas particles are known to transfer angular momentum to the old stellar particles during the encounter, providing rotational support to them and flattening their spatial distribution. This might lead to the $n\sim 1$ bulges built out of old stellar material. One example is shown on Fig.\,\ref{fig:BLDdecomp}. The contribution of the new stars (light blue) in the final remnant of gSbgSbo9 (black) only dominates the light distribution at the core of the galaxy ($R<1$\,kpc), which has been excluded from the decomposition. The region of the $n\sim 1$ bulge (red dashed line) basically consists of old stellar material, from $R\sim 1$\,kpc out to $R\sim 5$\,kpc, where the lense component starts to dominate.

This situation would probably change if the models were allowed to relax for a longer time period, as relaxation usually entails  dynamical mixing and reduces rotation. The bulge could potentially puff up and raise its S\'{e}rsic index. Moreover, these inner discs made up of young stars are expected to fade by $\sim 1$\,mag in $K$ in $\sim 1$ --2 Gyr (see Sect. 3.1); in that case, they would negligibly contribute to the surface brightness profile at the centre. Most present-day massive S0s have passively evolved for much longer time periods than these particular models \citep{2013A&A...558A..23D,2014arXiv1403.4932C}, so this is a natural explanation of why such low $n$ are rare in the local universe.

Many S0s host pseudobulges, i.e., bulges with $B/T<0.4$ and $n\sim 1$, with high levels of star formation and disc-related phenomena, such as spiral patterns or bars \citep[see][L10]{2006AJ....132.2634L}. The properties of these bulges have usually been attributed to a secular origin, mainly to bar evolution.  Our models demonstrate that a major merger can give rise to an S0 galaxy hosting a pseudobulge without requiring any bars (none of our major merger remnants develops a significant bar, see Sect.\,\ref{Sec:identification}).

In the right panel of Fig.~\ref{fig:n_BT} we plot $\re/\hd$ as a function of the S\'{e}rsic index. Although many of our remnants exhibit too low $n$ values compared to real S0s, the large scatter in $\re/\hd$ at each $n$ value of observational data masquerades the offset of these models in $n$ with respect to real S0s. However, there is a clump of models with $n\sim 1$ and $\re/\hd>0.4$ that clearly deviates from the rest of models in this plane. All these S0-like remnants come from gas-rich major encounters with very inclined orbits. The existence of very young (and thus bright) inclined inner discs in the centres of these remnants biases \re\ towards higher values than expected for their \hd. Cosmological simulations indicate that encounters with very inclined
orbits have been rare in the Universe, so it is understandable that these
encounters populate a region of the plane only sparsely covered by observations \citep{2005ApJ...629..219Z,2010hsa5.conf..295G,2011hsa6.conf..148B}.

\begin{figure*}[ht]
\center
\includegraphics[trim=0cm 0.4cm 0.15cm 0cm, clip=true,width = 0.48\textwidth]{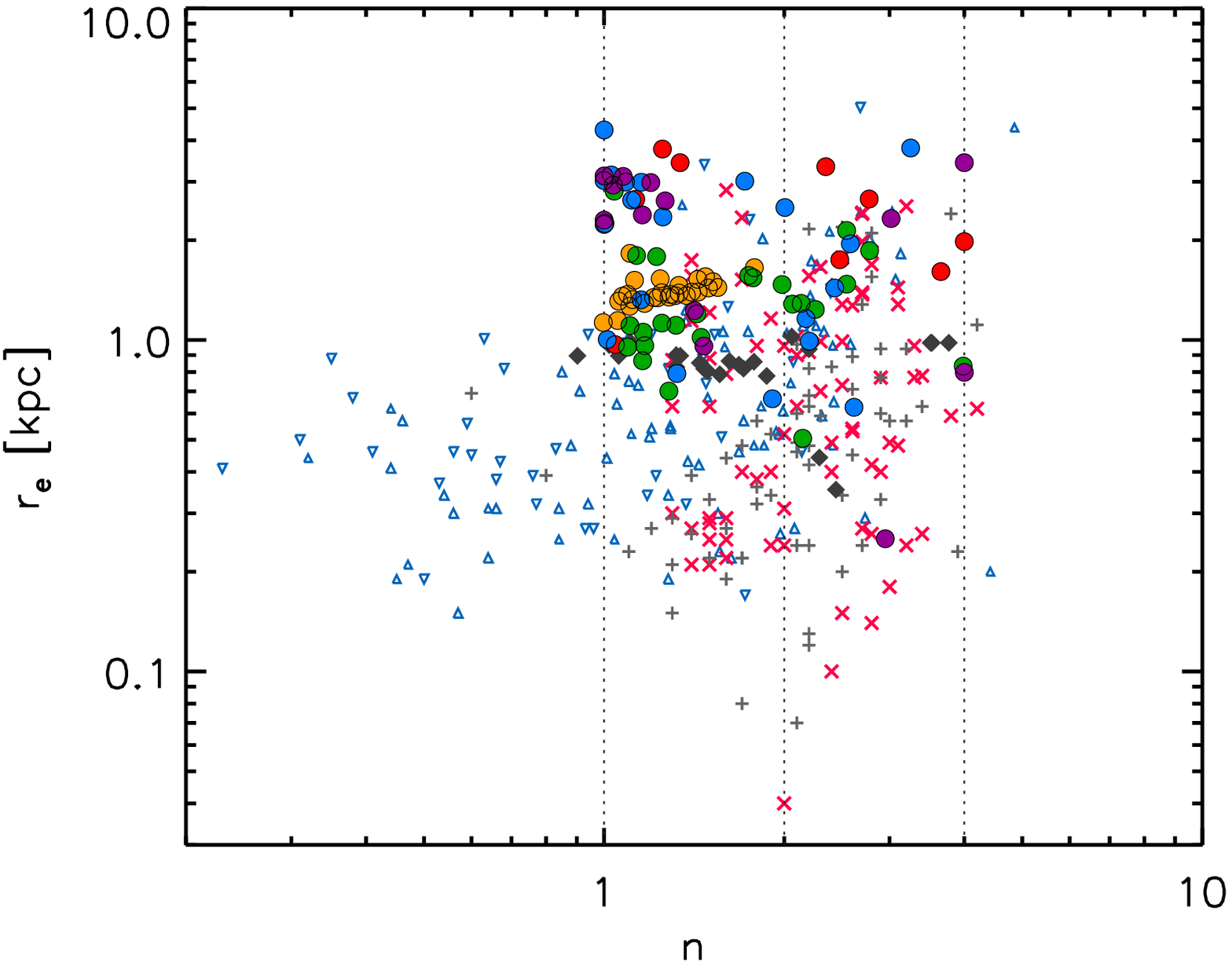} 
\includegraphics[trim=0cm 0.4cm 0.15cm 0cm, clip=true,width = 0.48\textwidth]{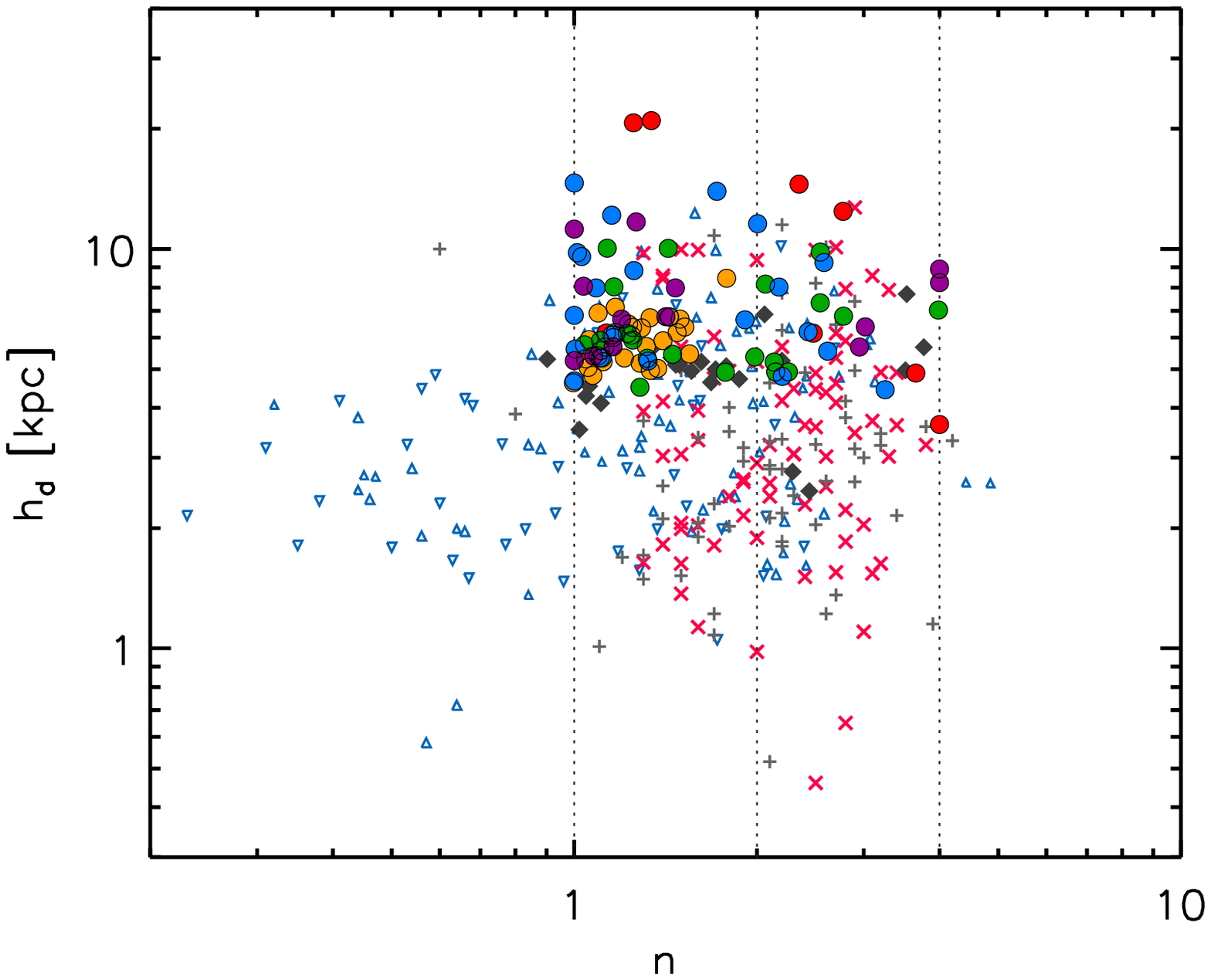} 
\caption{Distribution of our S0-like remnants in the log\,($r_\mathrm{e}$)-log\,($n$) and log\,($h_\mathrm{d}$)-log\,($n$) planes (left and right panels, respectively), compared to the observational distributions of S0s and spirals (L04; W09; L10). The location of previous collisionless simulations of minor mergers is also shown (A01; EM06; EM12; EM13). The vertical dotted line indicates the location of $n=1$, 2, and 4 in each diagram. The symbols represent the same models and observations as in the previous figures; consult the legend in Fig.~\ref{fig:h_re}.
[\emph{A colour version is available in the electronic
edition.}]}
\label{fig:ren_hdn}
\end{figure*}

Figure~\ref{fig:ren_hdn} is an alternative way of checking that our mergers reproduce the upper end of the observed S0 mass spectrum, and that, for such sizes, our remnants preferentially correspond to the lowest observed S\'{e}rsic indices. When the disc and bulge scalelengths are plotted against the S\'{e}rsic index, we see that the remants reproduce the upper half of the observational cloud in the \re\ -- $n$ and \hd\ -- $n$ planes. Our results agree well with previous collisionless simulations of dry minor mergers in both diagrams. The lower half of the observational distributions of S0s in these two photometric planes could be probably reproduced by mergers onto less massive progenitors than those used here. 

Finally, any trends with gas content and even mass ratios seem to be subdominant as to where galaxies end up in all the photometric diagrams shown in Figs.\,\ref{fig:h_re} -- \ref{fig:ren_hdn}. It is true that the most gas-rich major merger models disseminate more in the planes, whereas the minor mergers cover much tighter areas (just because they share the same main progenitor). The minor merger remnants are surrounded by those of our major merger events. Therefore, the mass ratio or gas content alone lack the predictive power to dictaminate the specific region of these planes into which remnants will fall. 

Summarizing, these simulations show that major mergers can build up S0s hosting pseudobulges without requiring the development of a bar, as already shown by EM13 for dry minor merger events. The presence of gas and star formation are, however, essential to explain the formation of S0s with disc-like bulges through major merger events.

\section{Limitations of the models}
\label{Sec:limitations}

GalMer models survey a wide set of initial conditions, but they are also limited. Here we comment on the limitations inherent to the present models.

\vspace{0.2cm}
\textit{1. Mass ratios of encounters and progenitor masses.}
\vspace{0.1cm}

GalMer models cover a significant range of mass ratios (1:1 to 20:1, see Table\,\ref{Tab:massratios}), but they are fixed for each pair of progenitor morphologies, so the effects of different mass ratios cannot be analysed for a given set of initial conditions (and in particular, for a given couple of progenitor types). In fact, intermediate mergers (with mass ratios from 4:1 to 7:1) are not present in the database, but many studies suggest that intermediate encounters and multiple minor mergers may have been as relevant for the evolution of S0 galaxies as major merger events \citep[see][and references therein]{2006ApJ...647..763M,2007A&A...476.1179B,2013MNRAS.433.2986W,2014A&A...565A..31T}.

More than the limitations in terms of mass ratios, the reason why our remnants only cover the largest observational S0s is that at least one of the progenitors is always a giant galaxy, of initial stellar mass $M_\mathrm{\star} \gtrsim 5 \times 10^{10} M_\odot$. Therefore, it would be interesting to complement the present study with examples from the intermediate-merger regime and with galaxies of more modest initial masses.

\vspace{0.2cm}
\textit{2. Gas and star formation effects.} 
\vspace{0.1cm}

Dissipative effects have been proven to be essential for establishing the global structure and kinematics of merger remnants  \citep[see e.g.][]{2007MNRAS.376..997J, 2007A&A...468...61D,2008A&A...492...31D}. A major advantage of GalMer models is that they include the dynamical effects of gas and star formation, providing a more realistic picture of the merger event than collisionless models, but it also entails a series of constraints. 

Firstly, we must bear in mind that our S0-like remnants correspond to the outcome of a merger between progenitor galaxies that are analogous to those found in our local Universe. At higher redshifts, where the actual encounters that led to present-day S0s took place, the gas fractions were even higher. 
For instance, the gas fraction can be up to $\sim 50$\% of the total stellar mass at $z\sim 1$ \citep{2005ApJ...631..101P, 2008ApJ...687...59G, 2008ApJ...680..246T, 2009ApJ...706.1364F, 2009ApJ...697.2057L}. As we will argue in the discussion, this increases the likelihood of forming an S0-like remnant, reinforcing the relevance of our results. In any case, expanding the present analysis to simulations with larger gas fractions would be helpful to interpret this point.


 Secondly, the conversion from mass into light is not a trivial issue. We have adopted a number of simplifications: (a) we assign a single age to every old stellar particle (10\,Gyr); (b) we assume a given SFH per morphological type, typical of each progenitor, and independent of the location of the particles within the galaxy; (c) for gas particles transforming into stars, we approximate the SFH by a SSP model. There is some observational evidence of variation in the age of stars contributing to $K$-band, both from galaxy to galaxy and across individual galaxies \citep{2001ApJ...553...90V, 2004ApJS..152..175M, 2012MNRAS.427..790S}; however, this is still a highly-debated issue from the observational point of view, and trying to adopt more complex distributions of stellar ages and/or SFHs would only introduce additional uncertainties and complicate further the interpretation of our results. In any case, with the present conversion of mass into light, we account for the morphological appearance of remnants, including the important effects of the recent starburts in the structure of the central bulge (see Sect.\,\ref{Sec:bulges}).

\vspace{0.2cm}
\textit{3. Total simulation time.} 
\vspace{0.1cm}

The models have been evolved for up to 3.5\,Gyr, involving relaxing periods of $\sim 2$\,Gyr at most. Even though we have checked that all the remnants analysed were dynamically relaxed at the end of the simulation, the stellar material acquired during the merger may have not had enough time to be properly mixed. This may be the reason for the over-structured bulge that we found in 2D decompositions in comparison to NIRS0S galaxies (Sect.\,\ref{Sec:profiles}). Moreover, the youth of these substructures made them brighter than usual even in the $K$ band. Stellar populations with these ages would decrease their flux by up to $\sim 1$\,mag in the $K$-band during the next $\sim 1$--2\,Gyr \citep{2013MNRAS.428..999P}, becoming completely smooth within the global bulge light distribution. Anyway, some nearby E-S0 galaxies of lower masses than those of our S0-like remnants exhibit blue nuclear structures, which are usually considered as traces of past merging activity \citep{2009AJ....138..579K,2010A&A...515A...3H,2010ApJ...708..841W}. Blue colours in massive S0s are also more common at intermediate redshifts \citep{2009MNRAS.393.1467F}.

\section{Discussion}
\label{Sec:discussion}

It is well known that mergers (even major events) can, under favourable conditions, preserve discs \citep{2005ApJ...622L...9S}, but major mergers in
particular are expected to produce remnants with decoupled bulge and disc structures. Contrary to this widespread belief, we have shown that S0-like remnants resulting from dissipative major and minor mergers exhibit bulge-disc structural coupling coherent with observations, extending the results obtained from dry minor merger simulations (EM12; EM13).

As commented in Sect.\,\ref{Sec:bulge-disc}, minor merger events directly preserve the bulge-disc coupling of the original main progenitor. EM13 showed that, even in the absence of gas and star formation, satellite accretions induce internal secular evolution in the progenitor disc that can even enhance this structural link. The addition of small amounts of gas to the satellites can only boost internal secular processes in the main progenitor, as observed in the minor merger models analysed here. On the contrary, major mergers destroy the original bulge and disc structures during the first phases of the encounter, \textit{but} the final bulge and disc structures are rebuilt (partially based on the relics of the original progenitor structures) to give rise to S0 remnants that overlap in the $n$ -- $B/T$ -- $\re/\hd$ parameter space with the distribution of bright S0s. Except for a few outliers, especially in terms of concentration (S\'{e}rsic index $n$), we otherwise reproduce the tail of the most massive lenticulars in NIRS0S in the various photometric planes as the consequence of both major or minor mergers. Therefore, S0s with realistically coupled bulge-disc structures may result from the relaxation process that follows certain mergers, even in such violent events as major mergers. This process occurs within timescales that are reasonable in a cosmological sense (just $\sim 2$--4 Gyr), and it is therefore a plausible mechanism to explain the formation of a fraction of the current population of S0s. The bulge-disc coupling observed in all disc galaxies may thus be the consequence of basic physical processes. This would explain the observational fact that both spirals and S0s present similar linear trends (slightly offset or tilted) in photometric planes such as $\hd$ -- \re, $M_K(\mathrm{disc})$ -- $M_K(\mathrm{bulge})$, or $\re/\hd$ -- $B/T$.

Our remnants span a relatively narrow range of final stellar masses, $\sim 1$ -- $3\times 10^{11}\Msun$, because one of the members of the encounter is always a giant galaxy, even if both major and minor mergers are considered.
In this context, one advantage of the merger origin picture is that, even considering such a narrow range in stellar masses, they can explain the trends and dispersions observed in the $n$ -- $B/T$ -- $\re$ -- $\hd$ planes for up to one dex. Mergers that differ only slightly in their initial conditions are capable of producing quite different sort of remnant systems, and all of them with bulge-disc structural coupling consistent with observations. If we also account for the observational and computational evidence that points to a merger origin for a significant fraction of S0s \citep[and in particular, for the most massive ones, see][]{2010A&A...519A..55E,2011MNRAS.412..684B,2011MNRAS.412L...6B,2012A&A...537A..25M,2013MNRAS.432..430B,2013MNRAS.433.2986W,2014A&A...565A..31T}, it seems unjustified to exclude major mergers from the current scenarios of S0 formation and evolution. 

Another relevant point is the role of gas fraction when producing S0s out of mergers. It has already been mentioned that our S0-like remnants arise from progenitor galaxies that are analogous to those found in our local Universe, with gas fractions typical of present-day spiral galaxies (see Table\,\ref{tab:morph_params}). However, many authors claim that massive S0s have evolved passively since $z\sim 0.8$, but they seem to have undergone strong star formation episodes at higher redshifts \citep{2013A&A...558A..23D,2014arXiv1403.4932C}. These results fit in a hierarchical formation picture in which a significant fraction of present-day massive S0s derive from mergers of spiral discs occurred at $z\sim 1$. In those early epochs, the typical gas fraction in spirals was higher than at present, reaching up to $\sim 50$\% of the total stellar mass at $z\sim 1$ \citep{2005ApJ...631..101P, 2008ApJ...687...59G, 2008ApJ...680..246T, 2009ApJ...706.1364F, 2009ApJ...697.2057L}. With such large amounts of gas, the formation of disc components in major merger events becomes more probable, and so does the probability of forming an S0-like remnant instead of an elliptical \citep{2009A&A...501..437Y, 2010ApJ...725..542H,2013MNRAS.431.3543H, 2009A&A...493..899P}. Even the lower gas amounts contemplated by the present models already point to major mergers producing S0 remnants with coupled bulges and discs; therefore, the buildup of a relevant fraction of present-day S0 galaxies through major mergers at $z\sim 1$, when gas fractions were higher, seems to be quite feasible. 

The fact that our results point to a merger origin of some S0s should not, of course, be overinterpreted. Naturally, it is not reasonable to claim that all lenticular galaxies derive from mergers, and we are far from being able to quantify what amount of the S0s in the local Universe are the direct consequence of one or several galaxy encounters. It is widely accepted that, within the densest environments, effects like ram-pressure stripping or galaxy harassment can explain the gas loss and consequent change of spirals into S0s \citep{2006A&A...458..101A}. Moreover, such mechanisms help explain the 
rise of the fraction of lenticulars with redshift, and the corresponding decline of spirals \citep{1980ApJ...236..351D}. However, here we would like to point out that, whatever that contribution has been, any S0s deriving from the major merger between two spirals would also contribute to the trend observed by \citeauthor{1980ApJ...236..351D} of transformation of spirals into S0s. It is beyond the scope of the present paper to quantify the relevance of major mergers in terms of creating S0s \citep[recent estimates indicate that they may have been essential in the buildup of $\sim 50$\% of present massive S0s at most, see][]{2014A&A...565A..31T}, but it is a mechanism that surely needs to be taken into account, and may especially explain the origin of the S0s that reside in groups and less dense environments. S0s are at least as common in groups as in clusters, and galaxy interactions are the dominant evolution mechanism in this regime \citep{2009ApJ...692..298W,2014ApJ...782...53M,2014AdSpR..53..950M}. Moreover, the role of merging in ``pre-processing'' galaxies in filaments before falling into a cluster and in ``post-processing'' them during their infall might also be underestimated at present \citep[see][]{2013MNRAS.435.2713V, 2014MNRAS.440.1690H}.

\section{Summary and conclusions}
\label{Sec:conclusions}

Because galaxy mergers are highly violent phenomena, it has often been claimed that they cannot possibly give rise to S0 galaxies in which a strong bulge-to-disc coupling holds.
Since such structural coupling has been measured observationally, this has led most authors to rule out mergers as a possible origin of S0 galaxies. We have thus studied the bulge-disc coupling in a set of major and minor merger simulations from the GalMer database that result in E/S0 or S0 types. We have simulated realistic surface brightness profiles of the remnants in $K$-band, mimicking the typical observational conditions, to perform structural photometric decompositions analogous to the ones that would have been obtained by observers. We have finally compared the distribution of the S0-like remnants with real S0 galaxies from the NIRS0S survey \citep{2011MNRAS.418.1452L} in photometric planes relating basic parameters of the bulges and discs.  In particular, we have found that:

\begin{enumerate}
\item S0-like remnants reproduce well the observed distribution of real bright S0s in the $B/T$ -- \re\ -- \hd\ parameter space. 

\item Although our remnants span a narrow range of stellar masses, they reproduce the observational values of \re, \hd, $\re/\hd$, and $B/T$ over an order of magnitude. Therefore, a wide variety of final structures consistent with observations can be achieved from mergers that differ only slightly in their initial conditions.

\item The majority of the experiments analysed ($\sim 64$\% of major mergers and $\sim 100$\% of minor events)  exhibit low bulge S\'{e}rsic indices ($1<n<2$), in agreement with the observed trend of real S0s to host $n<2$ bulges. 

\item However, nearly one half of the major-merger remnants present too low $n$ values compared to real analogues. These remnants host young inner discs formed in the starbursts induced by the encounters, biasing the bulge fit towards $n\sim 1$; the effects of these inner components are expected to vanish in $\gtrsim 2$ -- 3\,Gyr additional relaxation time.

\item The presence of young disc-like structures in the bulges of these major merger remnants and their global properties ($n<2$ and $B/T<0.4$) indicate that pseudobulges can also come out of major mergers without requiring bar phenomena (as none of these major merger experiments develops a relevant bar).

\item While minor mergers tend to preserve the original bulge-disc coupling of the main progenitor, major mergers are capable of rebuilding a bulge-disc coupling in the remnants after having destroyed the original structures of the progenitors. This suggests that the mechanisms after the bulge-disc coupling found in both S0 and spiral galaxies may be associated with fundamental physics.

\end{enumerate}

Therefore, these simulations demonstrate that realistic S0 galaxies with photometric parameters showing a bulge-disc coupling compatible with the one observed in real objects can emerge out of galaxy mergers in less than $\sim$3.5\,Gyr, and, in particular, from major mergers. Considering that mergers are complementary to other evolutionary mechanisms that probably operate preferentially over regions of different density, we conclude that mergers (and in particular, major ones) cannot be discarded from the formation scenarios of S0s on the basis of the strong bulge-disc coupling observed in these galaxies or their tendency to host pseudobulges.

\small  
%
\begin{acknowledgements}   

The authors would like to thank the anonymous reviewer for a very helpful and constructive referee report. We would also like to acknowledge I.\,Chilingarian, P.\,Di Matteo, F.\,Combes, A.-L.\,Melchior, and B.\,Semelin for creating the GalMer database. We also appreciate valuable comments by P.\,G.\,P\'{e}rez-Gonz\'{a}lez and E.\,Laurikainen, as well as the GALFIT code made publicly available by his author C.\,Peng. We also acknowledge the usage of the HyperLeda database (http://leda.univ-lyon1.fr). This research has made use of the NASA's Astrophysics Data System and NASA/IPAC Extragalactic Database (NED) which is operated by the Jet Propulsion Laboratory, California Institute of Technology, under contract with the National Aeronautics and Space Administration. 

\newline
\newline

M.Q. acknowledges financial support to the DAGAL network from the People Programme (Marie Curie Actions) of the European Union's Seventh Framework Programme FP7/2007- 2013/ under REA grant agreement number PITN-GA-2011-289313.
Supported by the Spanish Ministry of Economy and Competitiveness (MINECO) under projects AYA2006-12955, AYA2009-10368, AYA2012-30717, and AYA2012-31277, and by the Madrid Regional Government through the AstroMadrid Project (CAM S2009/ESP-1496, http://www.laeff.cab.inta-csic.es/projects/astromadrid/main/index.php). Funded by the Spanish MICINN under the Consolider-Ingenio 2010 Programme grant CSD2006-0070: "First Science with the GTC" (http://www.iac.es/consolider-ingenio-gtc/), and by the Spanish programme of International Campus of Excellence Moncloa (CEI). 
\end{acknowledgements}

\bibliographystyle{aa}
\bibliography{mq.bib}{}


\clearpage
\onecolumn
 \begin{landscape}

{
\footnotesize
\begin{center}

\begin{minipage}[t]{21cm}
\vspace{0.5cm}
\end{minipage}
  \begin{longtable}{cl l l   lll   r@{\,$\pm$\,}l r@{\,$\pm$\,}l r@{\,$\pm$\,}l r@{\,$\pm$\,}l r@{\,$\pm$\,}l c}
\caption{Characteristic photometric parameters of the bulges and discs of the S0-like relaxed remnants in the $K$ band}
\label{tab:parameters}
\\ \hline \\\vspace{-0.6cm}\\
\multirow{2}{*}{No.} & \multirow{2}{*}{Model} & \multicolumn{1}{c}{\multirow{2}{*}{Morph}}  & \multicolumn{1}{c}{\multirow{2}{*}{Fit}} & \multicolumn{1}{c}{$r_\mathrm{min}$} & \multicolumn{1}{c}{$r_\mathrm{max}$} & \multicolumn{1}{c}{$\chi^2$} & \multicolumn{2}{c}{ \re}  &  \multicolumn{2}{c}{\multirow{2}{*}{$n$}}  & \multicolumn{2}{c}{\hd}  & \multicolumn{2}{c}{\multirow{2}{*}{$B/T$}} & \multicolumn{2}{c}{\multirow{2}{*}{$D/T$}} & \multicolumn{1}{c}{$M_K(\mathrm{total})$}  \\
 &   &   &  & \multicolumn{1}{c}{[kpc]} &  \multicolumn{1}{c}{[kpc]} &  \multicolumn{1}{c}{[mag]} & \multicolumn{2}{c}{[kpc]} &  \multicolumn{2}{c}{} &  \multicolumn{2}{c}{[kpc]} &  \multicolumn{2}{c}{} &  \multicolumn{2}{c}{} &  \multicolumn{1}{c}{[mag]} \\ 
(1) & \multicolumn{1}{l}{(2)} & \multicolumn{1}{c}{(3)}  & \multicolumn{1}{c}{(4)} & \multicolumn{1}{c}{(5)}  & \multicolumn{1}{c}{(6)}  & \multicolumn{1}{c}{(7)}  & \multicolumn{2}{c}{(8)} & \multicolumn{2}{c}{(9)} & \multicolumn{2}{c}{(10)} & \multicolumn{2}{c}{(11)} & \multicolumn{2}{c}{(12)} & \multicolumn{1}{c}{(13)}  \vspace{0.1cm}\\\hline
\\\vspace{-0.5cm}\\
\endfirsthead
\caption{Characteristic photometric parameters of the bulges and discs of the S0-like remnants in the $K$ band \\\emph{(Continued)}}
\\ \hline
\\\vspace{-0.6cm}\\
\multirow{2}{*}{No.} & \multirow{2}{*}{Model} & \multicolumn{1}{c}{\multirow{2}{*}{Morph}}  & \multicolumn{1}{c}{\multirow{2}{*}{Fit}} & \multicolumn{1}{c}{$r_\mathrm{min}$} & \multicolumn{1}{c}{$r_\mathrm{max}$} & \multicolumn{1}{c}{$\chi^2$} & \multicolumn{2}{c}{ \re}  &  \multicolumn{2}{c}{\multirow{2}{*}{$n$}}  & \multicolumn{2}{c}{\hd}  & \multicolumn{2}{c}{\multirow{2}{*}{$B/T$}} & \multicolumn{2}{c}{\multirow{2}{*}{$D/T$}} & \multicolumn{1}{c}{$M_K(\mathrm{total})$}  \\
 &   &   &  & \multicolumn{1}{c}{[kpc]} &  \multicolumn{1}{c}{[kpc]} &  \multicolumn{1}{c}{[mag]} & \multicolumn{2}{c}{[kpc]} &  \multicolumn{2}{c}{} &  \multicolumn{2}{c}{[kpc]} &  \multicolumn{2}{c}{} &  \multicolumn{2}{c}{} &  \multicolumn{1}{c}{[mag]} \\ 
(1) & \multicolumn{1}{l}{(2)} & \multicolumn{1}{c}{(3)}  & \multicolumn{1}{c}{(4)} & \multicolumn{1}{c}{(5)}  & \multicolumn{1}{c}{(6)}  & \multicolumn{1}{c}{(7)}  & \multicolumn{2}{c}{(8)} & \multicolumn{2}{c}{(9)} & \multicolumn{2}{c}{(10)} & \multicolumn{2}{c}{(11)} & \multicolumn{2}{c}{(12)} & \multicolumn{1}{c}{(13)}  
\vspace{0.1cm}\\\hline
\\\vspace{-0.5cm}\\
\endhead
 -- & Original gE0 	& E	& B          &  0.5 & 13.4  & 0.069  &  3.67863 &  0.712 &   0.99 	&  0.15 &   \multicolumn{2}{c}{...}   & \multicolumn{2}{c}{1.0}  &   \multicolumn{2}{c}{0.0}    &	 -25.11\\
--  & Original gS0 	& S0	& B$+$C$+$D  &  0.5 & 16.1  & 0.063  &  0.88957 &  0.078 &   1.00	&  0.12 &  4.1 &  2.9   &  0.20 &   0.039  &  0.555027 &   0.000013        &	  -24.74 \\
 -- & Original gSa 	& Sa	& B$+$C$+$D  &  0.3 & 11.6  & 0.022  &  2.53212 &  0.577 &   0.89 	&  0.25 &  4.2 &  1.0   &  0.40 &   0.18  &  0.59 &   0.12        &	  -25.16\\
 -- & Original gSb 	& Sb	& B$+$C$+$D  &  0.5 & 15.0  & 0.173  &  0.96842 &  2.712 &   1.8 	&  1.0 &  4.2 &  2.3   &  0.27 &   0.12  &  0.72030 &   0.00073        &	  -24.55\\
 -- & Original gSd 	& Sd    & D          &  0.5 & 18.0  & 0.265  &  \multicolumn{2}{c}{...} &   \multicolumn{2}{c}{...}  &  4.677 &  0.032   &  \multicolumn{2}{c}{0.0} &    \multicolumn{2}{c}{1.0}   &	  -24.77\\
\hline\\\vspace{-0.5cm}\\
 1 & gE0gSao1	  &	S0     & B$+$C$+$D   &  0.5  &  21.4  &   0.082  &   1 &  16  	&   2.47 &  0.67  &	  6.15 &   0.95   &   0.48140 &   0.00026  &   0.27 &   0.33    &	 -26.01 \\
 2 & gE0gSao5	  &	S0     & B$+$C$+$D   &  0.5  &  23.5  &   0.071  &   0.9 &   7.8  	&   1.0 &  1.7  &	  5.30 &   0.65   &   0.24 &   0.20  &   0.46 &   0.17    &	 -25.99 \\
 3 & gE0gSao16	  &	S0     & B$+$C$+$D   &  0.5  &  26.5  &   0.676  &   1.60 &   0.33  	&   3.65 &  0.35  &	  4.8 &   2.6   &   0.452 &   0.018  &   0.4647 &   0.0015    &	 -26.02 \\
 4 & gE0gSao44	  &	S0     & B$+$C$+$D   &  1.0  &  14.4  &   0.261  &   1.98 &   0.10  	&   4.00 &  0.88  &	  3.633 &   0.086   &   0.594 &   0.082  &   0.363 &   0.048    &	 -26.00 \\
 5 & gE0gSbo5	  &	S0     & B$+$C$+$D   &  0.7  &  22.3  &   0.126  &   3.33 &   0.40  	&   2.34 &  0.40  &	 14.5 &   6.6   &   0.629 &   0.087  &   0.131 &   0.094    &	 -25.75 \\
 6 & gE0gSbo44	  &	S0     & B$+$C$+$D   &  1.0  &  23.5  &   0.205  &   3.42 &   0.18  	&   1.339 &  0.092  &	 20. &  18.   &   0.67 &   0.17  &   0.31 &   0.20    &	 -25.74 \\
 7 & gE0gSdo5	  &	S0     & B$+$C$+$D   &  1.0  &  20.0  &   0.286  &   3.76 &   0.24  	&   1.251 &  0.086  &	 20.6 &   1.7   &   0.696 &   0.091  &   0.261 &   0.013    &	 -25.94 \\
 8 & gE0gSdo16	  &	S0     & B$+$C$+$D   &  0.5  &  27.9  &   0.672  &   2.660 &   0.069  	&   2.773 &  0.071  &	 12.4 &   6.8   &   0.70 &   0.12  &   0.18 &   0.14    &	 -25.96 \\
 9 & gE0gSdo44	  &	S0     & B$+$C$+$D   &  0.7  &  28.7  &   1.692  &   2.65 &   0.80  	&   1.12 &  0.50  &	  6.1 &   2.8   &   0.54 &   0.13  &   0.37 &   0.14    &	 -25.95 \\
10 & gS0dE0o98	  &	S0     & B$+$D       &  0.5  &  19.0  &   0.032  &   1.552 &   0.016 	&   1.477 &  0.042  &	  6.175 &   0.058   &   0.5764 &   0.0045  &   0.4235 &   0.0045    &	 -24.89 \\
11  & gS0dE0o99	  &	S0     & B$+$C$+$D   &  0.5  &  19.0  &   0.018  &   1.39 &   0.36  	&   1.400 &  0.082  &	  5.8 &   1.7   &   0.533 &   0.031  &   0.414453 &   0.000007    &	 -24.89 \\
12  & gS0dE0o100    &	S0     & B$+$C$+$D   &  0.5  &  19.2  &   0.015  &   1.43 &   1.65  	&   1.49 &  0.82  &	  6.6 &  16.5   &   0.56 &   0.11  &   0.367982 &   0.000002    &	 -24.89 \\
13  & gS0dE0o101    &	S0     & B$+$C$+$D   &  0.5  &  19.6  &   0.018  &   1.65 &   0.10  	&   1.78 &  0.14  &	  8.44 &   0.45   &   0.630 &   0.033  &   0.3442 &   0.0096    &	 -24.89 \\
14  & gS0dE0o102    &	S0     & B$+$C$+$D   &  0.5  &  20.2  &   0.022  &   1.52 &   0.24  	&   1.43 &  0.20  &	  6.76 &   0.13   &   0.581 &   0.062  &   0.3935 &   0.0075    &	 -24.89 \\
15  & gS0dE0o103    &	S0     & B$+$C$+$D   &  0.5  &  18.1  &   0.015  &   1.384 &   0.056 	&   1.334 &  0.010  &	  4.9636 &   0.0037   &   0.52139 &   0.00075  &   0.468352 &   0.000001    &	 -24.89 \\
16  & gS0dE0o104    &	S0     & B$+$C$+$D   &  0.5  &  19.0  &   0.019  &   1.26 &   0.28  	&   1.1 &  1.1  &	  5.3 &  15.9   &   0.46 &   0.18  &   0.450404 &   0.000024    &	 -24.89 \\
17  & gS0dE0o105    &	S0     & B$+$C$+$D   &  0.5  &  18.2  &   0.016  &   1.346 &   0.029  	&   1.28 &  0.10  &	  5.17 &   0.28   &   0.528 &   0.019  &   0.427 &   0.034    &	 -24.89 \\
18  & gS0dE0o106    &	S0     & B$+$C$+$D   &  0.5  &  18.6  &   0.021  &   1.388 &   0.026  	&   1.247 &  0.063  &	  6.36 &   0.31   &   0.536 &   0.010  &   0.343 &   0.023    &	 -24.89 \\
19  & gS0dE0o109    &	S0     & B$+$C$+$D   &  0.5  &  19.4  &   0.018  &   1.49 &   0.61  	&   1.52 &  0.69  &	  6.3 &   3.7   &   0.575 &   0.075  &   0.408585 &   0.000012    &	 -24.89 \\
20  & gS0dE0o110    &	S0     & B$+$C$+$D   &  0.5  &  18.4  &   0.029  &   1.82 &   0.52  	&   1.104 &  0.010  &	  5.56 &   0.13   &   0.461 &   0.029  &   0.439322 &   0.000007    &	 -24.89 \\
21  & gS0dE0o111    &	S0     & B$+$C$+$D   &  0.5  &  19.5  &   0.020  &   1.37 &   0.19  	&   1.29 &  0.38  &	  6.3 &   6.3   &   0.512 &   0.060  &   0.394 &   0.053    &	 -24.89 \\
22  & gS0dE0o113    &	S0     & B$+$C$+$D   &  0.5  &  18.7  &   0.016  &   1.36 &   0.24  	&   1.31 &  0.61  &	  5.6 &   4.9   &   0.525 &   0.092  &   0.411068 &   0.000002    &	 -24.89 \\
23  & gS0dE0o115    &	S0     & B$+$C$+$D   &  0.5  &  19.2  &   0.034  &   1.28 &   0.50  	&   1.16 &  0.28  &	  7.14 &   0.39   &   0.475 &   0.091  &   0.365 &   0.011    &	 -24.89 \\
24  & gS0dE0o117    &	S0     & B$+$C$+$D   &  0.5  &  19.2  &   0.034  &   1.28 &   0.38  	&   1.16 &  0.31  &	  7.14 &   0.40   &   0.47 &   0.10  &   0.365 &   0.013    &	 -24.89 \\
25  & gS0dS0o97	  &	S0     & B$+$D       &  0.5  &  19.6  &   0.033  &   1.514 &   0.014 	&   1.124 &  0.021  &	  5.679 &   0.019   &   0.5519 &   0.0027  &   0.4480 &   0.0027    &	 -24.85 \\
26  & gS0dS0o98	  &	S0     & B$+$C$+$D   &  0.5  &  20.4  &   0.025  &   1.3 &   2.3  	&   1.435 &  0.010  &	  6.7 &   3.5   &   0.497 &   0.047  &   0.412 &   0.042    &	 -24.85 \\
27  & gS0dS0o99	  &	S0     & B$+$C$+$D   &  0.5  &  19.6  &   0.020  &   1.30 &   0.13  	&   1.057 &  0.010  &	  5.92 &   0.31   &   0.4721 &   0.0063  &   0.444035 &   0.000006    &	 -24.85 \\
28  & gS0dS0o100    &	S0     & B$+$C$+$D   &  0.5  &  18.8  &   0.015  &   1.4 &   1.2  	&   1.333 &  0.010  &	  6.72 &   0.35   &   0.582 &   0.019  &   0.389 &   0.030    &	 -24.85 \\
29  & gS0dS0o101    &	S0     & B$+$C$+$D   &  1.0  &  20.0  &   0.023  &   1.3 &   2.2  	&   1.094 &  0.010  &	  6.9 &   5.1   &   0.504 &   0.020  &   0.380 &   0.049    &	 -24.85 \\
30  & gS0dS0o102    &	S0     & B$+$C$+$D   &  0.5  &  19.6  &   0.021  &   1.339 &   0.037  	&   1.228 &  0.064  &	  6.481 &   0.017   &   0.495 &   0.053  &   0.402017 &   0.000014    &	 -24.85 \\
\hline\\
\pagebreak
31  & gS0dS0o103    &	S0     & B$+$C$+$D   &  0.3  &  18.0  &   0.029  &   1.12 &   0.39  	&   0.99 &  0.57  &	  4.623 &   0.032   &   0.405 &   0.051  &   0.485484 &   0.000004    &	 -24.85 \\
32  & gS0dS0o105    &	S0     & B$+$C$+$D   &  0.5  &  17.7  &   0.017  &   1.36 &   0.27 	&   1.37 &  0.69  &	  5.0 &   5.0   &   0.527 &   0.066  &   0.443 &   0.017    &	 -24.85 \\
33  & gS0dSao10	  &	S0     & B$+$C$+$D   &  0.5  &  18.4  &   0.023  &   1.34 &   0.14  	&   1.20 &  0.22  &	  5.332 &   0.011   &   0.505 &   0.010  &   0.449624 &   0.000004    &	 -24.91 \\
34  & gS0dSao103    &	S0     & B$+$C$+$D   &  0.5  &  18.4  &   0.017  &   1.43 &   0.35  	&   1.549 &  0.040  &	  5.458 &   0.017   &   0.532 &   0.020  &   0.430047 &   0.000009    &	 -24.91 \\
35  & gS0dSao105    &	S0     & B$+$C$+$D   &  0.5  &  18.3  &   0.021  &   1.32 &   0.16  	&   1.11 &  0.32  &	  5.219 &   0.043   &   0.476 &   0.015  &   0.457874 &   0.000006    &	 -24.91 \\
36  & gS0dSao106    &	S0     & B$+$C$+$D   &  0.5  &  17.9  &   0.018  &   1.14 &   0.15  	&   1.05 &  0.25  &	  5.04 &   0.74   &   0.404 &   0.012  &   0.451761 &   0.000014    &	 -24.91 \\
37  & gS0dSbo106    &	S0     & B$+$D       &  0.3  &  17.7  &   0.042  &   1.355 &   0.053  	&   1.073 &  0.010  &	  4.824 &   0.024   &   0.52 &   0.34  &   0.47 &  0.34    &	 -24.84 \\
38  & gS0dSdo100    &	S0     & B$+$D       &  0.3  &  18.6  &   0.032  &   1.529 &   0.015  	&   1.241 &  0.021  &	  6.082 &   0.036   &   0.5805 &   0.0026  &   0.4194 &   0.0026    &	 -24.87 \\
39  & gSagSao1	  &	S0     & B$+$C$+$D   &  0.5  &  26.9  &   0.229  &   2.139 &   0.034  	&   2.54 &  0.11  &	  7.329 &   0.040   &   0.674 &   0.011  &   0.263 &   0.011    &	 -26.10 \\
40  & gSagSao5	  &	S0     & B$+$D       &  0.3  &  23.9  &   0.264  &   1.540 &   0.062  	&   1.77 &  0.10  &	  4.9110 &   0.0057   &   0.54 &   0.17  &   0.45 &   0.17    &	 -26.13 \\
41  & gSagSao9	  &	S0     & B$+$D       &  0.5  &  23.1  &   0.328  &   1.235 &   0.086  	&   2.25 &  0.12  &	  4.922 &   0.017   &   0.637969 &   0.000002  &   0.362003 &   0.000001    &	 -26.13 \\
42  & gSagSbo1	  &	S0     & B$+$C$+$D   &  0.5  &  25.2  &   0.588  &   1.10 &   0.80  	&   1.3 &  1.6  &	  5.3 &   9.1   &   0.44 &   0.25  &   0.382913 &   0.000003    &	 -25.89 \\
43  & gSagSbo2	  &	S0     & B$+$C$+$D   &  0.5  &  20.6  &   0.262  &   1.46 &   0.73  	&   1.9 &  1.3  &	  5.3 &   9.8   &   0.56 &   0.16  &   0.38485 &   0.00018    &	 -25.90 \\
44  & gSagSbo5	  &	E/S0   & B$+$C$+$D   &  0.5  &  22.3  &   0.106  &   1.12 &   0.45  	&   1.2 &  1.2  &	  5.9 &  11.1   &   0.45 &   0.17  &   0.43 &   0.15    &	 -25.91 \\
45  & gSagSbo9	  &	S0     & B$+$C$+$D   &  1.7  &  24.3  &   0.132  &   1.797 &   0.097  	&   1.13 &  0.62  &	 10.04 &   0.81   &   0.497 &   0.075  &   0.331 &   0.031    &	 -25.92 \\
46  & gSagSbo21	  &	S0     & B$+$C$+$D   &  0.8  &  16.6  &   0.046  &   1.28 &   0.43  	&   2.13 &  0.62  &	  5.210 &   0.028   &   0.461 &   0.035  &   0.452291 &   0.000006    &	 -25.91 \\
47  & gSagSbo22	  &	S0     & B$+$C$+$D   &  0.5  &  22.7  &   0.266  &   0.50 &   0.21  	&   2.147 &  0.010  &	  4.920 &   0.023   &   0.398 &   0.010  &   0.403996 &   0.000001    &	 -25.92 \\
48  & gSagSbo24	  &	S0     & B$+$C$+$D   &  0.5  &  26.3  &   0.128  &   0.83 &   0.24  	&   3.983 &  0.010  &	  7.02 &   0.20   &   0.648 &   0.015  &   0.289063 &   0.000027    &	 -25.92 \\
49  & gSagSbo42	  &	E/S0   & B$+$C$+$D   &  1.1  &  25.1  &   0.138  &   0.86 &   0.39  	&   1.159 &  0.010  &	  6.001 &   0.065   &   0.306 &   0.010  &   0.450835 &   0.000001    &	 -25.89 \\
50  & gSagSbo43	  &	E/S0   & B$+$C$+$D   &  1.5  &  25.6  &   0.127  &   0.9 &   2.2  	&   1.16 &  0.58  &	  6.1 &   1.9   &   0.34 &   0.15  &   0.454 &   0.020    &	 -25.89 \\
51  & gSagSbo71	  &	E/S0   & B$+$C$+$D   &  0.5  &  20.7  &   0.114  &   0.701 &   0.089  	&   1.2830 &  0.0081  &	  4.5038 &   0.0098   &   0.3854 &   0.0052  &   0.435689 &   0.000001    &	 -25.88 \\
52  & gSagSdo2	  &	S0     & B$+$C$+$D   &  2.0  &  25.7  &   0.235  &   1.78 &   0.86  	&   1.2 &  1.5  &	  6.1 &   2.7   &   0.39 &   0.27  &   0.47 &   0.17    &	 -26.07 \\
53  & gSagSdo5	  &	S0     & B$+$C$+$D   &  0.7  &  30.9  &   3.771  &   0.95 &   0.10  	&   1.094 &  0.097  &	  5.4 &   6.9   &   0.393 &   0.059  &   0.469 &   0.058    &	 -26.07 \\
54  & gSagSdo9	  &	S0     & B$+$C$+$D   &  0.5  &  24.8  &   0.395  &   1.28 &   0.12  	&   2.065 &  0.405  &	  8.16 &   0.64   &   0.543 &   0.041  &   0.394 &   0.023    &	 -26.07 \\
55  & gSagSdo14	  &	E/S0   & B$+$D       &  1.0  &  28.3  &   1.085  &   2.70 &   0.49  	&   2.145 &  0.010  &	  16.4 &   1.0   &   0.651 &   0.019  &   0.348 &   0.019    &	 -26.03 \\
56  & gSagSdo18	  &	E/S0   & B$+$C$+$D   &  0.5  &  24.6  &   0.236  &   2.8 &   1.2  	&   1.03 & 0.44  &	 5.7 &   5.2   &   0.33 &   0.13  &	0.39 &   0.13	 &	-26.07 \\
57  & gSagSdo41	  &	E/S0   & B$+$C$+$D   &  1.0  &  25.0  &   0.346  &   1.47 &   0.27  	&   2.54 & 0.87  &	 9.81 &   0.70   &   0.621 &   0.096  &	0.280 &   0.087	 &	-26.04 \\
58  & gSagSdo42	  &	S0     & B$+$C$+$D   &  1.0  &  31.7  &   3.530  &   1.01 &   0.11  	&   1.452 & 0.010  &	 5.4482 &   0.0078   &   0.3864 &   0.0016  &	0.505225 &   0.000001	 &	-26.04 \\
59  & gSagSdo43	  &	E/S0   & B$+$C$+$D   &  1.2  &  27.9  &   0.714  &   1.10 &   0.75  	&   1.10 & 0.49  &	 5.8 &   2.5   &   0.353 &   0.052  &	0.530 &   0.055	 &	-26.06 \\
60  & gSagSdo70	  &	E/S0   & B$+$C$+$D   &  1.0  &  25.7  &   0.352  &   1.05 &   0.25  	&   1.16 & 0.81  &	 8.0 &   1.8   &   0.395 &   0.089  &	0.401 &   0.049	 &	-26.04 \\
61  & gSagSdo71	  &	E/S0   & B$+$C$+$D   &  1.5  &  30.1  &   0.434  &   1.19 &   0.45  	&   1.4 & 1.5  &	10.0 &   1.6   &   0.39 &   0.16  &	0.323 &   0.091	 &	-26.05 \\
62  & gSagSdo73	  &	E/S0   & B$+$D       &  1.5  &  26.5  &   0.364  &   1.8596 &   0.0085  &   2.7784 & 0.0083  &	 6.786 &   0.010   &   0.661 &   0.010  &	0.338 &   0.010	 &	-26.08 \\
63  & gSbgSbo9	  &	S0     & B$+$C$+$D   &  0.7  &  23.7  &   0.093  &   1.00 &   0.26  	&   1.012 & 0.010  &	 9.78 &   3.9   &   0.323 &   0.022  &	0.394 &   0.036	 &	-25.61 \\
64  & gSbgSbo16	  &	E/S0   & B$+$C$+$D   &  1.0  &  23.3  &   0.117  &   2.50 &   0.35  	&   2.00 & 0.36  &	11.568 &   0.99   &   0.56 &   0.10  &	0.361 &   0.019	 &	-25.63 \\
65  & gSbgSbo17	  &	E/S0   & B$+$D       &  1.5  &  22.5  &   0.121  &   2.641 &   0.047  	&   1.112 & 0.068  &	 5.335 &   0.20   &   0.609 &   0.025  &	0.390 &   0.025	 &	-25.62 \\
66  & gSbgSbo19	  &	E/S0   & B$+$C$+$D   &  0.5  &  20.6  &   0.080  &   3.79 &   0.71  	&   3.25 & 0.44  &	 4.4415 &   0.0019   &   0.29 &   0.10  &	0.391338 &   0.000004	 &	-25.62 \\
67  & gSbgSbo22	  &	S0     & B$+$D       &  1.0  &  22.6  &   0.196  &   1.43 &   0.11  	&   2.42 & 0.20  &	 6.2040 &   0.0051   &   0.61 &   0.23  &	0.38 &   0.23	 &	-25.63 \\
\hline\\
\pagebreak
68  & gSbgSbo41	  &	E/S0   & B$+$C$+$D   &  0.5  &  20.6  &   0.055  &   3.02 &   0.38  	&   1.00 & 0.14  	 &	 4.6704 &   0.0035   &   0.170 &   0.060  &	0.535280 &   0.000001	 &	-25.61 \\
69  & gSbgSbo42	  &	E/S0   & B$+$C$+$D   &  0.5  &  21.3  &   0.106  &   0.62 &   0.61  	&   2.61 & 0.54  &	 5.5 &   2.5   &   0.462 &   0.057  &	0.396057 &   0.000010	 &	-25.60 \\
70  & gSbgSbo69	  &	E/S0   & B$+$C$+$D   &  2.5  &  24.1  &   0.105  &   3.14 &   0.22  	&   1.02 & 0.45  &	 9.5 &   3.1   &   0.597 &   0.051  &	0.293 &   0.055	 &	-25.60 \\
71  & gSbgSbo70	  &	E/S0   & B$+$C$+$D   &  1.5  &  24.2  &   0.121  &   1.32 &   0.61  	&   1.1 & 1.8  &	12.1 &   2.4   &   0.37 &   0.18  &	0.349 &   0.094	 &	-25.59 \\
72  & gSbgSbo72	  &	S0     & B$+$D       &  2.0  &  24.2  &   0.108  &   1.9 &   1.0 	&   2.580 & 0.010  &	 9.2 &   2.9   &   0.57 &   0.16  &	0.42 &   0.16	 &	-25.65 \\
73  & gSbgSdo5	  &	S0     & B$+$C$+$D   &  0.5  &  22.6  &   0.456  &   0.791 &   0.040  	&   1.32 & 0.21  &	 5.23 &   0.20   &   0.402 &   0.026  &	0.426 &   0.019	 &	-25.84 \\
74  & gSbgSdo9	  &	S0     & B$+$C$+$D   &  0.5  &  24.6  &   0.181  &   0.66 &   0.74  	&   1.91 & 0.74  &	 6.6 &   2.9   &   0.39 &   0.11  &	0.442 &   0.063	 &	-25.82 \\
75  & gSbgSdo14	  &	S0     & B$+$C$+$D   &  1.5  &  22.8  &   0.453  &   2.2 &   3.3  	&   1.0 & 1.8  		 &	14.6 &   9.5   &   0.30074 &   0.00061  &	0.38 &   0.21	 &	-25.81 \\	
76  & gSbgSdo17	  &	S0     & B$+$C$+$D   &  0.5  &  21.8  &   0.659  &   0.991 &   0.095  	&   2.202 & 0.064  &	 4.795 &   0.035   &   0.524288 &   0.000002  &	0.475646 &   0.000001	 &	-25.83 \\
77  & gSbgSdo18	  &	S0     & B$+$C$+$D   &  1.5  &  24.9  &   0.560  &   2.3 &   1.4  	&   1.253 & 2.080  &	 8.8 &  12.1   &   0.47 &   0.28  &	0.41 &   0.19	 &	-25.83 \\
78  & gSbgSdo19	  &	S0     & B$+$C$+$D   &  0.3  &  26.6  &   0.395  &   2.24 &   0.49  	&   1.00 & 0.36  &	 5.6 &   4.3   &   0.511 &   0.094  &	0.467 &   0.094	 &	-25.81 \\
79  & gSbgSdo23	  &	E/S0   & B$+$C$+$D   &  1.0  &  26.4  &   0.399  &   4.29 &   0.52  	&   1.00 & 0.29  	 &	 6.8 &   2.3   &   0.40 &   0.10  &	0.361 &   0.042	 &	-25.85 \\
80  & gSbgSdo41	  &	S0     & B$+$D       &  0.3  &  24.4  &   0.311  &   2.989 &   0.082  	&   1.153 & 0.074  &	 6.0 &   1.3   &   0.607 &   0.015  &	0.392 &   0.015	 &	-25.81 \\
81  & gSbgSdo69	  &	S0     & B$+$C$+$D   &  0.3  &  26.0  &   0.356  &   2.993 &   0.075  	&   1.08 & 0.10  &	 7.98 &   0.39   &   0.470 &   0.042  &	0.392 &   0.010	 &	-25.81 \\
82  & gSbgSdo70	  &	S0     & B$+$C$+$D   &  2.5  &  23.8  &   0.423  &   3.0 &   1.0  	&   1.7 & 1.4  &	13.9 &  10.2   &   0.53 &   0.19  &	0.44 &   0.18	 &	-25.82 \\
83  & gSbgSdo71	  &	E/S0   & B$+$C$+$D   &  1.2  &  23.4  &   0.695  &   1.15 &   0.43  	&   2.1 & 1.7  &	 8.0 &   1.8   &   0.47 &   0.18  &	0.33 &   0.11	 &	-25.82 \\
84  & gSdgSdo1	  &	E/S0   & B$+$C$+$D   &  1.0  &  27.4  &   1.801  &   0.7 &   5.0  	& \multicolumn{2}{c}{4.0} &	 8.2 &   2.2   &   0.282 &   0.095  &	0.404461 &   0.000005	 &	-26.03 \\
85  & gSdgSdo2	  &	S0     & B$+$C$+$D   &  1.3  &  27.3  &   0.267  &   2.93 &   0.68  	&   1.03 & 0.26  &	 8.0 &   3.5   &   0.39 &   0.18  &	0.45 &   0.11	 &	-26.04 \\
86  & gSdgSdo5	  &	S0     & B$+$C$+$D   &  2.0  &  27.8  &   0.559  &   2.2 &   5.8  	&   1.0 & 1.2  	 &	11.2 &   3.1   &   0.24 &   0.28  &	0.42 &   0.21	 &	-26.02 \\
87  & gSdgSdo9	  &	S0     & B$+$C$+$D   &  1.3  &  28.3  &   0.654  &   2.32 &   0.34  	&   3.01 & 0.14  &	 6.380 &   0.042   &   0.4690 &   0.0088  &	0.496339 &   0.000002	 &	-26.03 \\
88  & gSdgSdo16	  &	E/S0   & B$+$D       &  0.3  &  31.8  &   2.354  &   2.98 &   0.16  	&   1.197 & 0.093  &	 6.67 &   0.42   &   0.466 &   0.033  &	0.533 &   0.033	 &	-25.97 \\
89  & gSdgSdo17	  &	S0     & B$+$C$+$D   &  1.0  &  29.4  &   2.561  &   3.11 &   0.50  	&   1.07 & 0.34  &	 5.3 &   7.9   &   0.36 &   0.10  &	0.446 &   0.063	 &	-26.00 \\
90  & gSdgSdo21	  &	E/S0   & B$+$C$+$D   &  1.5  &  29.5  &   0.411  &   0.9 &   1.4  	&   1.467 & 0.010  &	 7.9 &   2.6   &   0.381 &   0.042  &	0.460 &   0.039	 &	-25.99 \\
91  & gSdgSdo42	  &	S0     & B$+$C$+$D   &  1.0  &  25.5  &   0.639  &   0.251 &   0.010  	&   2.951 & 0.010  &	 5.68 &   0.23   &   0.401019 &   0.000013  &	0.466 &   0.025	 &	-25.99 \\
92  & gSdgSdo45	  &	E/S0   & B$+$C$+$D   &  1.0  &  24.4  &   0.286  &   2.383 &   0.048  	&   1.158 & 0.010  &	 5.6 &   1.8   &   0.345 &   0.022  &	0.439 &   0.029	 &	-25.99 \\
93  & gSdgSdo51	  &	E/S0   & B$+$C$+$D   &  1.5  &  26.1  &   0.257  &   1.2 &   1.0  	&   1.41 & 0.51  &	 6.7 &   1.1   &   0.333 &   0.036  &	0.51926 &   0.00088	 &	-26.07 \\
94  & gSdgSdo69	  &	S0     & B$+$C$+$D   &  1.0  &  28.3  &   1.122  &   3.12 &   0.17  	&   1.000 & 0.066  	 &	 5.2505 &   0.0055   &   0.121 &   0.017  &	0.562326 &   0.000001	 &	-25.97 \\
95  & gSdgSdo71	  &	S0     & B$+$C$+$D   &  0.3  &  27.1  &   0.689  &   2.627 &   0.021  	&   1.264 & 0.052  &	11.68 &   0.14   &   0.6520 &   0.0093  &	0.3090 &   0.0058	 &	-26.00 \\
96  & gSdgSdo74	  &	S0     & B$+$C$+$D   &  1.0  &  30.2  &   0.838  &   3.4 &   3.9  	& \multicolumn{2}{c}{4.0} &	 8.9 &   1.8   &   0.671 &   0.097  &	0.275 &   0.074	 &	-26.01 \\ \hline\\\vspace{-0.6cm}
 \end{longtable} 
\end{center}
}
\centering
\begin{minipage}[t]{24cm}
{\footnotesize
\emph{Columns}: (1) Number ID. (2) Model code: g[\emph{type1}]g[\emph{type2}]o[\emph{\#orbit}], see Sect\,\ref{Sec:models}. (3) Visual morphological type assigned according to realistic broad-band simulated images
(see Sect.\,\ref{Sec:identification}). (4) Multicomponent decomposition performed to the $K$-band surface brightness profile: bulge$+$disc (B$+$D) or bulge$+$[inner component]$+$disc (B$+$C$+$D). We have assumed only an spheroidal component to fit the gE0 progenitor (B) and just an exponential disc in the gSd progenitor (D). The additional inner component has been modelled with a S\'{e}rsic profile, and it may correspond to a lense, oval, bar, inner disc, or to several of them depending on the case (see Sect.\,\ref{Sec:decompositions}). (5) Minimum radius of data included in the fit, in kpc. (6) Maximum radius of data included in the fit, in kpc. (7) $\chi^2$ of the fit, in mag$^2$. (8) Bulge effective radius \re, in kpc. (9) Bulge S\'{e}rsic index $n$. (10) Disc e-folding scalelength \hd, in kpc. (11) Bulge-to-total luminosity ratio. (12) Disc-to-total luminosity ratio. (13) Total absolute magnitude of the galaxy in the $K$ band. All parameters have been derived from the multicomponent decompositions performed to the simulated $K$-band surface brightness profiles of the stellar progenitors and remnants. The bulge S\'{e}rsic index had to be fixed to $n=4$ in two models (no errors are thus indicated in these cases).
}
\end{minipage}

\end{landscape}

\clearpage
\twocolumn


\end{document}